\documentclass[letterpaper,11pt]{article}
\pdfoutput=1
\usepackage{jheppub}

\usepackage{amsmath,graphicx}
\usepackage{amssymb}
\usepackage{color}
\usepackage{xspace}
\usepackage[small]{subfigure}
\usepackage[utf8]{inputenc} 

\newcommand{\fms}[1]{{#1}\!\!\!/}

\newcommand{\n}{\overline{n}}

\newcommand{\mP}{\mathcal{P}}

\newcommand{\nbstwo}{\frac{\fms{\overline{n}}}{2}}

\newcommand{\SCETI}{\mathrm{SCET_I}}
\newcommand{\SCETII}{\mathrm{SCET_{II}}}
\newcommand{\SCETM}{\mathrm{SCET_M}}

\newcommand{\nb}{{\bar{n}}}

\newcommand{\df}{\mathrm{d}}
\newcommand{\msbar}{\overline{\textrm{MS}}}

\newcommand{\LambdaQCD}{\Lambda_{\textrm{QCD}}}

\def\bnslash{\bar n\!\!\!\slash}

\def\vslash{v\!\!\!\slash}
\def\OMIT#1{}

\newcommand{\nn}{\nonumber} 

\newcommand{\bn}{{\bar n}}
\newcommand{\beq}{\begin{equation}}
\newcommand{\eeq}{\end{equation}}
\newcommand{\bea}{\begin{eqnarray}}
\newcommand{\eea}{\end{eqnarray}}

\newcommand{\SCETa}{\mbox{${\rm SCET}_{\rm I}$ }}

\def\slashchar#1{\setbox0=\hbox{$#1$}           
  \dimen0=\wd0                                    
  \setbox1=\hbox{/} \dimen1=\wd1                  
  \ifdim\dimen0>\dimen1                           
    \rlap{\hbox to \dimen0{\hfil/\hfil}}            
    #1                                             
  \else                                          
    \rlap{\hbox to \dimen1{\hfil$#1$\hfil}}        
    /                                           
 \fi}                                           %

\begin{document}


\title{Effective field theory approach to heavy quark fragmentation}

\def\ARIZONA{Department of Physics, University of Arizona, Tucson, AZ 85721, USA}
\def\CHUL{Institute of Convergence Fundamental Studies and School of Liberal Arts,
Seoul National University of Science and Technology, Seoul 01811, Korea}
\def\MAINZ{PRISMA Cluster of Excellence \& Mainz Institute for Theoretical Physics, Johannes Gutenberg University, 55099 Mainz, Germany}
\def\LANL{Theoretical Division, Los Alamos National Laboratory Los Alamos, NM 87545, U.S.A.}

\author[a]{Michael Fickinger,}
\emailAdd{michael.fickinger@uni-mainz.de}
\affiliation[a]{\MAINZ}

\author[b]{Sean Fleming,}
\affiliation[b]{\ARIZONA}
\emailAdd{fleming@physics.arizona.edu}

\author[c]{Chul Kim,}
\emailAdd{chul@seoultech.ac.kr}
\affiliation[c]{\CHUL}

 
\author[d]{Emanuele Mereghetti}
\emailAdd{emereghetti@lanl.gov}
 \affiliation[d]{\LANL}


\abstract{
Using an approach based on Soft Collinear Effective Theory (SCET) and Heavy Quark Effective Theory (HQET) we determine the $b$-quark fragmentation function from electron-positron annihilation data at the $Z$-boson peak at next-to-next-to leading order with next-to-next-to leading log resummation of DGLAP logarithms, and  next-to-next-to-next-to leading log resummation of endpoint logarithms. This analysis improves, by one order, the previous extraction of the $b$-quark fragmentation function. We find that while the addition of the next order in the calculation does not much shift the extracted form of the fragmentation function, it does reduce theoretical errors indicating that the expansion is converging. Using an approach based on effective field theory allows us to systematically control theoretical errors. While the fits of theory to data are generally good, the fits seem to be hinting that higher order correction from HQET may be needed to explain the $b$-quark fragmentation function at smaller values of momentum fraction.
}

\maketitle


\section{Introduction} \label{intro}

The production of heavy flavored particles in collider experiments has been of great interest since the discovery of the charm quark. Because the heavy quark mass $m_Q$ is much larger than the hadronization scale $\LambdaQCD$, aspects of 
heavy quark production 
can be calculated within perturbation theory, which provides a clean test of QCD. 
A process of particular importance is the single inclusive production of heavy flavored mesons such as $B$ or $D$ mesons. 
At energies large compared to the meson mass, the single inclusive cross section is factored into the convolution of a short distance cross section and a  
fragmentation function \cite{Ellis:1978ty,Collins:1981ta}. The fragmentation function 
describes the probability of a parton produced in the hard scattering  to hadronize into a heavy meson with a fraction of the parton  momentum, and an inclusive sum of other particles. 
The important observation of Ref.~\cite{Mele:1990cw} was that the presence of the heavy quark mass allows the heavy meson fragmentation functions to be expressed in terms of partonic fragmentation functions,
which  describe the evolution of partons into heavy quarks and have a perturbative expansion in $\alpha_s(m_Q)$.
The partonic fragmentation functions are then convoluted with a universal factor for the hadronization of a heavy quark into a heavy meson of the same flavor. This feature greatly reduces the number of independent nonperturbative functions to be extracted from data \cite{Mele:1990cw,Cacciari:2005uk}. Once these heavy quark fragmentation functions (HQFFs) are determined from $e^+ e^-$ data, through reliable QCD factorization formulae we can predict the heavy meson production cross section at hadron colliders without further nonperturbative input. 

As HQFFs are an essential ingredient in calculations of inclusive heavy meson production at collider experiments, they must be determined with care. The importance of a precise extraction of the HQFF was made clear by the resolution of a fifteen year discrepancy between theory predictions~\cite{Nason:1987xz,Nason:1989zy,Beenakker:1990maa} and data on the transverse momentum spectrum of bottom quarks in hadronic collisions~\cite{Albajar:1986iu,Albajar:1990zu,Abe:1993sj,Abe:1993hr,Abe:1994qk,Acosta:2002qk,Abe:1995dv,Acosta:2001rz,Abachi:1994kj,Abbott:1999wu,Abbott:1999se,Abbott:2000iv},
with data exceeding  theory by a factor of 2-3. In Refs.~\cite{Cacciari:2002pa,Cacciari:2003uh} a prediction based on a combined next-to-leading order (NLO) calculation of $b$-quark production with next-to-leading log (NLL) resummation of $p_T/m_{b}$ (with $p_T$ the $b$-quark transverse momentum)~\cite{Cacciari:1998it}, and a similarly precise (NLO+NLL) extraction of the heavy quark fragmentation function from $e^{+}e^{-}$ annihilation experiments~\cite{Nason:1999zj,Mele:1990cw,Colangelo:1992kh} was shown to agree quite well with the data. Furthermore, it was noted that the accuracy of the fragmentation function was important for obtaining agreement with data, and that the transverse momentum spectrum of bottom quarks in hadronic collisions was particularly sensitive to the $N=5$ moment of the fragmentation function, which previously had not been determined precisely.

Subsequent updates from D\O\ and CDF experiments~\cite{Lewis:2014cka} and new data from ATLAS~\cite{ATLAS:2013cia}, CMS~\cite{Khachatryan:2011mk} and LHCb~\cite{Aaij:2012jd} experiments on $B$-meson production continue to find good agreement between theory and experiment, as summarized in Ref.~\cite{Cacciari:2012ny}. 
However, over a wide range of the heavy meson transverse momentum ($0 \le p_T \lesssim 40$ GeV), error bars from experiments are still overwhelmed by theory errors.
This puts the onus on the theory community to provide a higher precision result, i.e., N${}^2$LO+N${}^2$LL 
improved calculation of $B$-meson production in hadronic collisions. While this is a daunting endeavor it is not beyond the realm of possibility. Actually, for other processes such as top pair production, important progress has been made on pieces of the calculation~\cite{Baernreuther:2012ws,Czakon:2012zr,Czakon:2012pz,Czakon:2013goa}. 

In this paper we focus on the extraction of the $b$-quark fragmentation function at N${}^2$LO.
A precise extraction of the HQFF from $e^+e^-$ annihilation data is complicated by the presence of a number of disparate energy scales. Away from the endpoint $x \to 1$ there are two relevant scales: the large center-of-mass energy of the collision $Q$, and the heavy quark mass, $m_Q$.
To achieve N${}^2$LO accuracy in this region we need the partonic fragmentation functions at $\mathcal O(\alpha_s^2)$, which have been computed in Refs.   \cite{Melnikov:2004bm,Mitov:2004du}.
Large single logarithms of $Q^2/m_Q^2$ are resummed via DGLAP evolution~\cite{Gribov:1972ri,Altarelli:1977zs,Dokshitzer:1977sg}. The calculation of the time-like splitting functions at three loops,
accomplished in Refs. \cite{Mitov:2006ic,Moch:2007tx,Almasy:2011eq}, 
makes possible the   resummation of these large logarithms at N${}^2$LL.

The HQFF, however, is dominated by contributions from the endpoint region $x\sim 1$, where the heavy quark carries most of the energy of the parton emerging from the hard scattering. 
For $x\sim1$, three additional scales become relevant, the jet scale $Q\sqrt{1-x}$, which describes the invariant mass of the particles against which the heavy quark recoils, the soft scale $Q(1-x)$, and the hadronic scale $\LambdaQCD$, which describe the hadronization of the heavy quark into a meson. 
In order to provide accurate theoretical predictions in this region it is necessary to resum double logarithms of $(1-x)$ that appear both in the fragmentation function
and in the partonic cross section \cite{Cacciari:2001cw,Cacciari:2005uk}. Furthermore, since the HQFF contains both perturbative and nonperturbative effects, a systematic approach is needed not only to separate them but also to maintain universality.

In this work we use an effective field theory (EFT) approach to study the HQFF,
and to derive a factorization formula, valid in the full $x$ range, that allows us to extract the HQFF from data on $e^+e^- \to B+X$ at the $Z$ pole.
In Section~\ref{SCET} we begin with a review of the EFTs we use in our analysis; namely 
Soft Collinear Effective Theory~(SCET)~\cite{Bauer:2000ew,Bauer:2000yr,Bauer:2001ct,Bauer:2001yt,Bauer:2002nz,Leibovich:2003jd,Chay:2005ck},
and boosted Heavy Quark Effective Theory (bHQET)~\cite{Fleming:2007qr,Fleming:2007xt}.
In Section~\ref{sec:fullx} we consider the single inclusive cross section away from the endpoint region $x \sim1$. In this regime we rederive the established  perturbative QCD result that the differential cross section can be expressed as a convolution of a hard coefficient and the heavy meson fragmentation functions for various partonic species. The bHQET expansion at leading power in $\LambdaQCD/m_Q$ then lets us express the HQFF as a product of perturbative functions originating from physics at the heavy quark mass scale and an overall nonperturbative coefficient.

In Section~\ref{sec:xto1} we consider the factorization theorem in the endpoint region. We use a three step procedure. In the first step, which is discussed in Section~\ref{sec:SCET1}, we match QCD onto $\SCETI$ integrating out virtualities of order $Q^2$. Near the endpoint, the heavy quark recoils against a collimated spray of particles  
of invariant mass squared $Q^2(1-x)$, which is still dynamical in $\SCETI$. 
In the second step, in Section~\ref{sec:SCETM}, we match $\SCETI$ onto $\SCETM$~\cite{Leibovich:2003jd,Chay:2005ck}.
This integrates out the jet at virtuality $Q^2(1-x)$, and introduces dependences on the heavy quark mass $m_Q$.
After discussing the crossing of the $b$ threshold in Section~\ref{sec:flavorthr}, in Section~\ref{sec:bHQET} 
we integrate out the heavy quark mass $m_Q$, by matching $\SCETM$ onto bHQET. We thus arrive at the final factorization formula in the endpoint,
and express the cross section as the product of a hard coefficient $H_Q$, a jet function $J$, a mass coefficient $C_m$, and a shape function $S_{H/Q}$.
Each of these objects depends on a single scale, namely the hard scale $Q$, the jet scale $Q\sqrt{1-x}$, the mass scale $m_Q$ and the soft scale $Q(1-x)$ which in the rest frame of the $B$-meson becomes $m_{Q}(1-x) \sim \LambdaQCD$.
In Section \ref{resum} we describe the renormalization group equations (RGEs) that govern the scale dependence, and resum large logarithms of $1-x$ and $m_Q/Q$ by solving the RGEs.  
The results in  Section~\ref{sec:xto1}  complete the analysis of Ref. \cite{Neubert:2007je}, which first 
derives the factorization of the HQFF using SCET.

In Section \ref{sec:np} we discuss how to separate the perturbative and nonperturbative components of the shape function $S_{H/Q}$, which describes the hadronization of the heavy quark into an heavy meson.
In Section~\ref{Merge}, to give a full description of the differential cross section, we combine the two factorization theorems for moderate and large $x$.
Crucial to this is the notion of profile functions first introduced in Ref.~\cite{Abbate:2010xh}. 
Finally, in Section~\ref{sec:err} we perform a fit to data from the LEP experiments ALEPH \cite{Heister:2001jg}, OPAL \cite{Abbiendi:2002vt},  and DELPHI \cite{Abdallah:2011ep}, and from the SLAC experiment SLD \cite{Abe:2002iq}. 
We discuss in detail the impact of theoretical uncertainties on the fits. 
We conclude in Section~\ref{conclude}. 
In Appendix~\ref{Equations} we collect the fixed order expressions of the various functions that enter the factorized resummed cross section, the 
anomalous dimensions, and give the solution of the RGEs.

\section{Effective Field Theories}\label{SCET}

One of our main goals is to derive factorization theorems for the inclusive production cross section of a heavy hadron $H$ using a series of EFTs with degrees of freedom of progressively smaller off-shellness.
In this section we briefly summarize the most important ingredients of each EFT, establish our notation, and refer to the original literature for more details. 

\subsection{Soft Collinear Effective Theory}

Soft Collinear Effective Theory (SCET) \cite{Bauer:2000ew,Bauer:2000yr,Bauer:2001ct,Bauer:2001yt,Bauer:2002nz},
and its generalization to massive quarks ($\SCETM$)   \cite{Leibovich:2003jd,Chay:2005ck},
is an effective theory for fast moving, almost light-like, quarks and gluons, and their interactions with soft degrees of freedom. 
It has been successfully applied to a variety of processes, from $B$ decays to jet physics, 
with recent applications being to the fragmentation of light and heavy hadrons, mostly in the context of fragmentation inside a jet \cite{Procura:2009vm,Jain:2011xz,Bauer:2013bza,Baumgart:2014upa,Chien:2015ctp,Bain:2016clc,Kang:2016mcy,Kang:2016ehg,Dai:2016hzf}.

In high energy collisions a hard scattering process is sensitive to several, well separated, physical scales. 
The short distance dynamics is governed by a hard scale $Q$, for example the center-of-mass energy in 
$e^+ e^-$ annihilation. After the creation of high energy partons, their evolution into hadrons or jets of hadrons  happens on much longer distances,
and is sensitive to collinear and soft scales.
SCET takes advantage of this scale separation.  Degrees of freedom with virtuality of order  $Q^2$ are integrated out, 
leaving as dynamical degrees of freedom  collinear  quarks and gluons, with virtuality $p^2  \sim Q^2 \lambda^2$, and ultrasoft (usoft) quarks and gluons, 
with even smaller virtuality
$p^2 \sim Q^2 \lambda^4$. The SCET expansion is governed by power counting in the parameter $\lambda \sim Q_\textrm{\tiny LO}/Q \ll 1$, with $Q_\textrm{\tiny LO}$ the next relevant scale in the problem, e.g. a jet invariant mass.
In SCET different collinear sectors can only interact by exchanging  usoft degrees of freedom.
An important property of SCET is that usoft-collinear interactions can be moved from the SCET Lagrangian to matrix elements of external operators through a field redefinition~\cite{Bauer:2001yt},  which greatly simplifies derivations of factorized forms for observables.

We now summarize some SCET ingredients needed in the rest of the paper. For more details, we refer to the original papers \cite{Bauer:2000ew,Bauer:2000yr,Bauer:2001ct,Bauer:2001yt,Bauer:2002nz,Leibovich:2003jd,Chay:2005ck}.
We introduce two lightcone vectors   $n^{\mu}$ and $\bar n^{\mu}$, satisfying $n^2 =\bar n^2 = 0$, and $\bar n \cdot n  =2$.
The momentum of a particle can be decomposed in lightcone coordinates according  to
\begin{equation}
p^{\mu} = p^- \frac{n^{\mu}}{2} + p^+ \frac{\bar n^{\mu}}{2} + p_{\perp}^{\mu}\,.
\end{equation}
Particles collinear to the jet axis have $(p^+,p^-,p_{\perp}) \sim Q ( \lambda^2, 1,\lambda)$, while
usoft quarks and gluons have all components of the momentum roughly of the same size $(p^+,p^-,p_{\perp}) \sim Q (\lambda^2,\lambda^2,\lambda^2)$.

The SCET Lagrangian can be written as
\begin{equation}
\mathcal L_{\textrm{SCET}} = \sum_i \mathcal L_{n_i} + \mathcal L_{\textrm{us}}\,,
\end{equation}
where $\mathcal L_{\textrm{us}}$ is the usoft Lagrangian which has the same form as the QCD Lagrangian. 
Each collinear sector is described by a copy of the collinear Lagrangian $\mathcal L_n$, which
for  massless quarks is
\begin{equation}\label{SCET1}
\mathcal L_{n} = \bar \xi_{n} \left( i n \cdot D_n + g  n \cdot A_{us} + \left( \slashchar{\mathcal P}_{ \perp} + g \slashchar{A}_{n\perp} \right) W_n \frac{1}{\bar n \cdot \mathcal P} W_n^{\dagger} 
(\slashchar{\mathcal P}_{\perp} + g \slashchar{A}_{n\perp}) \right )\frac{\slashchar{\bar n}}{2}\xi_n\,,
\end{equation}
where $\xi_n$ and $ A_{n}$ are collinear quark and gluon fields, labeled by the lightcone direction $n$ and by the large components of their momentum $ \tilde p  = (p^-, p_{\perp})$.
We leave the momentum  label mostly implicit, unless needed.
The label momentum operator  $\mathcal P^{\mu}$ acting on collinear fields returns the value of the label, for example
\begin{equation}
\mathcal P^{\mu} \xi_{n,\tilde{p}}  = \left(  p^- \frac{n^{\mu}}{2} + p_{\perp}^{\mu}  \right) \xi_{n, \tilde{p}}\,.
\end{equation}
The collinear covariant derivative $D_n$ is defined as
\begin{equation}
i D_n^{\mu} = \left( \bar n \cdot  \mathcal P + g \bar n \cdot A_{n} \right)\frac{n^{\mu}}{2} 
+ \left( i n \cdot \partial + g n \cdot A_n \right) \frac{\bar n^{\mu}}{2} +   \mathcal P^{\mu}_{\perp} + g A^{\mu}_{n\perp} \, .
\end{equation}
The Wilson line $W_n$ in Eq.~(\ref{SCET1}) is constructed from collinear gluon fields, 
\begin{equation}
W_n (x) = \sum_{\textrm{perms}} \exp \left( -\frac{g}{\bar n \cdot \mathcal P} \bar n \cdot A_n (x)\right)\,,
\end{equation}
and obeys the equation of motion $[\bn\cdot D_n , W_n(x)] =0$. Finally, $A^{\mu}_{us}$ in Eq.~(\ref{SCET1}) is a usoft gluon field, and 
at leading order in $\lambda$ couples to collinear quarks only through the $n \cdot A_{us}$ term. This coupling between usoft and collinear fields can be eliminated from the Lagrangian via the BPS field redefinition \cite{Bauer:2001yt}:
\begin{eqnarray}
\xi_n^{(0)}(x) &=& Y_n(x) \xi_{n}(x)\, , \qquad 
A_n^{(0)}(x) = Y_n(x) A_{n}(x) Y^{\dagger}_{n}(x)\,,
\end{eqnarray}
 where $Y_n$ is a usoft Wilson line in the $n$ direction
\begin{align}\label{Ysoft}
Y_n(y)&=\bar{P} \left\{ \exp\!\left[-i\, g\int_0^\infty \!{\rm d}s\, n\cdot A(s n^\mu+y^\mu)\right] \right\}\,, \nn\\
Y_n^\dagger(y)&=P \left\{ \exp\!\left[i\, g\int_0^\infty \!{\rm d}s\, n\cdot A(s n^\mu+y^\mu)\right] \right\}\,,  
\end{align} 
with $P$($\bar{P} $) denoting path (anti-path) ordering. Note that we chose the integration path of the Wilson line to extend to positive infinity as this is the physical direction for hadronic production in $e^{+}e^{-}$ annihilation. As is discussed throughly in Ref.~\cite{Chay:2004zn,Arnesen:2005nk} the choice of boundary condition for the soft Wilson lines in the BPS field redefinition 
is arbitrary, however non-physical choice induce boundary Wilson lines, which ``force'' the physical direction as we have chosen. 
If one were to insist on non-physical directions of the Wilson lines in the factorization theorem (not just in the BPS field redefinition), then he would have to include nonzero Glauber contributions in the matching calculation~\cite{Rothstein:2016bsq}. 
The effect of the field redefinition is to eliminate the usoft gluon field in Eq.~\eqref{SCET1}, and to replace the collinear quark and gluon fields $\xi_n$ and $A_{n}$ with their noninteracting counterparts.
The same field redefinition also decouples usoft gluons from collinear gluons \cite{Bauer:2001yt}. From here on we always use decoupled collinear fields, and drop the superscript $(0)$.

Using the Wilson line $W_n$ it is possible to construct gauge invariant combinations of collinear fields. For example, the  gauge invariant quark and gluon fields 
for particles moving in the $n$ or $\bar n$ direction are defined as 
\begin{eqnarray}\label{chifield}
\chi_{ {n}, {\omega}}=\delta_{ {\omega},  \bar n \cdot \mathcal P   }W_n^\dagger\xi_n, \qquad & &   \qquad \chi_{\bar{n},\bar{\omega}}=\delta_{\bar{\omega},  n \cdot \mathcal P }W_\nb^\dagger\xi_\nb,  \nonumber \\
\mathcal B_{n \perp,\, \omega}  = \delta_{ {\omega},  \bar n \cdot \mathcal P   } \frac{1}{g} W^\dagger_n \, i D^\mu_{n\perp} W_n, 
\quad  & & \quad  \mathcal B_{\bar{n} \perp,\, \omega}  = \delta_{ {\bar\omega},   n \cdot \mathcal P   } \frac{1}{g} W^\dagger_{\bar n} \, i D^\mu_{\bar{n}\perp} W_{\bar n}.
\end{eqnarray}
These fields  serve as building blocks of gauge invariant operators, like jet or fragmentation functions, as we discuss later.

\subsection{SCET$_\textrm{M}$}

For fast moving massive particles  there are additional mass terms in SCET, which appear in the Lagrangian as~\cite{Leibovich:2003jd}
\begin{equation}
\mathcal L_{m} = m_Q \bar \xi_n \left[ \left(\slashchar{\mathcal P}_{\perp} + g \slashchar A_{n\perp}\right) , W_n \frac{1}{\bar n \cdot \mathcal P} W_{n}^{\dagger} \right]\frac{\slashchar{\bar n}}{2}\xi_n - m^2_Q \bar \xi_n W_{n} \frac{1}{\bar n \cdot \mathcal P} W_{n}^{\dagger} \frac{\slashchar{\bar n}}{2}\xi_n\,.
\end{equation}
Usually the theory with the mass terms is referred to as SCET$_\textrm{M}$, a useful shorthand which we will adopt as well.
We work with one massive quark  with mass $m_Q$, treat the remaining $n_f - 1$ flavors as massless, and assume that quarks heavier than $m_Q$ have been integrated out.  
We use $q$ to denote both heavy and light quarks when it is not necessary to specify the quark mass, and use 
$Q$ ($\bar Q$) exclusively for heavy quarks (antiquarks), while $l$ ($\bar l$) denotes the $n_l =  n_f - 1$ light quark flavors. Depending on the power counting, mass terms can be either leading or subleading in $\lambda$.

The combination of collinear and usoft degrees of freedom we have discussed so far is usually referred to as $\SCETI$. There are, however, additional degrees of freedom that can be included in the theory: those with soft momenta scaling as $k^{\mu} \sim \lambda Q$. Any interaction of soft and collinear particles would result in an object with momentum scaling as $p^{\mu} \sim Q(1,\lambda, \lambda)$. Such excitations are not part of the EFT since they have invariant mass $p^{2} \sim Q^{2} \lambda$, which is much greater than the invariant mass of soft or collinear particles. Thus no soft-collinear interaction can appear in the Lagrangian, nor can usoft-soft interactions appear. Soft degrees of freedom can, however, appear in operators as polynomials of matter or gauge fields. Any soft field must be accompanied by a soft Wilson line in such a way as to make the combination gauge invariant under soft gauge transformations. SCET formulated with collinear and soft (but no usoft) degrees of freedom is usually referred to as $\SCETII$. An interesting subtlety in $\SCETII$ is that a rapidity regulator must be introduced to maintain the separation between soft and collinear modes as both sit on the same invariant mass curve~\cite{Chiu:2011qc,Chiu:2012ir}.

\subsection{Boosted Heavy Quark Effective Theory}

HQET describes a heavy parton bound in a hadron. The heavy parton momentum can be decomposed as
\begin{align} 
\label{rmomentum}
p^{\mu} = m_Q\, v^{\mu} + k^{\mu}\,,
\end{align} 
where  in the heavy hadron rest frame $v=(1,{\bf 0})= \frac{n^{\mu}}{2} + \frac{\n^{\mu}}{2}$ is the velocity of the heavy hadron, $m_Q$ is the heavy quark mass, and the residual momentum scales as $k^{\mu}\sim \LambdaQCD$. 
HQET is an expansion in $ \LambdaQCD/m_{Q} \ll 1$, and the leading HQET Lagrangian is 
\begin{equation}
\mathcal L =  \bar{h}_{v} i v \cdot D h_{v} \,,
\end{equation}
where $h_{v}$ is a two component heavy quark field  which satisfies $\vslash h_{v} = h_{v}$. See Ref.~\cite{Manohar:2000dt} for a detailed treatment of HQET.

HQET is formulated in a frame independent manner, and while it has mainly been applied to the decay of heavy mesons in their rest frame, it is equally suited to describe a heavy hadron carrying high momentum. When the heavy quark in the hadron rest frame is boosted along the $n$-direction, 
the velocity becomes $v^{\mu}  = \n\cdot v~n^{\mu}/2 +  n\cdot v~ \n^{\mu}/2$,
where $\n\cdot v = (n\cdot v)^{-1} =\n\cdot \tilde{p}/m_Q$, 
with $\tilde p$ the large label momentum of the heavy quark. Thus the momentum of a heavy quark in a hadron moving in the $n$-direction is
\begin{align} 
\label{bmomentum}
p^{\mu} = \n\cdot \tilde{p} \frac{n^{\mu}}{2} + \frac{m_Q^2}{\n\cdot\tilde{p}} \frac{\n^{\mu}}{2} + k^{\mu} 
= m_Q v^{\mu} +k^{\mu}\,,\qquad v^{\mu} = \frac{\n\cdot \tilde{p}}{m_Q} \frac{n^{\mu}}{2} + \frac{m_Q}{\n\cdot\tilde{p}} \frac{\n^{\mu}}{2}\,,
\end{align}
where $\n\cdot \tilde{p} \sim Q$. The residual momentum of the boosted quark scales like $k^{\mu}= (m_Q\LambdaQCD/Q,$ $ Q\LambdaQCD/m_Q,\LambdaQCD)$ but still describes soft fluctuations of the quark in the hadron with virtuality $\LambdaQCD^2$. The application of HQET to heavy quarks in a highly boosted frame has been referred to as boosted-HQET (bHQET) in the literature~\cite{Fleming:2007qr,Fleming:2007xt}. Though bHQET is no different from HQET it is a convenient shorthand to refer specifically to the case of highly boosted heavy quarks. 

A new feature appears when considering highly boosted heavy quarks: the emergence of a Wilson line. This can be immediately seen when matching  the $\SCETM$ collinear heavy quark-Wilson line combination, $\chi_n$, onto bHQET degrees of freedom:
\begin{align} 
\label{quark} 
\sum_{\tilde{p}} \delta_{\tilde{p},m_Q v}\, e^{-i\tilde{p}\cdot x}\, \chi_n =\sum_{\tilde{p}} \delta_{\tilde{p},m_Q v}\, e^{-i\tilde{p}\cdot x}\, \Bigl(W_n^{\dagger} \xi_n\Bigr) \to  e^{-im_Q v\cdot x}\, \tilde{W}^\dagger_n h_{v,n}\,.
\end{align} 
The $\SCETM$ field $\xi_{n}$ matches onto the heavy quark spinor $h_{v,n}$ (where we add a second index $n$ explained below), and the subset of gluon fields in $W^{\dagger}_{n}$ that are soft and collinear match onto a bHQET Wilson line
\begin{equation}
\tilde{W}(x) = \textrm{P exp}\bigg( -i g \int^x_{-\infty} ds \, \bn\cdot A(\bn s)\bigg)\, 
\end{equation}
with the gluon momenta scaling as $k$ given above. The heavy quark field is indexed by the boosted velocity given in Eq.~(\ref{bmomentum}), which has a large component in the $n^{\mu}$ direction; hence the second index $n$ on the field.

\section{Factorization away from the endpoint $x\to 1$}\label{sec:fullx}

We consider the single inclusive cross section for the production of a  heavy hadron $H$ in $e^+ e^-$ annihilation: $e^+ e^- \rightarrow  H(p_H)  + X$. $H$ contains a heavy quark and has momentum $p_H$ which is measured, while all other final state particles are treated inclusively (as their properties are not measured) and are denoted by $X$. 
We consider the differential cross section with respect to the variable 
\begin{equation}
x = \frac{2 p_H \cdot q}{q^2} = \frac{2 E_H}{Q},
\end{equation}
where $q = p_{e^+} + p_{e^-}$ is the total momentum of the colliding  electron-positron pair. In the center-of-mass frame $q^{\mu} = (Q,\vec{0})$, where  $Q = \sqrt{q^2}$,
and $x$ equals the fraction of the beam energy carried away by the hadron $H$, so that $x \leq 1$.  The lower limit on $x$ depends on the hadron mass
\begin{equation}
x > \frac{4 m_H^2}{Q^2}.
\end{equation}
We are only considering fragmentation into $b$-flavored hadrons at the $Z$ pole, so $Q = m_Z$, and the lower limit is $x \gtrsim 0.05$.

A factorization theorem for single inclusive hadron production in $e^+ e^-$ annihilation was proven using QCD factorization methods in Refs.~\cite{Ellis:1978ty,Collins:1981ta} (see Ref.~\cite{Collins:2011zzd} for a recent discussion). Here we rederive the same factorization theorem using SCET.
We start with the expression for the differential cross section in terms of currents of quarks and leptons:
\begin{eqnarray}
\label{qcdcrosssection}
\frac{\df \sigma}{\df x \, \df\cos\theta } &=&\frac{1}{4} \sum_q \sum_{i={v,a,av}}
\sigma^{(0)}_i(Q,m_Z)  L_{\mu\nu}^{i} \ W^{\mu\nu}_i(x,Q)
\, ,
\end{eqnarray}
where $\cos\theta$ is the cosine of the angle between the momenta of the  identified hadron and the electron.
The first sum is over final state quarks, at LEP energies $q = u,\, d,\, c\,, s\,, b$, and 
\begin{align}\label{born}
\sigma^{(0)}_{v} &=  \frac{4 \pi {\alpha^2_{\textrm{em}}}}{Q^2} 
\bigg[\, e_q^2 - 
2 v_e v_q e_q \frac{2 Q^2 (Q^2 -m_Z^2)}{(Q^2-m_Z^2)^2 + m_Z^2 \Gamma_Z^2} + 
(v_e^2+a_e^2)v_q^2 \frac{Q^4 }{(Q^2-m_Z^2)^2 + m_Z^2 \Gamma_Z^2 }\, \bigg] \,,\nn\\
\sigma^{(0)}_{a} &=  \frac{4 \pi \alpha_{\textrm{em}}^2}{Q^2}  \bigg[ (v_e^2 + a_e^2) a_q^2  \frac{Q^4}{(Q^2 - m_Z^2)^2 + m_Z^2 \Gamma_Z^2}\, \bigg] \,,\nn\\
\sigma^{(0)}_{av} &=  \frac{4 \pi \alpha_{\textrm{em}}^2}{Q^2}   a_e a_q \bigg[ v_e v_q  \frac{Q^4}{(Q^2 - m_Z^2)^2 + m_Z^2 \Gamma_Z^2} - e_q \frac{Q^2 (Q^2-m_Z^2)}{(Q^2 - m_Z^2)^2 + m_Z^2 \Gamma_Z^2}\, \bigg] \,.
\end{align}
Here $m_Z$ and $\Gamma_Z$ are the $Z$ boson mass and width, $\alpha_{\textrm{em}}$ is the fine structure constant,  $e_q$ is the quark charge in units of $e$, and the axial and vector couplings of a fermion to the $Z$ boson are
\begin{eqnarray}
v_f = \frac{T_3^f-2 e_f \sin^2\theta_W}{2\sin\theta_W \cos\theta_W}\,,
\qquad\qquad
a_f = \frac{T_3^f}{2\sin\theta_W \cos\theta_W} \,,
\end{eqnarray}
where $T_3^f$ is the third component of weak isospin, and $\theta_W$ is the weak mixing angle. 
The total tree level cross section for the production of a $q \bar q$ pair is given by 
\begin{equation}
\sigma^{(0)} = \sigma^{(0)}_{v} + \sigma^{(0)}_{a},
\end{equation}
as the parity-odd axial-vector interference term vanishes when integrated over $\cos\theta$.

The leptonic tensor $L^{i}_{\mu \nu}$ is 
\begin{eqnarray}
L^{v}_{\mu \nu} = L^{a}_{\mu \nu} = \frac{ p_{e^+}^\mu p_{e^-}^{\nu} +  p_{e^+}^\nu p_{e^-}^{\mu} }{Q^2} - \frac{g^{\mu \nu}}{2}, \qquad L^{av}_{\mu \nu} = i \varepsilon_{\mu \nu \alpha \beta} \frac{p_{e^+}^\alpha p_{e^-}^\beta}{Q^2},
\end{eqnarray}
where $p_{e^-}$ ($p_{e^+}$) is the electron (positron) momentum. 
The hadronic tensor is
\begin{eqnarray}
\label{qcdwmunu}
W^{\mu\nu}_{a,v}(x,Q) = \frac{x}{4\pi} \int \df^4 y \, e^{iq\cdot y}\langle 0| {\cal J}^{\nu\dagger}_{a,v}(y) \sum_X |X H(p_H)\rangle
\langle X H(p_H) | {\cal J}^\mu_{a,v}(0) |0\rangle  \nonumber\, , \\
W^{\mu\nu}_{av}(x,Q) = \frac{x}{4\pi}\int \df^4 y \, e^{iq\cdot y}\langle 0| {\cal J}^{\nu\dagger}_{a}(y) \sum_X |X H(p_H)\rangle
\langle X H(p_H) | {\cal J}^\mu_{v}(0) |0\rangle  + \textrm{h.c.}
\, ,
\end{eqnarray}
where the vector and axial currents are given by
 \begin{align}
 \label{QCDcurrents}
 {\cal J}^\mu_v(y) &=  \bar{\psi}_q(y) \gamma^\mu \psi_q(y) \,,
 & {\cal J}^\mu_a(y) & =  \bar{\psi}_q(y) \gamma^\mu \gamma_5 \psi_q(y) \,.
 \end{align}
For convenience we will adopt the short-hand notation ${\cal J}^\mu_i
 =\bar\psi(y) \Gamma_i^\mu \psi(y)$, leaving the flavor label $q$ and the Dirac structure implicit. The sum over the polarizations of the final state hadron $H$ (if $H$ has spin) is also left implicit.

If we restrict ourselves to the region of phase space away from the endpoint $x \to 1$ then the final state has invariant mass of order $Q^2$, and the hadronic tensor in Eq.~(\ref{qcdwmunu}) can be matched onto operators in $\SCETI$~\cite{Bauer:2002nz} that only involve collinear fields in the direction of the observed hadron. 
The hadronic tensor with the insertion of two vector or axial currents of flavor $q$ can be expressed in terms of a transverse and longitudinal component with respect to the hadron momentum,
\begin{equation}
W^{\mu\nu}_{i}(x,Q) =  -3 g^{\mu \nu}_{\perp} \, W^{}_{i,\, T}(x,Q)  +   3 \frac{(n^\mu - \bar n^\mu) (n^\nu - \bar n^{ \nu})}{2}\,  W^{}_{i,\, L}(x,Q) ,
\end{equation}
for $i={a,v}$. We introduced the light-cone vectors $n^{\mu}$ and $\bar{n}^{\mu}$, aligned with and opposite to the hadron momentum
\begin{equation}
n^{\mu} = \left(1, \sin\theta \cos\varphi, \sin\theta \sin\varphi, \cos\theta \right), \qquad \bar{n}^{\mu} = \left(1, -\sin\theta \cos\varphi, -\sin\theta \sin\varphi, -\cos\theta \right),
\end{equation}
and  $g^{\mu \nu}_{\perp}$ is the metric on the transverse plane,
$g_{\perp}^{\mu \nu} = g^{\mu \nu} - {(n^{\mu}\bar n^{\nu} + n^\nu \bar{n}^{\mu})}/{2}$.
The $av$ component of the hadronic tensor, with one axial and one vector current, violates parity and can be expressed as  
\begin{eqnarray}
W^{\mu\nu}_{av}(x,Q) = - i \varepsilon_{\perp}^{\mu \nu} 3 W^{}_{A}(x,Q)
\end{eqnarray}
where $\varepsilon^{\mu \nu}_{\perp} = \varepsilon^{\mu \nu \alpha \beta} \bar{n}_\alpha  n_{\beta}/2$.

At leading order in the SCET power counting, and taking into account current conservation and the CP properties of the vector and axial currents, only a limited number of operators can contribute
to the transverse, longitudinal and asymmetric functions. For a given quark flavor $q$, we can express $W_{T,L,A}$ as 
\begin{eqnarray}
\label{scetwmunu}
 W_{k}(x,Q) &=&  \int \df \omega_+ \df \omega_- \frac{2 \bar n \cdot p_H}{\omega_+^2} \nonumber \\
 &\times&  \bigg\{
\sum_f   H_{k,\, f}(\omega_+,\omega_-)   \frac{1}{2 N_c}\textrm{Tr}\langle 0| \chi^f_{n,\omega_1} \sum_{X_n} |X_n H(p_H)\rangle \frac{\bnslash}{2} \langle X_n H(p_H)| \bar{\chi}^f_{n,\omega_2}|0\rangle \nonumber \\
& & + \sum_f H^{}_{k, \bar{f}}(\omega_+,\omega_-) \frac{1}{2 N_c}\textrm{Tr}\langle 0| \bar{\chi}^f_{n,\omega_1} \sum_{X_n} |X_n H(p_H)\rangle \frac{\bnslash}{2} \langle X_n H(p_H)| \chi^f_{n,\omega_2}|0\rangle
\nonumber \\ & & - \frac{ \omega_+}{4} H_{k, \, g}(\omega_+,\omega_-) \frac{1}{N_c^2 - 1} \langle 0| \textrm{Tr}\big[ \mathcal B^{\lambda}_{n\perp,\omega_1} \sum_{X_n} |X_n H(p_H)\rangle 
\langle X_n H(p_H)|\mathcal B_{n\perp,\omega_2\, \lambda}\big] |0\rangle \bigg\}
\nonumber \\ & & + \ldots ,
\end{eqnarray}
where $k \in \{T,L,A\}$, $N_c$ is the number of colors, $\omega_{\pm} = \omega_1 \pm \omega_2$, the sum is extended over all active quark flavors, the trace is over spin and color, and the dots represent terms that are suppressed by ${\cal O}(m_H^2/Q^2)$ or higher. In Eq. \eqref{scetwmunu}, we pulled out factors of $\omega_+$ in such a way that  $H_{f, \bar f}$ and $H_g$ are dimensionless.

The vacuum matrix element of the operators in this  equation can be related to the standard unpolarized 
quark, antiquark and gluon fragmentation functions, that give the probability of finding in the parton a heavy meson state $H$ moving in the $n$ direction with large light-cone momentum $\bn\cdot p_H$:
\bea
& & \frac{1}{2N_c} \textrm{Tr} \langle 0|\chi^f_{n,\omega_1} \sum_{X_n} |X_n H(p_H)\rangle \frac{\bnslash}{2} \langle X_n H(p_H)| \bar{\chi}^f_{n,\omega_2} |0\rangle \label{quarkfrag} \\ 
 & & \hspace{7.5cm}=\int_0^1\frac{\df z}{z} \delta (\omega_-) \, \delta\!\left(z  - \frac{2 \bn\cdot p_H }{\omega_+} \right)D_{H/f}(z)  
\nn \\
& &  \frac{1}{2N_c} \textrm{Tr} \langle 0|\bar{\chi}^f_{n,\omega_1} \sum_{X_n} |X_n H(p_H)\rangle \frac{\bnslash}{2} \langle X_n H(p_H)| \chi^f_{n,\omega_2} |0\rangle  \label{antiquarkfrag}  \\
& & \hspace{7.5cm} =\int_0^1\frac{\df z}{z} \delta (\omega_-) \, \delta\!\left(z  - \frac{2 \bn\cdot p_H }{\omega_+} \right)D_{H/\bar f}(z)  
\nn \\
& & \frac{1}{N^2_c-1}\langle 0| \textrm{Tr}\big[ \mathcal{B}^{\mu }_{n\perp,\omega_1} \sum_{X_n} |X_n H(p_H)\rangle 
\langle X_n H(p_H)|\mathcal{B}_{n\perp,\omega_2\, \mu}\big] |0\rangle   \label{gluefrag} \\
& & \hspace{7.5cm}  = -\frac{4}{ \omega_+}
\int_0^1\frac{\df z}{z} \delta (\omega_-) \, \delta\!\left(z  -\frac{ 2 \bn\cdot p_H }{\omega_+} \right)D_{H/g}(z) \,.\nn
\eea
These definitions agree with those in Refs.~\cite{Collins:1981uw,Procura:2009vm,Jain:2011xz}.
Notice that the heavy quark, heavy antiquark, light quark and gluon fragmentation functions have the same scaling in the SCET power counting. 
In Eq. \eqref{scetwmunu} the fragmentation functions are weighted by the coefficient functions $H_{k, i}$, which depend only on the hard scale, and have a perturbative
expansion in $\alpha_s$. Different terms in the hadronic tensor  thus appear at different perturbative orders:
for a given flavor $q$, $H_{T, q}$ and $H_{T, \bar q}$ start at leading order, 
$H_{L, q}$, $H_{L, \bar q}$, and $H_{k, g}$
at $\mathcal O(\alpha_s)$, and $H_{k, f\neq q}$ and $H_{k, \bar f\neq \bar q}$ at $\mathcal O(\alpha_s^2)$ \cite{Curci:1980uw,Furmanski:1980cm,Rijken:1996vr,Rijken:1996npa,Rijken:1996ns}.
$H_{A,\, g}$ vanishes at all orders, because of the charge conjugation invariance of QCD \cite{Rijken:1996npa}.

Using the definitions of the quark and gluon fragmentation functions given above in the leading term of the $\SCETI$ hadronic tensor in Eq.~(\ref{scetwmunu}) and then inserting this into Eq.~(\ref{qcdcrosssection}) gives the leading $\SCETI$ differential cross section for inclusive heavy hadron production in $e^+e^-$ collisions
\begin{equation}
\frac{\df \sigma^H}{\df x\, \df \cos\theta} =  \frac{3}{8} (1 + \cos^2\theta) \frac{\df \sigma^H_T}{\df x} +  \frac{3}{4}\sin^2\theta\, \frac{\df\sigma^H_L}{\df x} + \frac{3}{4}\cos\theta \, \frac{\df\sigma^H_A}{\df x},
\end{equation}
with each term  expressed as the convolution of a short distance coefficient and a fragmentation function:
\begin{eqnarray}\label{QCDfactorization}
\frac{\df \sigma^H_k}{\df x} &=& \sum_q  \int \frac{\df z}{z} \sum_f \sigma^{(0)}_i \left(  H_{k,\, f} \left(\frac{x}{z},\mu\right)  D_{H/f}(z,\mu) + H_{k,\, \bar{f}}\left(\frac{x}{z},\mu\right)  D_{H/\bar f}(z,\mu)\right) \nonumber \\ & & +  
\left( \sum_q \sigma^{(0)}_i \right)  \int \frac{\df z}{z} H_{k,\,g }\left(\frac{x}{z},\mu\right)  D_{H/g}(z,\mu),
\end{eqnarray}
where $i = a,\, v$ for the longitudinal and transverse cross sections, $i = av$ for the asymmetric, and the tree level cross sections $\sigma^{(0)}_i$ are given in Eq. \eqref{born}.
Integrating over $\cos\theta$, we obtain the differential cross section with respect to $x$, which is given by the sum of the transverse and longitudinal components
\begin{equation}
\frac{\df \sigma^H}{\df x} =   \frac{\df \sigma^H_T}{\df x} +   \frac{\df\sigma^H_L}{\df x}.
\end{equation}
In the rest of the paper, we will focus on this observable.

The coefficient functions $H_{k, i}$ describe the short-distance cross section for the production of a parton of species $i$.
They are purely perturbative, and, for massless partons, they have been computed at  N${}^2$LO \cite{Rijken:1996vr,Rijken:1996npa,Rijken:1996ns}.
Mass corrections to the production of heavy quarks have been considered \cite{Nason:1993xx,Ravindran:1998qz,Nason:1998ug}, but are relevant only at small $x$, and we neglect them.
It is instructive to look at the well known NLO expressions of $H_{T,\, Q}$ and $H_{L,\, Q}$  \cite{Curci:1980uw,Furmanski:1980cm}
\begin{align} \label{eq:QCD.2} 
H_{T,\, Q} (z,Q,\mu) &= \delta(1-z) + \frac{\alpha_s}{2\pi}\, \hat{a}_{T,\, Q}^{(1)} (z,Q,\mu) \,, \\
H_{L,\, Q} (z,Q,\mu) &= \frac{\alpha_s }{2\pi}\, \hat{a}_{L,\, Q}^{(1)} (z,Q,\mu) ,
\end{align}
with
\begin{align}\label{eq:QCD.3}
\hat{a}_{T,\, Q}^{(1)} (z,Q,\mu) &= C_F\, \Biggl\{\ln\!\frac{Q^2}{\mu^2}\, \left[\frac{1+z^2}{1-z}\right]_+ + (1+z^2)\, \left[\,\frac{\ln (1-z)}{1-z}\,\right]_+  - \frac{3}{2}\, \left[\frac{1}{1-z}\right]_+ \nn\\
&\qquad  + 2\, \frac{1+z^2}{1-z}\,\ln z + \frac{3}{2}\,(1-z) +\Bigl(\,\frac{2}{3}\,\pi^2 -\frac{9}{2}\,\Bigr)\,\delta(1-z) \Biggr\}\,, \\
 \hat{a}_{L,\, Q}^{(1)} (z,Q,\mu) &= C_F, \label{eq:QCD.4}
\end{align} 
where $C_F = 4/3$.
The hard scattering cross section is purely transverse at LO, while it has both transverse and longitudinal components at NLO and N${}^2$LO.
Close to the endpoint $z \sim 1$, a new scale $Q^2 (1-z)$ appears in the hard coefficients, and logarithms of $1-z$ need to be resummed.  
The singular behavior is all encoded in the transverse coefficient $H_{T,\, Q}$, while up to N${}^2$LO $H_{L,\, Q}$ has at most integrable singularities for $z\rightarrow 1$  \cite{Rijken:1996vr,Rijken:1996npa}. 

The N${}^2$LO expressions are given in Ref. \cite{Rijken:1996vr,Rijken:1996npa,Rijken:1996ns}, and are too lengthy to be reproduced here.
At N${}^2$LO, it is convenient to separate the hard scattering coefficient into flavor singlet and non-singlet components. 
As we discuss in Section \ref{sec:err}, the experimental analyses of $b$ fragmentation at the $Z$ pole focus on the fragmentation of a primary heavy quark
into a heavy meson and reject events with more than one $b$-quark in a hemisphere (which are associated with gluon splitting). 
For this reason, we will consider only the contribution of $q=b$ in Eq. \eqref{QCDfactorization}, that is we will not consider either the gluon or the flavor singlet contributions, but will concentrate on the flavor non-singlet contribution.

The second set of ingredients in Eq. \eqref{QCDfactorization} are the fragmentation functions $D_{H/i}$.
In the case of light quarks, the hadronic matrix elements in Eqs. \eqref{quarkfrag}, \eqref{antiquarkfrag} and \eqref{gluefrag} are purely nonperturbative, and need to be fit to data.
Sets of fragmentation functions are available for pions, kaons, and other light hadrons  \cite{deFlorian:2007aj,deFlorian:2007hc,Albino:2008fy}, 
Recently, the first extraction of the light hadron fragmentation functions at N${}^2$LO has been performed \cite{Anderle:2015lqa}.

In the case of heavy quarks, we can take advantage of the presence of a large scale $m_Q$, and compute the fragmentation function in perturbation theory. Inverting the expression in Eq.~(\ref{quarkfrag}) gives~\cite{Procura:2009vm} 
\begin{align}\label{frqf} 
D_{H/Q} (z) = \frac{z}{2N_c} \sum_{X_n}\, \mathrm{Tr} \langle 0 |\, \delta \left(\frac{\n\cdot p_H}{z} - \n\cdot \mP \right)\, \chi_n\, | H X_n\rangle
\langle H X_n |\, \bar{\chi}_n \, \nbstwo\, | 0 \rangle\,, 
\end{align} 
and similar expressions for the antiquark, the gluon and the light quark fragmentation functions.
Integrating out degrees of freedom with invariant mass $\sim m_Q^2$, we can match the fragmentation functions onto bHQET operators.
At lowest order in the bHQET power counting, we have
\begin{equation}\label{matchHQET}
D_{H/i}(z) = d_{Q/i}(z) \, \frac{1}{4N_c} \sum_{X_n}\, \mathrm{Tr}  \langle 0 |\,  \tilde{W}^\dagger h_{v,n}(0)\, | H_{v} X_n\rangle \langle H_{v} X_n |\, \bar{h}_{v,n}\tilde{W}(0)\, | 0 \rangle\,.
\end{equation}
In bHQET, the emission of $b\bar b$ pairs is no longer possible, and only matrix elements with the heavy fields $h$ have nonzero overlap with the heavy-flavored state $H$.
Thus, the quark, antiquark, gluon and light quark fragmentation function all match onto the same bHQET matrix elements, with different coefficients $d_{Q/i}$, 
which are the partonic fragmentation functions for a parton $i$ to fragment into a heavy quark $Q$. 
At tree level, this can be understood directly from Eq. \eqref{frqf}. Away from the endpoint $m_Q (1-z)$ is much larger than the hadronization scale $\LambdaQCD$ and at 
the matching scale $\mu \sim m_Q$ the term $\n\cdot \mP$ in the delta function of Eq.~\eqref{frqf} becomes $m_Q\n\cdot v$, which is fixed in bHQET. Furthermore, employing the matching condition of the $\chi_n$  field, Eq.~\eqref{quark}, we obtain 
\begin{align} \label{eq:frqfbHQET} 
D_{H/Q} (z)  &\approx \frac{z}{2N_{c}} \sum_{X_n}\, \mathrm{Tr}  \langle 0 |\, \delta \left(\frac{m_H \n\cdot v}{z} - m_Q \n\cdot v\right)\,\tilde{W}^\dagger h_{v,n}(0)\, | H X_n\rangle
\langle HX_n |\, \bar{h}_{v,n} \tilde{W}(0)\, \nbstwo\, | 0 \rangle \nonumber \\
&\approx \delta(1-z)\, \frac{1}{4N_c} \sum_{X_n}\, \mathrm{Tr}  \langle 0 |\,  \tilde{W}^\dagger h_{v,n}(0)\, | H_{v} X_n\rangle \langle H_{v} X_n |\, \bar{h}_{v,n}\tilde{W}(0)\, | 0 \rangle\,.
\end{align} 
For the the second equality we used the Lorentz invariant relation $|H \rangle=\sqrt{m_H} |H_v \rangle$,
and, to simplify the Dirac structure, the property $(1+\slashchar{v}) h_{v,n} = 2 h_{v,n}$ of the HQET field. We ignored the mass difference $\bar{\Lambda} = m_H - m_Q$ between the heavy hadron and the heavy quark. The necessity of the Wilson lines in the equation above was first discussed in Refs.~\cite{Nayak:2005rw,Nayak:2005rt} in the context of color-octet operators in quarkonium fragmentation, however, the same arguments hold here for the color-triplet operator.
As mentioned before the matrix element in Eq.~\eqref{eq:frqfbHQET} describes purely nonperturbative effects and can be used as definition for a nonperturbative normalization factor
\begin{align}\label{chiH}
\chi_H = \frac{1}{4N_c} \sum_{X_n}\, \mathrm{Tr}  \langle 0 |\,  \tilde{W}^\dagger h_{v,n}(0)\, | H_{v} X_n\rangle \langle H_{v} X_n |\, \bar{h}_{v,n}\tilde{W}(0)\, | 0 \rangle,
\end{align}
which satisfies the sum rule 
$
\sum_H \chi_H = 1,
$
where the sum is extended to all $b$-flavored hadrons. 

The perturbative fragmentation functions $d_{Q/i}$ in Eq. \eqref{matchHQET},
describing the fragmentation of a parton $i$ into the heavy quark $Q$, are known at N${}^2$LO \cite{Melnikov:2004bm,Mitov:2004du}. 
It is again instructive to look at the NLO result. 
At NLO,  the only possible processes are the fragmentation of a heavy quark into a heavy quark,  ${d}_{Q/Q}$, and of a gluon into a heavy quark, $d_{Q/g}$, which are given by~\cite{Mele:1990yq}
\begin{eqnarray}\label{Wilq}
d_{Q/Q} (z,\mu) &=&\, \delta(1-z) + \frac{\alpha_s}{2\pi}\,C_F\, \Biggl[\frac{1+z^2}{1-z}\, \Bigl(\ln \frac{\mu^2}{m_Q^2(1-z)^2} -1\Bigr)\Biggr]_+ \,, \\
d_{Q/g} (z,\mu) & = &   \frac{\alpha_s}{2\pi}\,T_F\, (z^2 + (1-z)^2) \ln \frac{\mu^2}{m_Q^2}\,, \label{Wilg} 
\end{eqnarray}
where $T_F = 1/2$.
From Eq. \eqref{Wilq} we see that the scale that appears in ${d}_{Q/Q}$ is $m_Q (1-z)$, rather than $m_Q$,
which points to the need of resumming logarithms of $1-z$, as we discuss in the next section. On the other hand, the gluon fragmentation function has a regular behavior for $z \rightarrow 1$,
and the only logarithms that appear are logarithms of the heavy quark mass, which are resummed by the DGLAP evolution.    

The N${}^2$LO corrections to the perturbative fragmentation function were computed in Refs. \cite{Melnikov:2004bm,Mitov:2004du}. 
At this order, in addition to $\mathcal O(\alpha_s^2)$ corrections to Eqs. \eqref{Wilq} and \eqref{Wilg}, 
the first contributions to $d_{Q/l}$ and $d_{Q/\bar Q}$,
the fragmentation functions of a light quark $l$ and a heavy antiquark $\bar Q$
into $Q$, arise. $d_{Q/g}$, $d_{Q/l}$ and $d_{Q/\bar Q}$ have a regular behavior in the endpoint,
while the  N${}^2$LO expression of $d_{Q/Q}$ contains logarithms of $ \mu^2/m_Q^2$ up to  $\log^2 \mu^2/m_Q^2$, and
plus distributions up to  $[\log^3(1-z)/(1-z)]_+$.

To eliminate gluon splittings and events with more than one heavy quark in each hemisphere, as done by the experimental collaborations,
we considered the non-singlet distribution $d_{\textrm{ns}} =d_{Q/Q} - d_{Q/\bar Q}$. At all scales, $d_{\textrm{ns}}$ satisfies the flavor sum rule
\begin{equation}\label{flavsum}
\int_0^1 \df z \, d_{\textrm{ns}}(z,\mu) = \int_0^1 \df z (d_{Q/Q}(z,\mu) - d_{Q/\bar Q}(z,\mu))  = 1,
\end{equation}
that is, it keeps the number of heavy quarks fixed to 1. Furthermore, $d_{\textrm{ns}}$ does not mix with $d_{Q/g}$ or $d_{Q/l}$.

We thus arrive at the final expression for the differential cross section of a primary $b$ quark fragmenting into a heavy meson $H$  
\begin{equation}\label{QCDfinal}
\frac{1}{\sigma^{(0)}}\frac{\df \sigma^H}{\df x} = \chi_H  \int_x^1 \frac{\df z}{z}  H_{\textrm{ns}}\left(\frac{x}{z}, \mu \right)\, d_{\textrm{ns}} (z,\mu).  
\end{equation}
The non-singlet hard coefficient is given at $\mathcal O(\alpha_s)$ in Eq. \eqref{eq:QCD.3} and \eqref{eq:QCD.4}, and at $\mathcal O(\alpha_s^2)$ in Ref. \cite{Rijken:1996ns}.
The non-singlet fragmentation function coincides at one loop with the quark fragmentation function in Eq. \eqref{Wilq}, while the $\mathcal O(\alpha_s^2)$ correction
is given in Ref. \cite{Melnikov:2004bm}.

The hard function and the fragmentation function in the factorization formula \eqref{QCDfinal} depend
on the factorization scale $\mu$.
As we discussed, and as is explicitly demonstrated in Eqs. \eqref{eq:QCD.3} and \eqref{Wilq}, away from the endpoint the coefficient function
depends solely on the scale $Q$, while the fragmentation function only depends on $m_Q$. Since $Q \gg m_Q$, at fixed order, for any choice of $\mu$,  large logarithms appear in  Eq. \eqref{QCDfinal}. These are the standard ``logs of $Q$'' which can be resummed via 
 the DGLAP equations~\cite{Gribov:1972ri,Altarelli:1977zs,Dokshitzer:1977sg}:
\begin{equation}\label{DGLAP}
\frac{\df}{\df \log \mu} D_{H/i}(z,\mu) = 2  \int \frac{\df \xi}{\xi} P_{j i}(\xi) D_{H/j} \left(\frac{z}{\xi}, \mu \right)
\,,
\end{equation}
where $P_{j i}(\xi)$ are the time-like splitting functions. The splitting functions are computed in perturbation theory 
\begin{equation}
P_{j i}(z) =  \frac{\alpha_s}{2\pi} \sum_{n = 0}^{\infty} \left( \frac{\alpha_s}{2\pi} \right)^{n} P_{j i}^{(n)}(z)\,,
\end{equation}
with the one-loop expressions \cite{Gribov:1972ri,Altarelli:1977zs,Dokshitzer:1977sg}:
\begin{eqnarray}
P^{(0)}_{q_j q_i}(z) &=& \delta_{ij} C_F  \left[ \frac{1+z^2}{1-z}\right]_+ \, , \label{Pqq}\\
P^{(0)}_{g q}(z) &=&  C_F  \left( \frac{1 + (1-z)^2}{z}\right)  \label{Pgq} \, , \\ 
P^{(0)}_{q g}(z) &=&  T_F  \left( z^2 + (1-z)^2 \right)  \label{Pqg}  		\, , \\ 
P^{(0)}_{g g}(z) & =& 2  C_A \left(   z \left[\frac{1}{1-z}\right]_+ +    \frac{1-z}{z}+ z (1-z )   \right) + \frac{\beta_0}{2}\delta(1-z) \label{Pgg}\,.
\end{eqnarray}
The color factors in Eqs.~\eqref{Pqq}--\eqref{Pgg} are $C_F = 4/3$, $C_A = 3$, $T_F = 1/2$, while $\beta_0$ is the leading order coefficient of the beta function,
\begin{equation}
\beta_0 = \frac{11}{3} C_A -\frac{4}{3} T_F n_f\,.
\end{equation}
From Eqs. \eqref{DGLAP} and \eqref{Pqq} - \eqref{Pgg}, it is easy to see that the non-singlet fragmentation function $d_{\textrm{ns}}$ does not mix with the gluon fragmentation function, and its lowest order evolution 
is governed by $P^{(0)}_{qq}$ only. In particular, since the integral of $P^{(0)}_{qq}$ vanishes, the normalization of $d_{\textrm{ns}}$ is unchanged by the evolution, confirming Eq. \eqref{flavsum}.  
These statements extend beyond the leading logarithmic evolution.

The  time-like splitting functions at $\mathcal O(\alpha_s^2)$ have been known for some time~\cite{Furmanski:1980cm,Curci:1980uw}, and are nicely summarized in Ref.~\cite{Ellis:1991qj}.
The $\mathcal O(\alpha_s^3)$ corrections to 
the non-singlet components and the singlet splitting functions $P^{(2)}_{qq}$ and $P^{(2)}_{gg}$ are given in Refs. \cite{Mitov:2006ic,Moch:2007tx}, and the nondiagonal entries of the singlet matrix, $P^{(2)}_{g q}$  and $P^{(2)}_{q g}$, were determined in Ref. \cite{Almasy:2011eq}.
With the calculation of the $\mathcal O(\alpha_s^2)$ corrections to the fragmentation function \cite{Melnikov:2004bm,Mitov:2004du}
and of the non-singlet splitting function at $\mathcal O(\alpha^3_s)$ \cite{Mitov:2006ic} all the ingredients for the resummation  of the non-singlet distribution $(d_{Q/Q}(z) - d_{Q/\bar Q}(z))$ 
at N${}^2$LL accuracy are now available.

We solved the DGLAP equation for $d_{\textrm{ns}}$ directly in $z$ space, by discretizing Eq. \eqref{DGLAP}, following the approach described in Ref. \cite{Miyama:1995bd}. Some detail on the solution of the DGLAP equation are given in Appendix \ref{python}.


\section{Formalism in the endpoint $x\to 1$}\label{sec:xto1}

The discussion in Section \ref{sec:fullx} anticipates that care has to be taken in describing the endpoint region where the momentum fraction $x$ approaches one. In this regime the heavy quark has large energy of order $Q$ and is accompanied by a jet-like spray of particles with energy of order $Q(\LambdaQCD/m_{Q}) \ll Q$, all of which recoils against a jet of energy of order $Q$ and invariant mass squared of order $(1-x)Q^2 \ll Q^2$.
This occurs in a background of soft interactions which change momenta on the order of $(1-x)Q$ or less. As a consequence the heavy quark carries almost all of the energy in one hemisphere, while the jet carries most of the energy in the other hemisphere. The soft interactions can not change the order of the jet invariant mass squared, so that the invariant mass squared of all final state particles beside the heavy hadron is $p_X^2 \sim (1-x)Q^2 \ll Q^2$. 

The observable ${\rm d}\sigma/{\rm d}x$ is not sensitive to the details of the final state $X$, which can be composed of one or more jets, but it 
is sensitive to the momentum squared ($p_X^2$) of all final state particles beside the heavy hadron.
As $x$ increases, the final state becomes more and more collimated, leaving only one jet in the endpoint region.  
The sensitivity to $p_X^2$ gives rise to large logarithms of $(1-x)$ in the perturbative expansion, and the jet in the endpoint must be included in our description. 
 
In a similar way, for $x\sim 1$ large endpoint logarithms appear in the perturbative fragmentation function as well, as evidenced by Eq. \eqref{Wilq}. These logarithms are
associated with the appearance in the endpoint of a new soft scale $m_Q (1-x) \sim \LambdaQCD$, and threaten the convergence  
of the perturbative fragmentation function unless they are resummed. 

We thus divide the $x$ spectrum into two regions with different power counting:
\begin{align}\label{eq:scalesnregions}
  &  \text{peak region:} & & Q\gg \sqrt{1-x}\,Q\gg \, m_Q\gg m_Q (1-x) \sim  \LambdaQCD \,, \nonumber\\
  & \text{tail region:} & & Q\,\sim\, \sqrt{1-x}\,Q\sim (1-x)Q\gg m_Q\gg \LambdaQCD \,. \nonumber
\end{align}
The tail region was described in Section \ref{sec:fullx}. In this section we study the peak region.

A factorization formula for the peak region  can be derived in the framework of SCET.
Here we state the final result
\begin{eqnarray}\label{eq:finalfact}
\frac{{\rm d}\sigma}{{\rm d}x} & = & \sigma^{(0)}\, H_Q(Q,\mu)\, C_m\left(m_Q,\mu\right) \,  m_H \, Q \int\! {\rm d}r \, {\rm d}\ell \, J_\nb(Q r,\mu)\,  
\nonumber\\
& & S_{Q/Q}(m_Q \bar n \cdot v (1-x) -r-\ell, \mu)  S^{hadr}_{H/Q}(\ell) 
 + {\cal O}\left(\frac{\LambdaQCD}{m_Q}\right) + {\cal O}\left(\frac{m_Q}{Q}\right)\,,
\end{eqnarray}
which we derive in Sections \ref{sec:SCET1} -- \ref{sec:bHQET}.
The cross section is thus factored into a product of four different functions, each dependent on a single scale:
$H_Q(Q,\mu)$ is the hard function, which  encodes hard momentum fluctuations
with scaling $\mu_{Q}\sim Q$; $J_{\bar n}(Q r,\mu)$ is the jet function describing the collinear final state that recoils against the heavy hadron, with typical scale  $\mu^2_J \sim Q^2 (1-x)$;
$C_m\left(m_Q,\mu\right)$ includes all the dependence on the heavy quark mass and thus has typical scaling $\mu_{M} \sim m_Q$; 
the shape function $S$ captures physics at the scale of the residual momentum of the heavy quark inside the heavy meson,
$\omega \sim  m_Q \bar n \cdot v (1-x)$, which, in the heavy quark rest frame, is the nonperturbative scale  $\mu_S \sim  m_Q (1-x)$.
\footnote{
Notice that the shape function $S$ is boost invariant, and the boost factor $\bar n \cdot v$ is irrelevant for the dynamics, as can be explicitly seen from the perturbative
expression of the shape function in Eq.~\eqref{shape.1}.}
As explained in Section \ref{sec:np}, we further divide the shape function in a perturbative ($S_{Q/Q}$) and nonperturbative ($S^{hadr}_{H/Q}$) component. 

We obtain Eq.~(\ref{eq:finalfact}) by using a tower of EFTs. Hard momentum fluctuations of order $\mathcal O(Q)$ are removed by matching QCD onto $\SCETI$, as detailed in  Section~\ref{sec:SCET1}. At lower momentum the heavy quark mass becomes nonnegligible whereas the invariant mass of the jet can be treated as a high energy scale and can be removed from the theory. This is done by switching to $\SCETM$. In Section~\ref{sec:SCETM} the factorization in $\SCETM$ and its matching onto $\SCETI$ are discussed. 
The separation of the dynamics at the scale $m_Q$ and $\LambdaQCD\sim m_Q (1-x)$ is achieved by matching $\SCETM$ onto bHQET, as we discuss in Section~\ref{sec:bHQET}. 
Furthermore, bHQET allows the factorization of the  nonperturbative dynamics of the heavy quark inside the hadron. 

The factorization formula in Eq.~\eqref{eq:finalfact} is equivalent to that derived in Refs. \cite{Cacciari:2005uk,Cacciari:2001cw} in the framework of perturbative QCD.
As discussed in more detail in Section 6, an important check of Eq.~\eqref{eq:finalfact} is that, at fixed order, the product of the hard function $H_Q$ and the jet function $J_{\bar n} (Q r)$
and the product of the mass coefficient $C_m(m_Q)$ and the shape function $S_{Q/Q}$ reproduce, respectively, the soft, $z\rightarrow 1$, limit of the coefficient function $H_{\textrm{ns}}(z)$ , 
and of the fragmentation function $d_{\textrm{ns}}(z)$ in Eq.~\eqref{QCDfinal}.

\subsection{Matching onto $\SCETI$}\label{sec:SCET1}

To derive the SCET$_{\textrm{I}}$ factorization formula, we go back to the hadronic tensor introduced in Eq.~(\ref{qcdwmunu}),
\begin{eqnarray}
\label{qcdwmunu2}
W^{\mu\nu}_i(x,Q) = \frac{x}{4\pi} \int \df^4 y \, e^{iq\cdot y}\langle 0| {\cal J}^{\nu\dagger}_i(y) \sum_X |X H(p_H)\rangle
\langle X H(p_H) | {\cal J}^\mu_i(0) |0\rangle 
\, .
\end{eqnarray}
In the endpoint, the final state $X$ has virtuality $p^2_X \ll Q^2$, and is still dynamical after the hard scale is integrated out.
Thus, we match the QCD currents ${\cal J}^\mu_i$ given in Eq.~(\ref{QCDcurrents}) onto the SCET current for the production of two back-to-back jets,
\begin{equation}\label{SCET1current}
\mathcal J_i^\mu(y) = \sum_{\omega,\bar{\omega}} {\cal C}(\omega,\bar{\omega})\, e^{i(\omega n\cdot y-\bar{\omega}\bar{n}\cdot y)/2}\, \bar{\chi}_{n,\omega}(y)\, Y_n^\dagger(y)\, \Gamma^{\mu}_i\, Y_\nb(y)\, \chi_{\bar{n},\bar{\omega}}(y)\,.
\end{equation}
$\bar{\chi}_{n,\omega}$ and $\chi_{\bar{n},\bar{\omega}}$ are collinear gauge invariant fields, defined in Eq.~\eqref{chifield}, and $Y_{n, \bar n}$ are soft Wilson lines, defined in Eq.~\eqref{Ysoft}.
$\Gamma^{\mu}_i$ encodes the Dirac structure of the current, for the vector and axial currents  $\Gamma^{\mu}_{v,a} = \{ \gamma^{\mu}_{\perp},\, \gamma^{\mu}_{\perp} \gamma_5   \}$. 
The matching coefficient $\mathcal C$ has been computed to $\mathcal O(\alpha_s^3)$ \cite{Baikov:2009bg}, and it is the same for vector and axial current. In this work, we will need the two loop expression derived in Refs. \cite{Matsuura:1987wt,Matsuura:1988sm,Gehrmann:2005pd,Moch:2005id}, which we quote in Appendix \ref{Equations}.

Using Eq. \eqref{SCET1current} in the hadronic tensor gives
\begin{eqnarray}\label{HadTensSCET.1}
W^{\mu \nu}_i &=& \sum_{\omega, \bar{\omega}} \, \mathcal C^*(Q, -Q) \mathcal C(\omega, \bar\omega) \, \frac{x}{4\pi}\int \df^4  y \, 
\langle 0 | \overline{\textrm{T}} \{ \bar{\chi}_{\bar n}( y)  Y_\nb^\dagger(y)\, \Gamma^\nu_i\, Y_n(y) {\chi}_{n} (y) \}| H X \rangle \nonumber 
\\ & &  \langle H X | \textrm{T} \{  \bar \chi_{n, \omega}(0)  Y_n^\dagger(0)\, \Gamma^\mu_i \, Y_\nb(0) \chi_{\bar n, \bar\omega}(0) \}  | 0 \rangle,
\end{eqnarray}
where we have implicitly split the position integral in Eq. \eqref{qcdwmunu2} into a sum over labels and an integral over the residual part of the coordinates.
The time-ordering (T) and anti-time-ordering ($\overline{\rm{T}}$) are relevant for the proper ordering of the usoft field in the Wilson lines $Y_{n, \bar n}$ and $Y^{\dagger}_{n,\bar n}$ \cite{Fleming:2007qr}.
The sum over labels fixes the label momenta of the currents to be $\pm Q$.

Next, we decompose the final state $|H X \rangle$  into its collinear, anticollinear and soft components, $|H X \rangle = | X_{\bar n} \rangle | X_s \rangle \, | H X_n \rangle$,
and rearrange the color and spin indices in Eq. \eqref{HadTensSCET.1} to be singlets in each sector:
\begin{eqnarray}\label{HadTensSCET.2}
W^{\mu \nu} &=& - \frac{1}{2}g^{\mu \nu}_{\perp} \sum_{\omega, \bar{\omega}} \, \mathcal C^*(Q, -Q) \mathcal C(\omega, \bar\omega) \frac{x}{N_c^2}  \int \frac{\df^4  y}{4\pi} 
\sum_{X_{\bar n}}\langle 0 | \bar{\chi}_{\bar n}(y) | X_{\bar n} \rangle \langle  X_{\bar n} | \frac{\slashchar{n}}{2} \chi_{\bar n, \bar\omega}(0) | 0 \rangle \,   \nonumber \\
& \times & \sum_{X_{n}}\textrm{Tr} \langle 0 | \frac{\slashchar{\bar n}}{2} {\chi}_{n}(y) | H X_{n} \rangle \langle H X_{n} | \bar \chi_{n, \omega}(0) | 0 \rangle \, \nonumber \\
& \times & \sum_{X_s} \langle 0 | \textrm{Tr}\left[ \overline{\textrm{T}}\{ Y_\nb^\dagger(y)\, Y_n(y)\} | X_s \rangle \langle X_s | \, \textrm{T}\{ Y_n^\dagger(0)\, Y_\nb(0) \}\right]  | 0 \rangle .
\end{eqnarray}
In the $\bar n$-collinear and soft sector, the sum is over a complete set of states, and can be replaced by the identity.  

The $\bar n$-collinear matrix element can be expressed in terms of the inclusive jet function \cite{Bauer:2002ie} 
\begin{equation}\label{jetfunction}
\frac{1}{2N_c}\sum_{X_{\bar n}} \langle 0 | \bar \chi_{\bar n}( y) | X_{\bar n} \rangle \, \langle X_{\bar n} | \frac{\slashchar{n}}{2} \chi_{\bar n, \bar \omega} | 0 \rangle \equiv 
Q \delta_{-\bar\omega,Q}\delta(y^-) \delta^{2}(y_{\perp}) \int \df r \, e^{-\frac{i}{2} r y^+} J_{\bar n}(Q r,\mu).
\end{equation} 
The jet function depends only on the off-shellness $p^2_X$ of the jet in the $\bar n$ direction, and we choose to work in a frame where the jet moving in the $\bar n$ direction has no transverse momentum with respect to $\bar n$, $r_{\perp} = 0$. In this frame, $p^2_X = Q r$, and the dependence of the jet function on $y^-$ and $y_{\perp}$ reduces to  delta functions. 
Label momentum conservation forces the label $\bar \omega = -Q$. 
The jet function in Eq. \eqref{jetfunction} is the same as the one that appears in the thrust distribution \cite{Schwartz:2007ib},
and it is known to two loops \cite{Becher:2006qw}. 

The delta functions in Eq. \eqref{jetfunction} force the remaining two matrix elements to depend only on $y^+$. 
We define the soft function as 
\begin{align}\label{eq:softfct}
S(\ell,\mu)=\frac{1}{N_c}\int\frac{\df y^+}{4\pi}\, e^{\frac{i}{2}\ell y^+}\, \langle 0\!\mid {\rm Tr}\!\left[ \overline{\textrm{T}}\{ Y_\nb^\dagger(y^+)\, Y_n(y^+)\} \, \textrm{T} \{Y_n^\dagger(0)\, Y_\nb(0)\} \right]\mid\! 0\rangle\,,
\end{align}
and using this definition along with the definition for the jet function, Eq.~(\ref{jetfunction}), in Eq.~\eqref{HadTensSCET.2} we arrive at
\begin{eqnarray}\label{HadTensSCET.3}
W^{\mu \nu} &=& -g^{\mu \nu}_{\perp}Q |\mathcal C(Q, -Q)|^{2} {N_c}  \int dr  J_{\bn}(Qr,\mu) \int d \ell S(\ell,\mu) \int \frac{d  y^{+}}{4\pi} e^{-\frac{i}{2} (r+\ell)y^{+}}\nonumber \\
& \times & \frac{x}{2N_{c}}\sum_{X_{n}}\textrm{Tr} \langle 0 | \frac{\slashchar{\bar n}}{2} {\chi}_{n}(y^{+}) | H X_{n} \rangle \langle H X_{n} | \bar \chi_{n}(0) | 0 \rangle \, .
\end{eqnarray}
However, as we noted previously, there can be no radiation that is collinear (with any soft-collinear overlap removed) to the heavy quark as this would result in a final state with invariant mass of order $Q^{2}$ which is not part of the endpoint regime. Thus in the endpoint the sum over $X_{n}$ only has a nonvanishing contribution from $X_{n} = 0$, and we can make further simplifications:
\begin{eqnarray}\label{collinearfactor}
\frac{x}{2N_{c}}\sum_{X_{n}}\textrm{Tr} \langle 0 | \frac{\slashchar{\bar n}}{2} {\chi}_{n}(y^{+}) | H X_{n} \rangle \langle H X_{n} | \bar \chi_{n}(0) | 0 \rangle 
 & = & \frac{x}{2N_{c}}\textrm{Tr} \langle 0 | \frac{\slashchar{\bar n}}{2} {\chi}_{n}(y^{+}) | H  \rangle \langle H  | \bar \chi_{n}(0) | 0 \rangle \\
 & &   \hspace{-1cm} = e^{-i\bn\cdot p^{\textrm{res}}_{H}y^{+} }\frac{x}{2N_{c}}\textrm{Tr} \langle 0 | \frac{\slashchar{\bar n}}{2} {\chi}_{n}(0) | H  \rangle \langle H  | \bar \chi_{n}(0) | 0 \rangle \nonumber \\
 & &  \hspace{-1cm} \approx e^{iQ(1-x)y^{+}} C_{H/Q}(\mu)\,, \nonumber
 \end{eqnarray}
where we use $\bn\cdot p^{\textrm{res}}_{H} = -Q(1-x)$, and expanded $x$ around 1 in the last line. 
Using this in Eq.~\eqref{HadTensSCET.3} we arrive at the factorized expression for the hadronic tensor in \SCETa
\begin{eqnarray}\label{HadTensSCET.4}
W^{\mu \nu} &=& - g^{\mu \nu}_{\perp}N_{c}   H_{Q}(Q,\mu) C_{H/Q}(\mu) \, Q \int \!\df r\,J_\nb(Q r,\mu)\,S( Q(1-x)-r,\mu)\,
\end{eqnarray}
where $H_{Q}(Q,\mu) \equiv  \, |\mathcal C(Q, -Q)|^2$ is the hard matching coefficient.

\subsection{Matching onto $\SCETM$}\label{sec:SCETM}

Matching $\SCETI$ onto $\SCETM$ removes virtualities of order $Q^2(1-x)$. Since the ultra-soft contributions from $\SCETI$ describe fluctuations of order $Q^2(1-x)^{2}\ll Q^2(1-x)$ they are still dynamical degrees of freedom within $\SCETM$, and the partonic ultra-soft function defined in Eq.~(\ref{eq:softfct}) becomes the partonic soft function in $\SCETM$. The collinear degrees of freedom in $\SCETI$ have virtualities of  order $Q^2(1-x)$, while the collinear degrees of freedom in $\SCETM$ have virtualities of order $m_{Q}^2 \ll  Q^2(1-x)$. However the collinear factor defined in Eq.~(\ref{collinearfactor}) remains unchanged because it is only sensitive to the minus component of the residual momentum which is the same in both  $\SCETI$ and $\SCETM$. Thus the partonic collinear function of $\SCETI$ becomes the partonic collinear function of $\SCETM$, and only the jet function, which involves degrees of freedom with  virtualities of ${\cal O}((1-x)Q^2)$, is integrated out. As a result 
the factored form of the differential cross section in $\SCETM$ looks identical to the one in $\SCETI$:
\begin{eqnarray}\label{eq:facM}
W^{\mu \nu} &=& - g^{\mu \nu}_{\perp}  N_c H_{Q}(Q,\mu) C_{H/Q}(\mu) \, Q \int \!\df r \,J_\nb(Q r,\mu)\,\tilde{S}( Q(1-x)- r,\mu)\,.
\end{eqnarray}
There is no ultra-soft function in $\SCETM$ since all ultra-soft Wilson lines are contracted to the same point in space-time and therefore cancel. 
While the factored form of the differential cross section in $\SCETM$ looks nearly identical to the one in $\SCETI$ the power counting in the two theories is different. In $\SCETM$, $\lambda=m_Q/Q$, and in  the definition of the soft function $\tilde{S}$ the ultra-soft Wilson lines $Y_n$ in Eq.~\eqref{eq:softfct} have to be replaced by soft Wilson lines $S_n$ in which the gluon momenta scale as $(\lambda,\lambda,\lambda)\,Q\sim (m_Q,m_Q,m_Q)$ and are completely decoupled from the collinear degrees of freedom.

While Eq.~(\ref{eq:facM}) is formally correct it is neither convenient for calculating the running of the $\SCETM$ differential cross section, nor for calculating the matching coefficient when the scale $m_{Q}$ is integrated out. While both the running and matching can be determined indirectly, it is both edifying and gratifying to have a direct calculation of each. Towards this end we reorganize $\SCETM$ by explicitly separating out the collinear quark mode that has label momentum $p^{\mu}= m_{Q}v^{\mu}$, where $v^{\mu}$ is the heavy quark velocity. We will call this mode the massive-bin. In addition, we will separate out soft-collinear modes from soft modes. After subtracting UV divergences from which the anomalous dimension can be extracted, the combination of the massive-bin and soft-collinear modes matches directly onto the HQET shape function, while the remainder gives the matching coefficient.

To be specific, once again consider the scaling of the $\SCETM$ degrees of freedom: collinear momenta scale as $p^{\mu}_{c}\sim ( m^{2}_{Q}/Q, Q,m_{Q})$ and soft momenta scale as $k^{\mu}_{s}\sim (m_{Q},m_{Q},m_{Q})$. Below the scale $m_{Q}$, the correct EFT is HQET in a boosted frame, where the heavy quark has momentum $p_{h}^{\mu}
=m_{Q}v^{\mu}+k^{\mu}$ with $v^{\mu}=(m_{Q}/Q,Q/m_{Q},0)$, and residual momentum scaling as $k^{\mu}\sim (\LambdaQCD m_{Q}/Q, \LambdaQCD Q/m_{Q},\LambdaQCD)$. The residual momentum sets the scaling for the gluonic and light quark degrees of freedom. Note that the heavy-quark degree of freedom has $p^{-}_{h}\approx m_{Q}v^{-}\approx Q$, so it is contained within the collinear degrees of freedom in $\SCETM$. How about the residual momentum, which appears to be both soft and collinear: is it a subset of the collinear or of the soft degrees of freedom of $\SCETM$? This can be determined by comparing the largest component of the residual momentum $k^{-}\sim \LambdaQCD Q/m_{Q}$ with the soft scaling in $\SCETM$. At the energies we are considering, for $b$ quarks, taking $\LambdaQCD \approx 0.25$ GeV we find $\LambdaQCD Q/m_{Q} \approx 4.5 \, \textrm{GeV} \approx m_{b}$. This implies that the residual momenta in HQET are a subset of the soft modes of $\SCETM$. 

Now we will calculate the different contributions in Eq.~(\ref{eq:facM}) while separating out the massive-bin from the collinear contribution and the soft-collinear part of the soft contribution. In the endpoint there can be no real collinear radiation into the final state due to momentum conservation, and as a consequence the collinear factor $C_{H/Q}(\mu)$ is purely virtual. Using dimensional regularization (DR) the naive one-loop virtual collinear contribution (including the heavy quark self-energy contribution) is:
\bea
\label{naivevirtcollm}
\tilde V_{n} = 
\delta(1-z) \frac{\alpha_{s}C_{F}}{2 \pi}\bigg[ \frac{1}{\epsilon^{2}} + \frac{1}{\epsilon} \ln\bigg(\frac{\mu^{2}}{m^{2}_{Q}}\bigg)+ \frac{1}{2 \epsilon}+\frac{1}{2} \ln^{2}\bigg(\frac{\mu^{2}}{m^{2}_{Q}}\bigg)
+\frac{1}{2}  \ln\bigg(\frac{\mu^{2}}{m^{2}_{Q}}\bigg)+2+\frac{\pi^{2}}{12}\bigg]\,.
\eea 
From this we have to subtract the zero-bin (the overlap of soft and collinear) $V^{\slash \!\!\! 0}_{n}$ and the massive-bin (overlap of heavy and collinear) $V^{m\slash \!\!\! 0}_{n}$, and we have to add back the overlap of all three $V^{\slash \!\!\! 0,m\slash \!\!\! 0}_{n}$. So the subtracted virtual collinear contribution is
\beq
V_{n}= \tilde V_{n} -V^{\slash \!\!\! 0}_{n}-V^{m\slash \!\!\! 0}_{n}+V^{\slash \!\!\! 0,m\slash \!\!\! 0}_{n}\,.
\eeq
The massive bin $V^{m\slash \!\!\! 0}_{n}$ is zero in DR while $V^{\slash \!\!\! 0}_{n}=V^{\slash \!\!\! 0,m\slash \!\!\! 0}_{n}$, 
leaving $V_{n}= \tilde V_{n}$. To find the virtual soft piece $V_{s}$ we take the naive soft contribution $\tilde V_{s}$ and subtract the 
overlap with the soft-collinear contribution $V_{s}^{m\slash \!\!\! 0}$. However $\tilde V_{s} = V_{s}^{m\slash \!\!\! 0}$ so $V_{s} = 
0$. Finally we need to include the massive-bin virtual contribution $V_{m}$. At one-loop in DR this is zero. Thus the total virtual 
contribution is
\beq
V_{\textrm{tot}}=V_{n}+V_{s}+V_{m}=  \tilde V_{n}. 
\eeq

The collinear contribution to real radiation is zero, which leaves only soft and soft-collinear radiation. Once again in DR the real soft contribution is zero. Thus the total real contribution is given by real soft-collinear radiation in the massive-bin, $R_{\textrm{tot}}=R_{m}$
:
\bea
\label{realsoftcollinear}
R_{m}&=& 
\frac{\alpha_{s}C_{F}}{2 \pi}\bigg\{-\frac{1}{\epsilon^{2}}\delta(1-z)+ \frac{1}{\epsilon} \frac{2}{(1-z)_{+}}  - \frac{1}{\epsilon} \bigg[\ln\bigg(\frac{\mu^{2}}{m^{2}_{Q}}\bigg)-1\bigg]\delta(1-z) \\
&& \hspace{-7ex}
+\bigg[ \ln\bigg(\frac{\mu^{2}}{m^{2}_{Q}}\bigg)-\frac{1}{2}\ln^{2}\bigg(\frac{\mu^{2}}{m^{2}_{Q}}\bigg)-\frac{\pi^{2}}{12} \bigg]\delta(1-z) + \frac{2}{(1-z)_{+}}\bigg[ \ln\bigg(\frac{\mu^{2}}{m^{2}_{Q}}\bigg)-1\bigg] -4 \bigg[ \frac{\ln(1-z)}{(1-z)}\bigg]_{+} \bigg\}\,.\nn
\eea
Adding Eq.~(\ref{naivevirtcollm}) to Eq.~(\ref{realsoftcollinear}) gives the total $\SCETM$ amplitude. First let us consider the pieces that are singular in the $\epsilon \to 0$ limit:
\beq
{\cal A}_{\textrm{tot}}^{\textrm{sing}}= V_{\textrm{tot}}^{\textrm{sing}}+R^{\textrm{sing}}_{\textrm{tot}}
= \frac{1}{\epsilon} 
\frac{\alpha_{s}C_{F}}{2 \pi} \bigg[\frac{2}{(1-z)_{+}}+\frac{3}{2}\delta(1-z)\bigg].
\eeq
These divergences are canceled by the $\SCETM$ counterterm $Z^{-1}_{M}(z)=\delta(1-z)-\delta_{M}(z)$, with 
\beq
\delta_{M}(z) = \frac{1}{\epsilon} \frac{\alpha_{s}C_{F}}{2 \pi} \bigg[\frac{2}{(1-z)_{+}}+\frac{3}{2}\delta(1-z)\bigg].
\eeq
This expression is the $z\rightarrow 1$ limit of $P_{qq}$ in Eq. \eqref{Pqq}, and  satisfies the consistency condition
\beq
Z^{-1}_{M}(z)=Z_{J}(z)Z_{H}\,,
\eeq
where~\cite{Manohar:2003vb}
\begin{equation}
Z_{J}(z) = \delta(1-z) +\frac{\alpha_{s} C_{F}}{\pi}\bigg[ \bigg( \frac{1}{\epsilon^{2}} +\frac{3}{4 \epsilon} -\frac{1}{\epsilon}\ln \frac{Q^{2}}{\mu^{2}}\bigg) \delta(1-z) -\frac{1}{\epsilon}\frac{1}{(1-z)_{+}}\bigg]\,,
\end{equation}
is the jet-function counter-term, and 
\begin{equation}
Z_{H} = 1+\frac{\alpha_{s} C_{F}}{\pi}\bigg( - \frac{1}{\epsilon^{2}} -\frac{3}{2 \epsilon} +\frac{1}{\epsilon}\ln \frac{Q^{2}}{\mu^{2}}\bigg)
\end{equation}
 is the counter-term for the hard coefficient $H_Q(Q,\mu)$.

Including the counterterm results in a finite expression for the NLO $\SCETM$ amplitude, and,
as we will see in Section \ref{sec:bHQET}, the finite part of the  real diagrams $R_m$ in Eq. \eqref{realsoftcollinear} is exactly reproduced by the bHQET shape function.
Thus what is left over in the matching between $\SCETM$ onto bHQET is the finite part of Eq. \eqref{naivevirtcollm}, 
which gives the matching coefficient $C_m(m_{Q},\mu)$.

\subsection{Flavor threshold}\label{sec:flavorthr}

In $\SCETM$ we can run the theory down to the $b$ mass $m_b$. At this scale we match the theory with five active flavors to a theory with four flavors, where $b$ quarks are frozen out.
The matching can be done at any scale $\mu_M \sim \mathcal O(m_b)$, and, in our error analysis, we varied the flavor threshold $\mu_M$ between $m_b/2$ and $2 m_b$.  
To implement the flavor threshold we set $n_f=4$ at scales smaller than $\mu_M$, including the HQET matching coefficient at $\mu_M$\footnote{The matching to HQET is formally done after the matching at the flavor threshold. This can be reversed, but one has to be very careful in recalculating the matching coefficients.}, and $n_f=5$ above. Especially for the running it is crucial to run with $n_f=4$ below $\mu_M$ and $n_f=5$ above $\mu_M$ to preserve the consistency relations.

In addition, two loop diagrams in which the heavy quark emits a gluon, and the gluon splits in a $Q \bar Q$ pair are not present in the four flavor theory, and are reproduced by including a matching coefficient $C_{thr}(z)$ \cite{Neubert:2007je}  
\begin{align}\label{ma}
C_{thr}(z) = \delta(1-z) + \left(\frac{\alpha_s(\mu_M)}{2\pi}\right)^2 C_FT_F c_2^{thr}(z)
\end{align}
starting at ${\cal O}(\alpha_s^2)$. The two-loop coefficient $c_2^{thr}$ is given in Eq.~\eqref{eq:c2thr} and was obtained from Ref.~\cite{Melnikov:2004bm} by taking the most singular terms of the expression $F_Q^{C_F T_F}$.

The expression in Eq. \eqref{eq:c2thr} suggests that the threshold coefficient  $c_2^{thr}(z)$ depends both on the scales $m_Q$ and $m_Q (1-z)$, giving rise to unresummed large logarithms of $1-z$. 
As shown in Refs. \cite{Pietrulewicz:2014qza,Hoang:2015iva,Hoang:2015vua}, these large logarithms are rapidity logarithms, that arise because of the rapidity separation of collinear and soft secondary massive quark modes,
and can be resummed by solving a rapidity RGE. Because of the limited numerical impact of the threshold matching coefficient, we chose not to resum this class of logarithms.

The switch from $n_f=5$ to $n_f=4$ at $\mu_M$ has also to be implemented in the running of $\alpha_s$. We follow the procedure described in Ref.~\cite{Schroder:2005hy}: we run with five flavors from $m_Z$ to $\mu_M$, then 
we use the pole mass decoupling relation
given in Refs.~\cite{Schroder:2005hy, Chetyrkin:2005ia}, and continue the running with four flavors below $\mu_M$\footnote{Even though we utilize the pole mass relations, the mass used in the calculation is the $1S$-mass. This leads to logarithms of $m_{pole}/m_{1S}$, which can be neglected.}. This procedure leads to a discontinuity in $\alpha_s(\mu)$ at the flavor threshold.  We neglect the effects of other flavor thresholds. Note that for programming reasons we choose to use $\alpha_s(\mu_M, n_f=5)$ in the HQET matching coefficient, even though $n_f=4$ should be used. We have checked that the error introduced through this approach is negligible.

\subsection{Matching onto bHQET}\label{sec:bHQET}

The final step in obtaining the factorization formula in Eq.~\eqref{eq:finalfact} requires integrating out the mass of the heavy quark, $m_Q \gg \LambdaQCD$.
This is achieved by matching the product of the soft and collinear functions in $\SCETM$
onto a bHQET shape function
\begin{equation}\label{eq:matchbhqet}
 C_{H/Q}(\mu)\tilde{S}(\omega,\mu)  = m_H \,  C_m(m_Q,\mu) S_{H/Q}(\omega,\mu),
\end{equation}
where the fragmentation shape function $S_{H/Q}(\omega,\mu)$ is defined as~\cite{Neubert:2007je}
\begin{equation}\label{eq:fracbHQET}
S_{H/Q}(\omega,\mu) = \frac{1}{2 N_c} \sum_X  \langle 0 |\, \delta \left( \omega - i \n\cdot  \partial\right)\, \tilde{W}_n^\dagger h_{v,n}\, | H_v X\rangle
\langle H_v X |\,  \bar{h}_{v,n} \tilde{W}_n \, \nbstwo\, | 0 \rangle.
\end{equation}
Here $\omega$ corresponds to the residual momentum of the heavy meson {\it and} of the soft particles moving collinear to its direction, and is of order ${\cal O}(\bar n \cdot v \, \LambdaQCD)$.
The factor of $m_H$ in Eq. \eqref{eq:matchbhqet} arises from the normalization of bHQET states.
In order to create a heavy meson $H$ with mass $m_H > m_Q$, the residual momentum needs to be larger than $ \bar n \cdot v\, \bar\Lambda $,
with $\bar\Lambda = m_H - m_Q$, implying that the shape function has support in the region $\omega/\bar n \cdot v \in  [\bar\Lambda,+\infty)$. For simplicity, in what follows we will use the variable 
$\hat\omega =  \omega - \bar n \cdot v\, \bar{\Lambda}$, with support in $[0,+\infty)$.
The residual momentum  $\hat\omega$ is related to the momentum fraction by  $\hat{\omega} = m_Q \bar n \cdot v (1-z)/z \sim m_Q \bar n \cdot v(1-z)$. (Recall the boost from the center-of-momentum frame to the heavy quark rest frame induces the factor $\bar n \cdot v = Q/m_{Q}$.)

\begin{figure}
\center
\includegraphics[width=10cm]{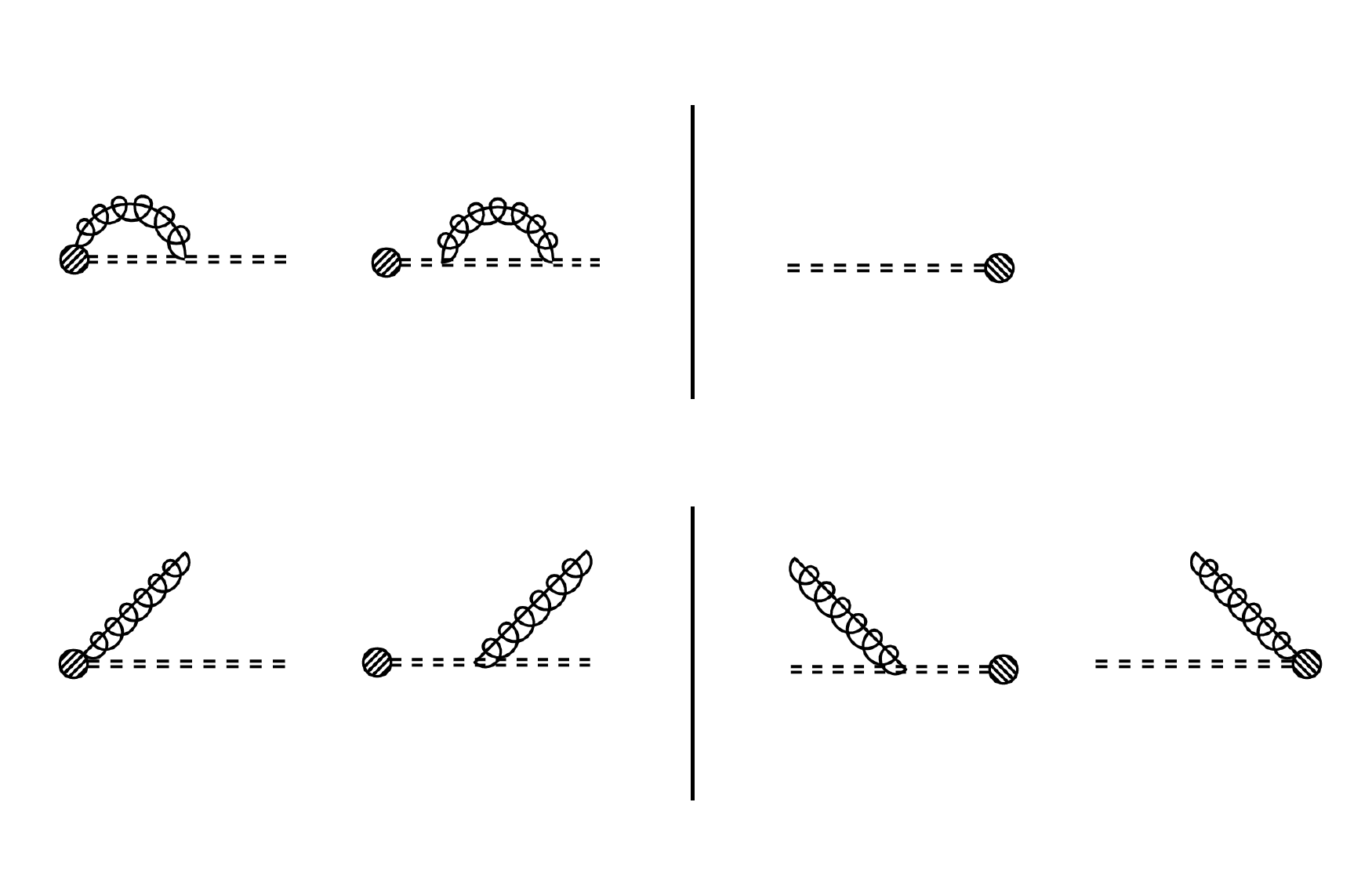}
\caption{Virtual and real $\mathcal O(\alpha_s)$ corrections to the bHQET shape function.
Double-dashed lines denote the boosted heavy quark. Springs denote soft-collinear gluons. }\label{fig:bhqet}
\end{figure}

The one-loop matching onto $\SCETM$ requires the computation of the diagrams in Fig. 1, which, subtracting the poles in the $\overline{\textrm{MS}}$ scheme, yield
\begin{eqnarray}\label{kernel2}  
m_Q S_{Q/Q}(\mu)  &=& m_Q \left( \delta\left( \frac{\omega}{\bar n\cdot v}\right)   - \frac{\alpha_s}{2\pi}  C_F  \left[
\delta\left( \frac{\omega}{\bar n\cdot v}\right)\frac{\pi^2}{12}  +   \left[  \frac{\theta(\omega) \bar n \cdot v }{\omega} \left(2 + 4 \log \frac{\omega}{\bar n \cdot v \mu}\right)\right]_+^{(\mu)}
 \right] \right) \nn \\
&= & \delta(1-z) + \frac{\alpha_s}{2\pi}\,C_F\, \left[\delta(1-z)\, \left(\,\ln\!\frac{\mu^2}{m_Q^2} - \frac{1}{2}\,\ln^2\!\frac{\mu^2}{m_Q^2}-\frac{\pi^2}{12}\,\right)  \right. \nn\\
& & \left. +\frac{2}{(1-z)_+}\,\left(\ln\!\frac{\mu^2}{m_Q^2} -1\right) - 4\, \left[\,\frac{\ln(1-z)}{1-z}\,\right]_+\right] \,. 
\end{eqnarray}
The plus distributions of the dimensionful variable $\omega$ are defined in Eq. (A.5), and, in the second line, we neglected terms of $\mathcal O(1-z)$.
Comparing this result to the finite part of the massive bin result in Eq.~\eqref{realsoftcollinear}, we see that the two are identical. The matching coefficient is then the finite part of Eq.~(\ref{naivevirtcollm}) 
\cite{Neubert:2007je}
\beq 
C_{m}(m_Q,\mu)= V_{\textrm{tot}}^{\textrm{finite}} = \frac{\alpha_{s}C_{F}}{2 \pi}\bigg[ \frac{1}{2} \ln^{2}\bigg(\frac{\mu^{2}}{m^{2}_{Q}}\bigg)
+\frac{1}{2}  \ln\bigg(\frac{\mu^{2}}{m^{2}_{Q}}\bigg)+2+\frac{\pi^{2}}{12}\bigg]\,. \label{matchingCM}
\eeq
The only physical scale appearing in $C_m$ is the scale of the heavy quark mass. The one loop calculation also allows us to extract the anomalous dimension of the shape function  $S_{H/Q}(\hat{\omega},\mu)$ and of the matching coefficient $C_m(m_Q,\mu)$, which we give in Appendix \ref{Equations}. 
In Ref. \cite{Neubert:2007je} it was shown that the perturbative expression of the fragmentation shape function equals the shape function that appears in $B$ decays at all orders.   
Using this result, the two-loop shape function can be extracted from Ref. \cite{Becher:2005pd}, and the two-loop mass coefficient $C_m$ is obtained by subtracting the two-loop shape function from the $z\rightarrow 1$ limit of the 
perturbative fragmentation function in Ref. \cite{Melnikov:2004bm}.

Substituting  Eqs. \eqref{eq:matchbhqet} and \eqref{eq:facM} in the differential cross section \eqref{qcdcrosssection} and integrating over $\cos\theta$, we arrive to an expression that closely resembles our final factorization formula \eqref{eq:finalfact}. The final step consists in expressing the bHQET shape function as a convolution of a perturbative piece $S_{Q/Q}$ and a nonperturbative hadronization model, $S^{hadr}_{H/Q}$. We discuss this step in Section \ref{sec:np}.

\subsection{Resummation}\label{resum}

The endpoint factorization formula, Eq. \eqref{eq:finalfact}, expresses the single inclusive heavy hadron production cross section in terms of four functions,
each of them dependent on a single scale and containing double logarithms of the ratio of this scale and the factorization scale $\mu$, 
as can be explicitly seen in the fixed order expressions, Eqs. \eqref{kernel2},  \eqref{matchingCM}, 
and Eqs. \eqref{hard.1}, \eqref{mass.1}, \eqref{jet.1}, and \eqref{shape.1}.
The $\mu$ dependence of the hard, jet, soft and mass functions is governed by RGEs
that resum these large logarithms, a resummation that, as we will see in Section \ref{sec:err}, is crucial to achieve a good description of the data.

The hard and mass coefficients satisfy the renormalization group equations
\begin{eqnarray}
\frac{\df}{\df \ln \mu} H_Q(Q,\mu) &= &- 2 \left[ \Gamma_{\textrm{cusp}}(\alpha_s) \, \ln \frac{\mu^2}{Q^2} + 2 \gamma_H(\alpha_s) \right]  H_Q(Q,\mu),   \label{rgeH.0}\\
\frac{\df}{\df \ln \mu} C_m(m_Q,\mu) &=& \left[ \Gamma_{\textrm{cusp}}(\alpha_s) \ln \frac{\mu^2}{m_Q^2} + 2 \gamma_M(\alpha_s) \right] C_m(m_Q,\mu), \label{rgeM.0}
\end{eqnarray}
where $\Gamma_{\textrm{cusp}}(\alpha_s)$ is the quark cusp anomalous dimension \cite{Korchemsky:1987wg,Korchemskaya:1992je,Moch:2004pa}, and 
$\gamma_H(\alpha_s)$ and $\gamma_M(\alpha_s)$ are the non-cusp anomalous dimensions.

The jet and shape function have convolution RGEs of the form
\begin{eqnarray}
\frac{\df}{\df \ln \mu} J_{\bar n} (Q r,\mu) &=& Q \int {\rm d} r^\prime \gamma_J(Q r - Q r^\prime, \mu)\, J_{\bar n}(Q r^\prime,\mu),  \label{rgeJ.0} \\
\frac{\df}{\df \ln \mu} S_{H/Q} (\omega, \mu) &=&  \int {\rm d} \omega^\prime \gamma_S(\omega - \omega^\prime, \mu)\, S_{H/Q}(\omega^\prime,\mu), \label{rgeS.0}
\end{eqnarray}
with anomalous dimensions given by
\begin{eqnarray}
\gamma_J(r,\mu) &=& - 2 \Gamma_\textrm{cusp}(\alpha_s) \left[\frac{\theta(r)}{r}\right]_+^{(\mu^2)}  + \gamma_J(\alpha_s)  \delta(r),   \label{anomJ}\\
\gamma_S(\omega,\mu) &=&  2 \Gamma_\textrm{cusp}(\alpha_s)\left[\frac{\theta(\omega)}{\omega}\right]_+^{(\mu)}  + 2 \gamma_S(\alpha_s)  \delta(\omega). \label{anomS} 
\end{eqnarray}
The plus distributions are defined in Eq. \eqref{dist.3}. Once again, the leading logarithmic structure is determined by the universal quark cusp anomalous dimension $ \Gamma_\textrm{cusp}$, while
$\gamma_J(\alpha_s)$ and $\gamma_S(\alpha_s)$ are the non-cusp components of the anomalous dimension.

The solutions of the RGEs have the form 
\begin{eqnarray}
H_{Q}(Q,\mu) &=& H_{Q}(Q,\mu_H) U_H(\mu,\mu_H), \quad J_{\bar n}(Q r, \mu) = Q \int {\rm d} r^\prime J_{\bar n}(Q r^\prime, \mu_J) U_J(Q r - Q r^\prime, \mu,\mu_J), \nonumber\\
C_{m}(m_Q,\mu) &=& C_{m}(m_Q,\mu_M) U_M(\mu,\mu_M), \quad
S_{H/Q}(\omega, \mu) =  \int {\rm d} \omega^\prime S_{H/Q}(\omega^\prime, \mu_S) U_S(\omega - \omega^\prime, \mu,\mu_S)\,, \nonumber
\label{solutions} 
\end{eqnarray}
where the evolution factors $U_H(\mu,\mu_H)$, $U_M(\mu,\mu_M)$, $U_J(r,\mu,\mu_J)$,  and $U_S(\omega,\mu,\mu_S)$ are given in Eqs. \eqref{solH}, \eqref{solM}, \eqref{kernJ} and \eqref{kernS}.
To minimize the logarithms in the fixed order expressions of $H_{Q}$, $J_{\bar n}$, $C_m$ and $S_{Q/Q}$ the initial scales $\mu_H$, $\mu_J$, $\mu_M$
and $\mu_S$ should be of order $Q$, $Q \sqrt{1-x}$, $m_Q$ and $m_Q (1-x)$, respectively.
We describe our choice of scales, and the scale variations we used to assess the residual scale dependence in  Section \ref{Merge}.

\begin{table}
\center
\begin{tabular}{|| c |c | ccc  c || c c||}
\hline \hline  
endpoint  & logs  & cusp & non-cusp & matching & $\beta(\alpha_s)$           \\
\hline 
LL 	& $\alpha_s^n L^{n+1}$ &  1 &  - & tree & $\beta_0$            \\
NLL 	& $\alpha_s^n L^{n}$   &  2 &  1 & tree & $\beta_1$            \\
N$^2$LL & $\alpha_s^n L^{n-1}$   &  3 &  2 & 1 & $\beta_2$             \\
N$^2$LL$^\prime$ & $\alpha_s^n L^{n-1}$ & 3 &  2 & 2 & $\beta_2$       \\
N$^3$LL & $\alpha_s^n L^{n-2}$  &  4 &  3 & 2 & $\beta_3$              \\
\hline \hline
\end{tabular}
\caption{Order counting for the resummation in the endpoint, $x\sim 1$. $L$ denotes the resummed logarithms, e.g. $\ln (1-x)$. In the third
to sixth columns we indicate the loop order at which each ingredient is needed to achieve a given logarithmic accuracy.
}  \label{ResumIngredients}
\end{table}

In Table \ref{ResumIngredients} we summarize the ingredients needed to achieve approximate N${}^3$LL accuracy in the endpoint.
We count logarithms in the exponent of the RGE kernels $U_I(\mu,\mu_I)$, with $I \in \{H,J,M,S\}$, and in the second column of Table \ref{ResumIngredients} we indicate the logarithmic series that we resum at each order.
In the third to sixth columns we show the loop order at which the cusp and non-cusp anomalous dimensions, the fixed order expressions of the hard, jet, mass and shape functions, and the QCD $\beta$ function are needed. 
The difference between the primed and the unprimed counting schemes is that in the primed scheme all fixed order series are considered at one order higher with respect to the unprimed \cite{Abbate:2010xh,Almeida:2014uva}.
All ingredients to achieve N${}^2$LL and N${}^2$LL$^\prime$ resummation, namely the three-loop cusp anomalous dimension and QCD $\beta$ function, the two-loop non-cusp anomalous dimension and
the two-loop fixed order expression of each function, are known \cite{Moch:2004pa,Becher:2006qw,Melnikov:2004bm,Becher:2005pd}. 
Most ingredients for N${}^3$LL resummation are also known, in particular the four-loop QCD $\beta$ function, and the three-loop non-cusp anomalous dimension of the hard and jet function  \cite{Moch:2004pa,Becher:2006qw}. The missing ingredients are the four-loop cusp anomalous dimension $\Gamma_3$, and the three-loop non-cusp anomalous dimension of the mass and shape functions,  $\gamma^S_2$ and $\gamma^M_2$, of which only the sum $\gamma^S_2+\gamma^M_2$ is known.
In our analysis, we use the Pad\'{e} approximation  for the unknown coefficient  $\Gamma_3$,
\begin{align}
\Gamma_3 = (1+e_\Gamma)\frac{\Gamma_2^2}{\Gamma_1}\,,
\end{align}
where $\Gamma_2$ and $\Gamma_1$ are the three- and two-loop cusp anomalous dimension,
and we vary $e_\Gamma$ between $-2$ and $+2$. 
For $\gamma_2^S$ and $\gamma_2^M$, we use
\begin{align}
\gamma_2^S = (1+e_\gamma)c_S\frac{(\gamma_1^S)^2}{\gamma_0^S}\,, \\
\gamma_2^M = c_M\frac{(\gamma_1^M)^2}{\gamma_0^M} - e_\gamma c_S\frac{(\gamma_1^S)^2}{\gamma_0^S}\,,
\end{align}
and fix the coefficients $c_M$ and $c_S$ by imposing that $\gamma_2^M +\gamma_2^S$ equals the $z\rightarrow 1$ limit of the anomalous dimension of the fragmentation function, Eq. \eqref{gamma2MS},
both for $n_f=5$ and $n_f=4$.
In our theory error budget, we vary the parameter $e_\gamma$  between $-2$ and $2$. 
As discussed in Section \ref{sec:err}, by varying $e_\Gamma$ and $e_\gamma$ and by comparing with the exact N${}^2$LL$^\prime$ results
we find the errors induced by missing orders in the cusp and non-cusp anomalous dimensions to be negligible.

The counting discussed so far applies to the resummation of double logarithms that appear in the endpoint.  Away from the endpoint the only relevant logarithms
are single logarithms of $m_Q/Q$, which are resummed by solving the DGLAP equation. For these logarithms, we adopt the standard nomenclature, that is we denote by N${}^k$LL 
the resummation of terms of the form $\alpha_s^n L^{n-k}$, achieved with $(k+1)$-loop splitting functions and $k$-loop initial condition. 
The maximum order we work at is N${}^2$LL, which requires three-loop time-like splitting functions, and two-loop fragmentation function.


\section{Nonperturbative effects}\label{sec:np}

The shape function \eqref{eq:fracbHQET} encodes physics at the scale $\LambdaQCD$, and is  a nonperturbative object,
which, like the parton distributions or the light quark fragmentation functions, needs to be extracted from data.
Following Ref. \cite{Ligeti:2008ac}, we express the HQET shape function  Eq.~\eqref{eq:fracbHQET} as a convolution of a partonic piece $S_{Q/Q}$, which is computed perturbatively,  and a hadronic nonperturbative piece $S_{H/Q}^{hadr}$
\begin{align}\label{eq:npfac1}
S_{H/Q}(\hat{\omega} + \bar n \cdot v \bar{\Lambda},\mu)=\int_{0}^{\infty}\!\df \hat{\omega}'\, S_{Q/Q}(\hat{\omega}-\hat{\omega}',\mu)\, S_{H/Q}^{hadr}(\hat{\omega}')\,.
\end{align}
Note that in the partonic picture the hadron and heavy parton masses are equal, $m_H=m_Q$. 
This means that $\bar{\Lambda}=0$ and $S_{Q/Q}$ has support on $[0,\infty)$. On the other hand $S_{H/Q}$ still has support on $[\bar{\Lambda},\infty)$, as we 
made explicit by using the variable $\hat\omega =\omega- \bar n \cdot v \,\bar{\Lambda}$. 

The nonperturbative function 
$S_{H/Q}^{hadr}(\hat\omega)$ is then expanded in a complete set of orthonormal functions, as described in Ref.~\cite{Ligeti:2008ac}: 
\begin{align}\label{eq:Shadr}
S_{H/Q}^{hadr}(\hat{\omega},\lambda,\{c_i\})=
 \frac{N_H}{\bar n \cdot v\, \lambda}\, \bigg[  \sum_{n=0}^N
c_n \, f_n\bigg(\frac{\hat\omega}{\bar n \cdot v\lambda}\bigg)
\bigg]^2,
\end{align}
where $N_H$ is the overall normalization of the shape function, the parameters $c_i$ are normalized $\sum_i c_i^2=1$ and the basis functions are orthonormal.
It is convenient to express $f_n$ in terms of the Legendre polynomials $P_n$,
\begin{eqnarray}
f_n(x) = \sqrt{ \frac{2 n +1}{2}\, y^\prime(x)} \, P_n(y(x)), 
\end{eqnarray}
where the change of variables $y(x)$ maps the interval $[0,\infty)$ into $[-1,1]$,
\begin{equation}
y(x) = - 1 + 2 \int_0^x \df x^\prime\, Y(x^\prime), \qquad Y(x,p) = \frac{(p+1)^{p+1}}{\Gamma(p+1)} x^p e^{-(p+1)x}.
\end{equation}
The shape function \eqref{eq:Shadr} is thus parametrized by the dimension one parameter $\lambda$,
the dimensionless parameter $p$, and $N$ independent coefficients $c_n$. For $N=0$,
\begin{equation}\label{model}
S_{H/Q}^{hadr}(\hat{\omega},p, \lambda,\{c_i\}) = \frac{N_H}{\bar n \cdot v \lambda} \frac{(p+1)^{p+1}}{\Gamma(p+1)} \left(\frac{\hat\omega}{\bar n \cdot v \lambda}\right)^p e^{-(p+1)\frac{\hat\omega}{\bar n \cdot v \lambda}},
\end{equation}
which is the model studied in Ref. \cite{Neubert:2007je}.  

The advantages of Eqs. \eqref{eq:npfac1} and \eqref{eq:Shadr} are many, and are discussed in detail in Ref. \cite{Ligeti:2008ac},
where this representation  was devised for the $B$ meson shape function that appears in $B$ decays. 
The most important are:
\begin{itemize}
\item 
the shape function has, by construction, the correct dependence on the renormalization scale $\mu$. This is captured by the perturbative function $S_{Q/Q}(\hat{\omega},\mu)$, which manifestly
satisfies the RGE,
\item the moments of $S_{H/Q}^{hadr}(\hat{\omega})$ are finite, and are related to matrix elements of local HQET operators,
\item after renormalon subtraction, the perturbative and nonperturbative components of the shape function are clearly factorized,
\item the uncertainty related to the unknown functional form of the shape function can be estimated by increasing 
the number $N$ of terms in the basis.
\end{itemize}

The arguments of Ref. \cite{Ligeti:2008ac} were developed for the decay shape function, but can be easily extended to the fragmentation shape function.
Here we briefly discuss the relation between the moments of the shape function and the matrix elements of local HQET operators, and the subtraction of renormalon ambiguities.

For $\hat{\omega}$ in the perturbative range, $\hat \omega/ \bar n \cdot v \gg \LambdaQCD$, the shape function can be expanded
as a series of local operators
\begin{equation}\label{OPE.1}
S_{H/Q}(\hat{\omega},\mu) = \sum_n C_n(\hat\omega,\mu) O_n.
\end{equation}
For $ n \leq 2$, the only possible operators in the operator product expansion are  of the form
\begin{equation}\label{opdef}
\mathcal O_n = \frac{1}{4 N_c}\sum_X \langle 0 | \left( \frac{i \bar{n} \cdot \partial}{\bar n \cdot v} -  \bar{\Lambda} \right)^n W^\dagger_n h_{v, n} | H_v X  \rangle \langle H_v X | \bar{h}_{v, n} W_n | 0 \rangle,
\end{equation}
while more complicated structures, like four-fermion operators with heavy and light quark fields, can appear for higher $n$.
The matching coefficients $C_n(\hat{\omega},\mu)$ can be computed by taking the matrix element of  both sides of Eq. \eqref{OPE.1} between quark states with zero residual momentum.
For $n \leq 2$, one can prove that
\begin{equation}\label{OPE.2}
C_n(\hat\omega,\mu) = \frac{(- \bar n \cdot v)^n}{n!}  \frac{\df^n}{\df\hat\omega^n} S_{Q/Q} (\hat\omega,\mu).
\end{equation}
In a similar way, for $\hat\omega \gg \hat\omega^\prime$ we can expand Eq. \eqref{eq:npfac1} as 
\begin{align}\label{eq:npfac2}
S_{H/Q}(\hat{\omega} + \bar n \cdot v \bar{\Lambda},\mu)= \sum_n \left( \frac{(-\bar n \cdot v)^n}{n!} \frac{\df^n}{\df \hat\omega^n} S_{Q/Q}(\hat{\omega},\mu) \right)  \int_{0}^{\infty}\!\df \hat\omega^\prime\, \left(\frac{\hat{\omega}'}{\bar n \cdot v}\right)^n\,  S_{H/Q}^{hadr}(\hat{\omega}')\,.
\end{align}
Thus, using the expression of Eq. \eqref{OPE.2} in Eq.\eqref{eq:npfac2}  and comparing to Eq. \eqref{OPE.1}
 we can see that  the moments of the nonperturbative shape function are related to local matrix elements
\begin{equation}\label{OPE.3}
\mathcal O_n = \int_{0}^{\infty}\!\df \hat\omega^\prime\, \left(\frac{\hat{\omega}'}{\bar n \cdot v}\right)^n\,  S_{H/Q}^{hadr}(\hat{\omega}').
\end{equation}

The relation to matrix elements of local operators is extremely useful in $B$ decays,  since it relates the first two nontrivial moments 
of the hadronic shape function to well known matrix elements, $\bar\Lambda$ and the matrix element of the kinetic operator $\lambda_1= \langle B | \bar h_v (i D)^2 h_v  | B \rangle $ \cite{Ligeti:2008ac}. 
In the case of fragmentation, Eq. \eqref{OPE.3} fixes the normalization of $S_{H/Q}^{hadr}$ to the nonperturbative parameter $\chi_H$ defined in Eq. \eqref{chiH}, 
\begin{equation}
N_H = \chi_H =  \frac{1}{4 N_c}\sum_X \langle 0 |  W^\dagger_n h_{v, n} | H_v X  \rangle \langle H_v X | \bar{h}_{v, n} W_n | 0 \rangle.
\end{equation}
Since we will be fitting to data for the production of all possible $b$-flavored hadrons, we can use the sum rule $\sum_H \chi_H =1$ and normalize the hadronic shape function to 1. 
The first moment of $S_{H/Q}^{hadr}$ is related to the matrix element
\begin{equation}\label{OPE.3b}
\mathcal O_1 = \frac{1}{4 N_c}\sum_X \langle 0 | \left( \frac{i \bar{n} \cdot \partial}{\bar n \cdot v} -  \bar{\Lambda} \right) W^\dagger_n h_{v, n} | H_v X  \rangle \langle H_v X | \bar{h}_{v, n} W_n | 0 \rangle,
\end{equation}
which is an unknown nonperturbative quantity. Thus, Eq. \eqref{OPE.3} does not put strong constraints on the fit to $e^+ e^-$ data that we perform in Section \ref{sec:err},
but rather the fit allows to extract unknown matrix elements like $\mathcal O_1$.

Eq. \eqref{eq:npfac1} hints at a separation of perturbative and nonperturbative contributions to the shape function. The former are captured by $S_{Q/Q}$, whose expansion in $\alpha_s$ we give in Appendix \ref{Equations},
the latter by the parameters of $S_{H/Q}^{hadr}$, which are fit to data.
However, it is known that in schemes like the pole mass scheme the shape function and its moments suffer from renormalon ambiguities \cite{Bosch:2004th,Ligeti:2008ac}.
For example, in the pole mass scheme the heavy quark mass $m_Q^{pole}$, and thus the first moment of the decay shape function $\bar{\Lambda}_{pole}=m_H-m_Q^{pole}$,
are sensitive to infrared dynamics \cite{Beneke:1994sw,Bigi:1994em}.
Using the pole mass in the perturbative calculations therefore introduces an ambiguity into the factorization of long- and short-distance physics in Eq.~\eqref{eq:npfac1}, which makes the position of the peak in the shape function, and the parameters in $S_{H/Q}^{hadr}$, unstable with respect to the perturbative expansion in $\alpha_s$. 
To remove this ambiguity one has to switch to a suitable short-distance mass scheme. This is done by introducing an additional scale at which perturbative short-distance and nonperturbative long-distance physics are separated, the subtraction scale $R$. The pole mass can then be replaced by $m_Q^{pole}=\hat{m}_Q(R,\mu)+\delta m_Q(R,\mu)$ 
with $\hat{m}_Q(R,\mu)$ independent of long-distance effects below $R$. 

The renormalon subtraction amounts to shifting perturbative corrections between $S_{Q/Q}$ and $S_{H/Q}^{hadr}$.
Here we closely follow  the  prescription of Ref.~\cite{Ligeti:2008ac}, in which a renormalon-free perturbative kernel $\hat{S}_{Q/Q}$
is achieved by demanding that the moments of $S_{H/Q}^{hadr}$ are free of renormalon
ambiguities. At $\mathcal O(\alpha_s^2)$, this can be accomplished by defining $m_Q$ and $\lambda_1$
in short distance schemes. We use the 1S scheme for the heavy quark mass \cite{Hoang:1998ng,Hoang:1998hm}, and the ``invisible scheme''
introduced in Ref.~\cite{Ligeti:2008ac} for the $B$ meson kinetic energy. We summarize the relevant formulae in Appendix \ref{renormalon}.

Note that in Ref.~\cite{Ligeti:2008ac} the renormalon subtraction was derived for $B$ decays, not fragmentation. However, Neubert~\cite{Neubert:2007je} argues that the perturbative expression of the HQET fragmentation shape function is identical to the perturbative shape function in $B$ decays, at all orders of perturbation theory.
Since the renormalon is subtracted by a shift of perturbative terms between $S_{Q/Q}$ and $S_{H/Q}^{hadr}$, the same subtraction terms which fix the renormalon ambiguity in $B$ decays also fix it in fragmentation. 

The effect of the renormalon, and of using a consistent short distance scheme for $m_Q$, is most important in the endpoint. Away from the endpoint, the perturbative fragmentation function
of Ref. \cite{Melnikov:2004bm,Mitov:2004du} was computed in the  pole mass scheme. In the numerical evaluations we nevertheless use $m_Q = m_Q^{1S} = 4.66$ GeV \cite{Agashe:2014kda}. 
This formally induces an error at $\mathcal O(\alpha_s^2)$, which we checked and found to be extremely small.

Finally, we notice that,  in order to combine the theoretical predictions in the endpoint and the tail, it is important to use the same nonperturbative model for both regions.
This involves some ambiguity, since the convolutions in the endpoint are naturally done in momentum space, while for $x$ away from the endpoint  it is convenient to retain a form 
in which the Mellin moments of the perturbative cross section and of the model factorize.
To achieve this, we replace Eq. \eqref{matchHQET} with 
\begin{equation}\label{FragFuncFull}
D_{H/i}(z) = \int_z^1 \frac{\df \xi}{\xi} \, d_{Q/i}\left(\frac{z}{\xi}\right) \tilde{S}_{H/Q}^{hadr}(\xi), 
\end{equation}
with $\tilde{S}_{H/Q}^{hadr}(\xi) = Q S_{H/Q}^{hadr}(Q(1-\xi))/\bar n \cdot v$. For $z\sim 1$, this convolution is  equivalent to \eqref{eq:npfac1}, as we illustrate in the next section.
As discussed in Section \ref{sec:fullx}, away from the endpoint the nonperturbative physics should be described by a single parameter, not a shape function. 
However, for $x \ll 1$, replacing Eq. \eqref{matchHQET} with Eq. \eqref{FragFuncFull} amounts to include a series of power corrections, suppressed by powers of $\LambdaQCD/m_Q$. Since we are working at leading power in $\LambdaQCD/m_Q$, using Eq. \eqref{FragFuncFull} is justified. 

In our analysis we set all $c_{i\ge 1}$ to 0, and use Eq. \eqref{model} as the model function. Thus we fit for only one parameter: $\lambda$. We checked that adding more coefficients $c_i$ has a negligible  effect  and choose to use variations in $p$, instead of $c_i$, to estimate the hadronic uncertainty since this appears to be the more conservative choice. We vary $p$ between $3$ and $5$ with a default value of $4$.
The first moment of the model function \eqref{model} is given by $\bar n \cdot v \lambda$, and, using  Eq. \eqref{OPE.3}, we find 
\begin{equation}
\lambda = \frac{1}{4 N_c}\sum_X \langle 0 | \left( \frac{i \bar{n} \cdot \partial}{n \cdot v} -  \bar{\Lambda} \right) W^\dagger_n h_v | H_v X  \rangle \langle H_v X | \bar{h}_v W_n | 0 \rangle.
\end{equation}
Since the residual momentum is always greater than $\bar \Lambda$, $\lambda$ is a positive number of $\mathcal O(\LambdaQCD)$. As discussed in Section \ref{sec:err}, we find good fits to $e^+ e^-$ data for $\lambda \sim 0.5$ GeV.



\section{Extending the description to the full $x$ spectrum}\label{Merge}

In Sections \ref{sec:fullx} we discussed the factorization of  single inclusive hadron production in $e^+ e^-$ annihilation away from the endpoint,
and how large logarithms of the ratio of the quark mass and center of mass energy $Q$ are resummed by the DGLAP evolution of the fragmentation function.
We then discussed  how for $x\sim 1$ two new scales arise, the jet scale $Q \sqrt{1-x}$ and the nonperturbative scale $m_Q (1-x)\sim \LambdaQCD$. 
An accurate description of the endpoint region thus requires the resummation of logarithms of $1-x$, and the inclusion
of a nonperturbative component of the fragmentation function.

In this section, we discuss how to combine the two descriptions, in order to describe the differential cross $\df \sigma/ \df x$ in the full $x$ range.
We express the differential cross section as:
\begin{equation}\label{eq:ext.1}
\frac{\df \sigma}{\df x} = \frac{\df \sigma_{QCD}}{\df x}  - \left. \frac{\df \sigma_E}{\df x} \right|_{\textrm{fix}}  + \left.  \frac{\df \sigma_E}{\df x} \right|_{\textrm{resummed}}.
\end{equation}
The first term  in Eq. \eqref{eq:ext.1} is the QCD cross section discussed in Eq. \eqref{QCDfinal}, and includes the resummation of single logarithms of $m_Q/Q$ through the DGLAP evolution.
This term gives  an accurate description of the intermediate $x$ region, but it lacks the resummation of logarithms of $1-x$ and the nonperturbative effects that are needed to
reproduce the peak. The last term is the endpoint cross section of Eq. \eqref{eq:finalfact}. In this expressions, single logarithms of $m_Q/Q$ and double logarithms of $1-x$ are correctly resummed, 
and a nonperturbative shape function describes the hadronization of the heavy quark into a hadron. However,  ${\df\sigma_E}/{\df x}$ does not contains powers of $1-x$, which, as one moves away from the peak region, 
become more and more important. 
In order to obtain a correct description in the whole range, and to avoid double counting between the endpoint and the QCD regions, we have to subtract the second term in Eq. \eqref{eq:ext.1}, which is equal to 
the endpoint cross section, with the resummation of logarithms of $1-x$ turned off. 
In practice turning off the resummation is accomplished by setting the soft scale equal to the mass scale $\mu_S = \mu_M$, and the jet scale equal to the hard scale $\mu_J = \mu_H$.

As is possible to explicitly verify using the formulae in App. \ref{Equations},
when $\mu_J = \mu_H$ the product of the hard coefficient $H_Q$ and the jet function $J_{\bar n}$ reproduces the $x\rightarrow 1$ limit of the QCD hard coefficient,
given at one loop in Eqs. \eqref{eq:QCD.3} and \eqref{eq:QCD.4}, and at two loops in Ref. \cite{Rijken:1996ns}.
The anomalous dimension for the product $H_Q \times J_{\bar n}$
is the $x\rightarrow 1$ limit of the QCD splitting functions. In a similar way, when $\mu_S = \mu_M$  the product of the mass coefficient $C_m$ and the bHQET shape function gives the $z\rightarrow 1$ limit of the QCD fragmentation
function, and the anomalous dimension of $C_m \times S	$ equals the $z\rightarrow 1$ limit of $P_{qq}$. These properties guarantee that the subtraction term in Eq. \eqref{eq:ext.1} will exactly cancel $\df\sigma_{QCD}/\df x$
as $x$ approaches 1, leaving only the resummed endpoint cross section.

\begin{figure}[ht]
\centering
\subfigure[]{%
  \includegraphics[width=6.5cm]{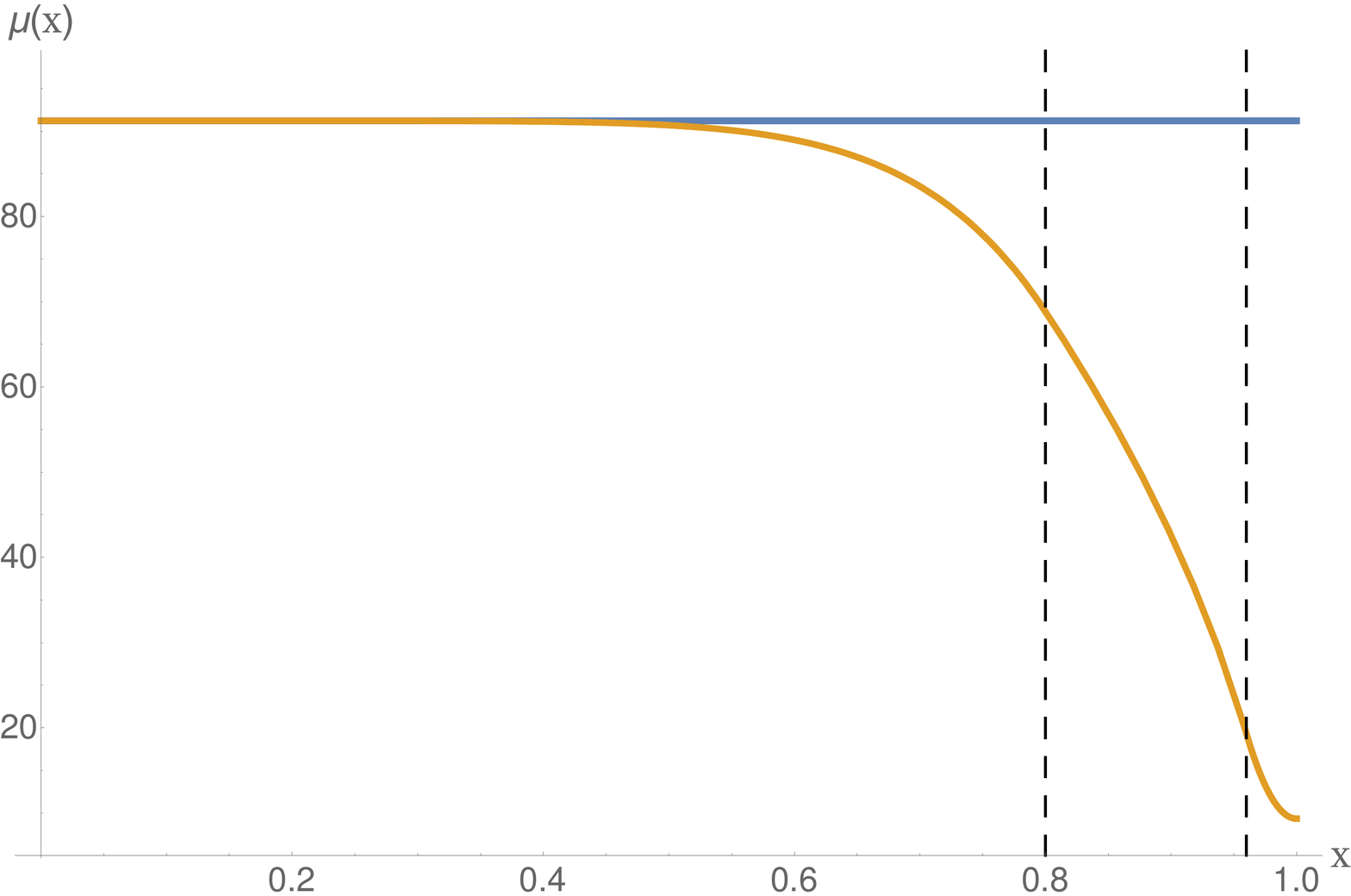}
  \label{profileHJ}}
\quad
\subfigure[]{%
  \includegraphics[width=6.5cm]{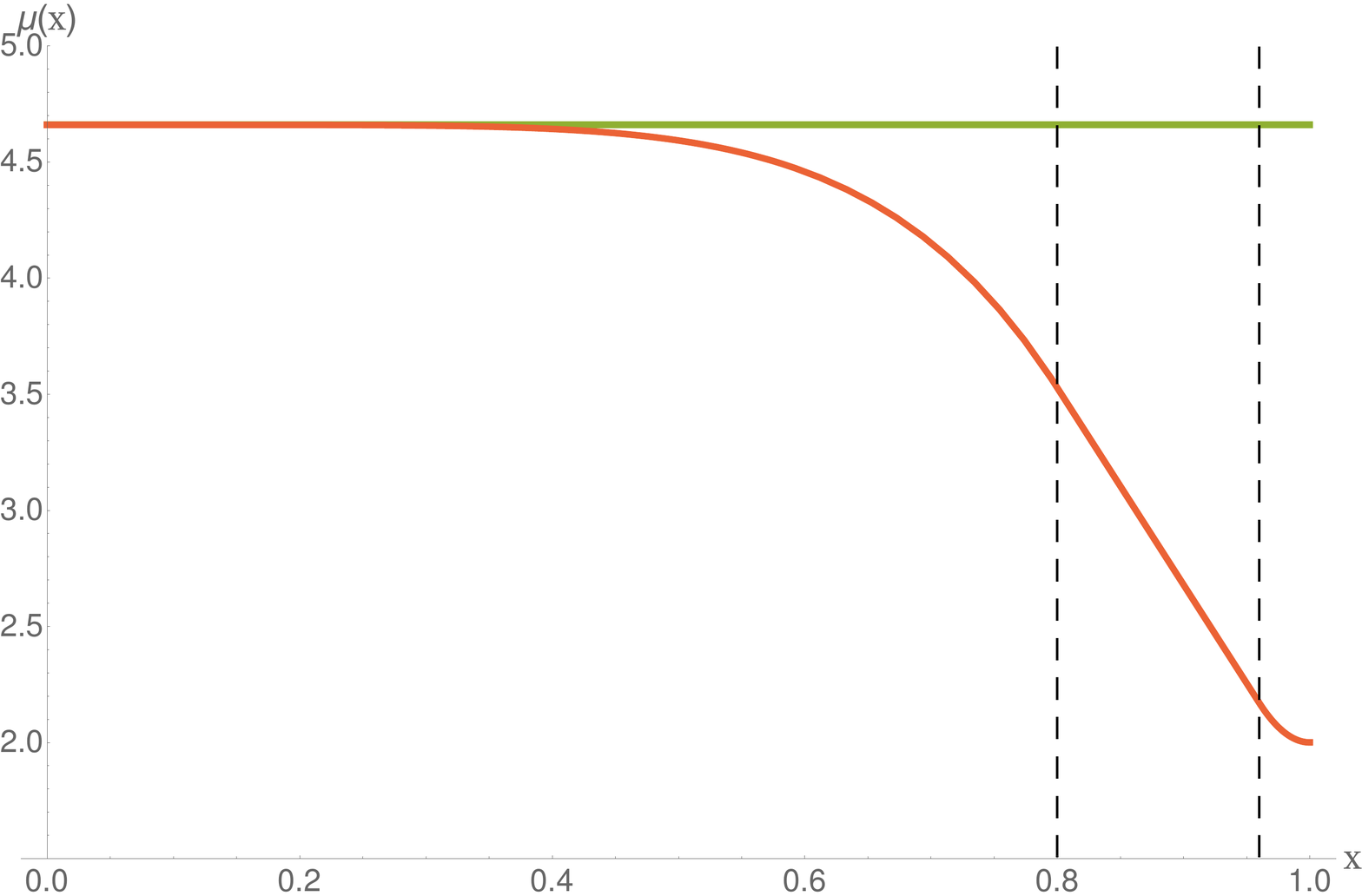}
  \label{profileMS}}
\caption{Profile functions. In the left panel (a), the blue and orange lines denote $\mu_H$ and $\mu_J$. In the right panel (b) the green and red lines denote $\mu_M$ and $\mu_S$. We plotted the profile function for the default values of the parameters $e_H$, $e_M$, $\mu_0^j$, $\mu_0^s$, $x_1$ and $x_2$. The points $x_1$ and $x_2$, which mark the transition between different regions, are denoted by vertical dashed lines.}
\label{profile}
\end{figure}

In order to ensure that away from the endpoint only the QCD cross section contributes we need the subtraction and endpoint terms to cancel for small $x$.
We accomplish this by using profile functions; that is by choosing $x$-dependent jet and soft scales. 
To determine the profile functions, we first choose the $x$-independent hard and mass scale as follows
\begin{equation}\label{pro1}
\mu_H = e_H Q, \qquad \mu_M = e_M  m_Q,
\end{equation}
where $e_H$ and $e_M$ are free parameters that we vary between 1/2 and 2.
The jet and soft scales must approach $\mu_H$ and $\mu_M$ in the limit $x \ll 1$,
and must have the correct scaling, namely $\mu_J \sim Q \sqrt{1-x}$ and $\mu_S \sim m_Q (1-x)$, 
in the resummation region. At very large $x$, we also need to make sure that all scales remain perturbative.
To satisfy these requirements, we choose the jet and soft scale as 
\begin{equation}\label{pro2}
\mu_J(x)=  
\begin{cases}
 \mu_H -c_j x^8 & 0\leq x\leq  x_{1} \\
 d_j + \sqrt{1-x} \, b_j  & x_{1}\leq x\leq x_{2} \\
 a_{j} (1-x)^2+\mu^j_{0} &  x_{2}\leq x\leq 1 \\
\end{cases} \qquad 
\mu_S(x) = 
\begin{cases}
 \mu_M- c_s x^6 & 0\leq x\leq x_{1} \\
 d_s+ b_s (1-x) &  x_{1}\leq x\leq x_{2} \\
 a_{s} (1-x)^2+\mu^s_{0} & x_{2}\leq x\leq 1, \\
\end{cases}
\end{equation}
where the coefficients $a_{j,s}$, $b_{j,s}$, $c_{j,s}$ and $d_{j,s}$ are chosen so that the profile functions are continuous, and with continuous derivatives.
An illustration of the behavior of the profile functions is shown in Fig. \ref{profile}.

Our choice of scales is thus determined by six parameters, $e_H$, $e_M$, $x_{1}$, $x_2$, $\mu_{0}^s$, and $\mu_0^j$, that we vary in order to assess the dependence of the theoretical  
cross section on missing orders of the perturbative expansion. 
The scales $\mu_0^j$ and $\mu_0^s$ represent the minimum values of the jet and soft scales, and are reached for $x$ very close to 1.
We chose $\mu_0^j = 9.32$ GeV, about twice the heavy quark mass, and in the error analysis we varied $\mu_{0}^j$ between $4.66$ and $18.64$ GeV, with the condition $\mu_{0}^j > \mu_M$.
For the soft scale, we chose $\mu_0^s = 2$ GeV as central value, and varied it between $1$ and $4$ GeV, with the condition $\mu_0^s < \mu_M$.
The range $x \in \{x_1, x_2\}$ is the range in which resummation is the most important. Our profile functions guarantee that in that interval $\mu_S$ and $\mu_J$ scale according to the power counting $\mu_J \sim  Q \sqrt{1 -x}$
and $\mu_S \sim m_Q (1-x)$. 
For $x<x_1$ the resummation is quickly turned off, and $\mu_J$ and $\mu_S$ become equal to $\mu_H$ and $\mu_M$ for $x \sim 0.5$. 
In the fits to data, we choose  $x_1 = 0.8$, close to the peak of the heavy quark fragmentation function, and vary $x_1$ between $0.7$ and $0.9$.
Our default choice for $x_2$, is $x_2 = 0.96$, and vary it between $0.9$ and $1$.

\begin{figure}
\center
\includegraphics[width=10cm]{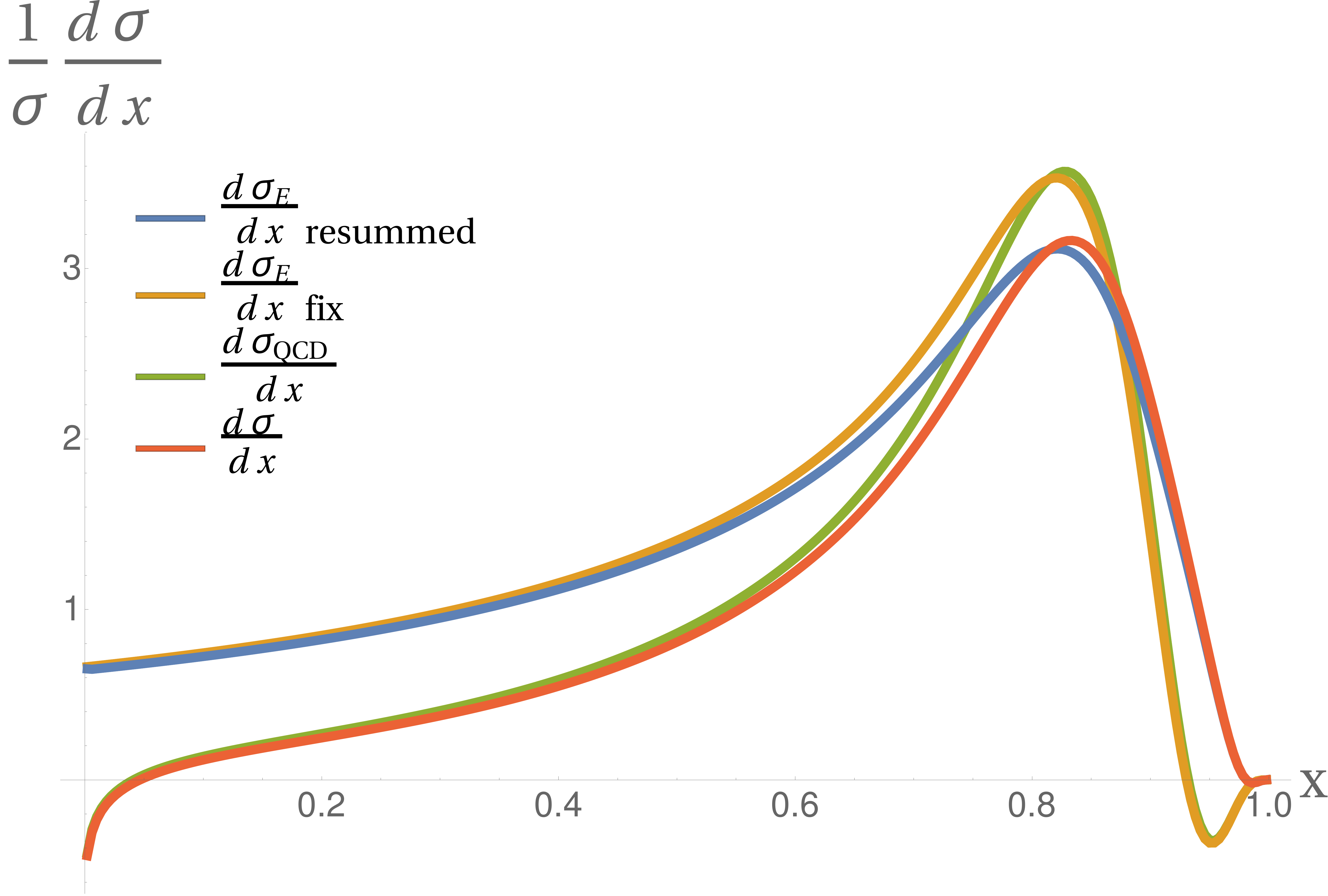}
\caption{The differential cross section as a function of $x$: the blue curve is the N${}^2$LO endpoint cross section with N${}^3$LL resummation, the orange curve is the endpoint result at fixed order ({\it i.e.} no resummation), the green curve is the N${}^2$LO QCD cross section, and the red curve is the combined cross section, Eq. \eqref{eq:ext.1}.}\label{combine}
\end{figure}

In Fig. \ref{combine} we plot: the N${}^2$LO endpoint cross section $\df\sigma_E/\df x$ with N${}^3$LL resummation of logarithms of $1-x$ (blue curve) and without resummation of logarithms of $1-x$ (orange curve),
the N${}^2$LO QCD cross section $\df\sigma_{\textrm{QCD}}/\df x$ (green curve), which includes only the resummation of DGLAP logs, and the combined cross section, $\df \sigma/\df x$,  given by Eq. \eqref{eq:ext.1} (red curve).
One can see that for $x \sim 0.5$, the resummation is turned off, and the contribution of $\df\sigma_E/\df x$ is completely canceled by the subtraction term in Eq. \eqref{eq:ext.1}, leaving only the QCD contribution.
On the other hand, at large $x$, the QCD cross section is dominated by terms singular in $1-x$, which are captured by  $\df \sigma_E/\df x_\textrm{fix}$, so that the combined cross section lines up with the endpoint resummed cross section.
These curves are produced using the model function in Eq. \eqref{model}, with $p=4$ and $\lambda = 0.5$ GeV. As discussed in Section \ref{sec:np}, in our calculations we  apply the nonperturbative model to all $x$. 
While this prescription is correct in the endpoint, it introduces an error in the tail region where the nonperturbative shape function should reduce to a single normalization constant $\chi_H$.
However, the size of the mistake we are making by multiplying the nonsingular terms  by the model is of order $\mathcal O(\LambdaQCD/m_Q)$. This is of the same size as other  power corrections which we neglect, and therefore justifies our treatment.

\section{Fits to $e^+ e^-$ data at the $Z$ pole}\label{sec:err}

The inclusive production of $b$-flavored hadrons in $e^+ e^-$ annihilation at the $Z$ pole has been measured by ALEPH~\cite{Heister:2001jg}, SLD~\cite{Abe:2002iq}, OPAL~\cite{Abbiendi:2002vt} and DELPHI~\cite{Abdallah:2011ep}.
All experimental collaborations give the normalized distribution $1/N \df N/ \df x$, where $x$ is the momentum fraction of the weakly decaying $B^+$, $B^0_d$ and $B^0_s$ mesons. 
The weakly decaying mesons are either produced directly after the hadronization phase, or result from the decay of a primary $B^{**}$ and $B^*$ meson. All experiments require the $B$ meson to be measured in a $b\bar  b$ event, with one $b$ or 
$\bar b$ quark in each hemisphere (the hemispheres are defined with respect to the event thrust axis). This requirement effectively eliminates the contributions of gluon or light quarks splittings into 
$b \bar b$ pairs.  Some experiments (e.g. DELPHI) assign events with four $b$-quarks, which also  require  $g \rightarrow b \bar b$ splittings, to the background.

We fit the theoretical cross section  simultaneously to all available data. Our fits include the correlation matrices given in the experimental papers. SLD only provides statistical correlations, not systematic correlations~\cite{Abe:2002iq}. We therefore treat the SLD data as having no systematic correlations, though for the other experiments the systematic correlations are larger than the statistical ones.\footnote{Using correlation models like the minimal overlap or the maximal overlap model for the systematic correlations leads to worse fit results.} OPAL quotes asymmetric systematic errors and correlations~\cite{Abbiendi:2002vt}. To simplify the analysis we symmetrized the correlations by using the arithmetic mean of the positive and negative correlations. While the correlation matrices should be positive semidefinite, with zero eigenvalues in the case of complete correlations between two measurements, some of the experimental correlation matrices contain negative eigenvalues. Since the experimental bins are highly correlated especially in the far-tail of the distribution (in the case of the OPAL experiment the bins are completely correlated at small $x$), the errors due to rounding the entries of the correlation matrix can cause negative eigenvalues. In a situation where some bins are highly correlated it is not sensible to treat them as independent degrees of freedom. Thus we fit only to the most significant eigenvalues of the data. We follow the prescription of DELPHI~\cite{Abdallah:2011ep} and take the effective number of degrees of freedom for ALEPH, OPAL, and DELPHI as $7$, $5$ and $7$, respectively. For SLD we use all $22$ bin values since, as mentioned earlier, the systematic correlation matrix is not given and the statistical correlations are modest.
The inclusion of OPAL data, even with the reduced weight assigned to them by the DELPHI procedure, leads in general to worse fits.   

We perform fits to the theoretical cross section computed at different orders. We denote by N${}^2$LO + N${}^3$LL the cross section with 
$\mathcal O(\alpha_s^2)$  fixed order expressions of the hard coefficient and perturbative fragmentation function, N${}^2$LL resummation of DGLAP logarithms, and N${}^3$LL resummation 
of endpoint logarithms, $\ln(1-x)$. N${}^2$LO +  N${}^2$LL$^\prime$ differ from  N${}^2$LO + N${}^3$LL only in the endpoint, where the resummation is carried out at N${}^2$LL$^\prime$. 
Finally, NLO +  N${}^2$LL denotes the cross section with $\mathcal O(\alpha_s)$ matching, NLL resummation of DGLAP logs, and N${}^2$LL resummation in the endpoint.

The N${}^2$LO + N${}^3$LL theory description includes $10$ theoretical parameters, beside the nonperturbative parameter $\lambda$ that we fit to data. 
Table~\ref{tab:theoryparams} lists all theory parameters together with their default values used for the best fit result and the range of variation we used in the error analysis.
The six parameters $e_H$, $e_M$, $\mu_0^j$, $\mu_0^s$, $x_1$ and $x_2$ govern the scale dependence of the theoretical prediction. 
$\mu_H = e_H Q$, with $Q=91.2$ GeV at the $Z$ pole, is the hard scale in the process at which the hard scattering coefficient $H_Q$ is evaluated. $\mu_M = e_M m_Q$ 
is the scale associated with the heavy quark mass.
$\mu_0^j$, $\mu_0^s$, $x_1$ and $x_2$ enter the parameterization of the profile functions, discussed in Section \ref{Merge}.
$\mu_0^j$ and $\mu_0^s$ are the minimal values that the jet and soft scales can assume. $x_1$ and $x_2$ determine the region where the resummation of logarithms of $1-x$ is important.

A complete N$^3$LL resummation requires the knowledge of the four-loop cusp anomalous dimension $\Gamma_3$
and of the three-loop non-cusp anomalous dimension of the shape function. At the moment, these two quantities are not known. To estimate them, we use the Pad\'{e} approximation, and allow for 200\% variations.
For $\alpha_s$ we use as our central value the world average, $\alpha_s(m_Z)=0.1185$, and vary it within the error quoted by the PDG \cite{Agashe:2014kda}.
The final parameter in Table \ref{tab:theoryparams}, $p$, is used to gauge the dependence of the fits on the hadronization model, Eq. \eqref{model}. 

To estimate the theory uncertainty we vary the parameters in Table \ref{tab:theoryparams} and calculate the best fit for each parameter set. We do this for all possible combinations of each parameter's default value and up and down variation.\footnote{We did constrain $\mu^j_0$ to always be larger than $\mu_M = e_M m_Q$ and $\mu^s_0$ to always be smaller than $\mu_M$.} 
For the N$^2$LO + N$^3$LL  analysis, we consider 45927 theory settings. For the lower order fits, N$^2$LO + N$^2$LL$^{\prime}$ and NLO + N$^2$LL, 
it is not necessary to include $\Gamma_3$ and $\gamma_S^2$ which reduces the number of theory settings to  5103.

\begin{table}[t]
\center
\begin{tabular}{||c|c|c||}
\hline
parameter\ & \ default value\ & \ range of values \ \\
  \hline
  $e_H$ & 1 & 0.5 to 2.0\\
  $e_M$ & 1 & 0.5 to 2.0\\
  $\mu^j_0$ & 9.32\,{\rm GeV} & 4.66 to 18.64\, {\rm GeV}\\ 
  $\mu^s_0$ & 2\,{\rm GeV} & 1 to 4\, {\rm GeV}\\ 
  $x_1$ & 0.8 & 0.7 to 0.9\\
  $x_2$ & 0.96 & 0.900 to 0.999\\
  \hline
  $\Gamma_3(n_f=5)$ & $1553.06$ & $-1553.06$ to $+4569.18$ \\
  $\gamma_2^{S}(n_f=5)$ & $1551.42$ & $-1551.42$ to $+4654.25$ \\
  $\alpha_s(m_Z)$ & $0.1185$ & $0.1179$ to $0.1191$ \\ 
  \hline
  $p$ & 4 & 3 to 5 \\
  \hline
\end{tabular}
\caption{Parameters relevant for the estimate of the theory uncertainty. We give the
default values and the range of values used in  the fitting
procedure.}
\label{tab:theoryparams}
\end{table}

To make our theory comparable to the experimental results we normalize the theoretical distribution by the 
integral of the hard coefficient
\begin{align}
\int_0^1 \,\df z \, H_{\textrm{ns}}(z,\mu_H) = 1.04\,,
\end{align}
for $\mu_H = 91.2$ GeV.

For the N$^2$LO + N$^3$LL fits, we calculate the $\chi^2$ for  $\lambda \in \{ 0.350, 0.400, 0.425, 0.450, 0.475, \\0.500, 0.525, 0.550, 0.600 \}$ GeV 
and use this $\chi^2$-grid to construct a $\chi^2$ interpolating function.  To compute the  $\chi^2$,  we bin the theoretical distribution, that is we integrate the theory distribution over the extent of
bins used by the experimental collaborations, and divide by the size of the bins.
The best fit value of $\lambda$ is the minimum of the $\chi^2$ interpolating function. The $1\sigma$ statistical uncertainty is obtained by finding the values of  $\lambda$ for which $\chi^2 = \textrm{min}(\chi^2) + 1$.
When working at NLO + N$^2$LL, we have to extend the $\chi^2$-grid to include  $\lambda = \{0.575, 0.625, 0.650,0.700\}$ GeV.

\begin{figure}
\center
\includegraphics[width=10cm]{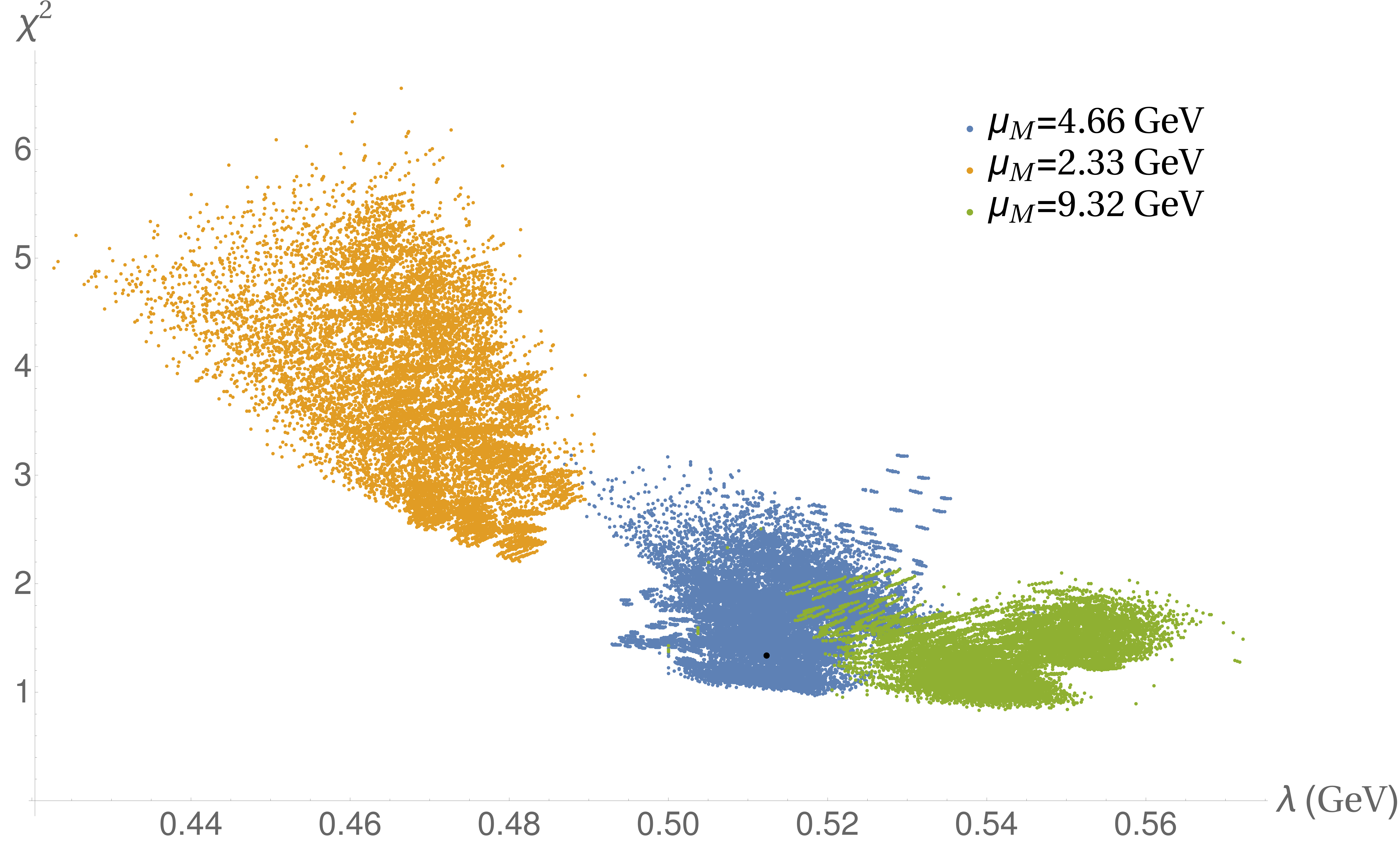}
\caption{$\chi^2/dof$ and best fit value of $\lambda$ for the N$^2$LO + N$^3$LL fits to the ALEPH, DELPHI, and SLD data. The black dot denotes the fit with default values of theory parameters.}
\label{chi2vslambda}
\end{figure}

\subsection{N${}^2$LO + N$^3$LL fits}\label{FitsA}

We obtain the central value of the parameter $\lambda$ by setting the theoretical
parameters to their default settings, summarized in Table \ref{tab:theoryparams}. 
We consider two cases: 
\begin{enumerate}
\item we perform simultaneous fits to all data over the complete $x$ range available,
\item we exclude data from OPAL.
\end{enumerate}
The best fit value of $\lambda$, the statistical error and the $\chi^2$ per degree of freedom ($\chi^2/dof$) in these two scenarios are 
\begin{eqnarray}
\lambda_1 & = 0.545 \pm 0.055 \, \textrm{GeV}  \qquad \chi^2/dof = 2.80,  \label{fitAll}\\
\lambda_{2} & = 0.512 \pm 0.070 \, \textrm{GeV}  \qquad \chi^2/dof = 1.33. \label{fitNoOPAL}  
\end{eqnarray}
Eqs. \eqref{fitAll} and \eqref{fitNoOPAL} make it clear that including data from the OPAL experiment  
pushes $\lambda$ to higher values and noticeably worsens the $\chi^2/dof$. A closer look at fits to individual experiments reveals that the fits to ALEPH, DELPHI and SLD
are good, with $\chi^2/dof$ between 1.5 and 0.8. However, the fit to OPAL data only is poor, with $\chi^2/dof = 13$, and does not ruin the  
simultaneous fits to all experiments only because of the small weight assigned to OPAL by the DELPHI prescription~\cite{Abdallah:2011ep}. 
The disagreement between the theoretical cross section and the OPAL data is mostly in the tail, and the effect on the $\chi^2$ is amplified by the 
small error on the data points in this region. 
As we will discuss in greater detail later in this section, we have indications that the theoretical setup 
we have adopted is overestimating the tail of the distribution, and that power corrections 
of $\mathcal O(\LambdaQCD/m_Q)$, which we have not included, could lead to a better description of the data. 
Since our theoretical cross section describes the OPAL data poorly, we will first focus our discussion on fits to ALEPH, DELPHI and SLD, and include OPAL data afterwards.  

The data from the LEP experiments ALEPH and DELPHI, and the SLAC experiment, SLD, show some tension in the peak region of the differential cross section.
We checked that excluding SLD data from the fit has a negligible effect on the value of $\lambda$ and on the quality of the fits.

We now turn to the discussion of the theoretical error, that we estimate by varying the theory parameters in the ranges described in Table \ref{tab:theoryparams}. 
In Fig. \ref{chi2vslambda} we show the distributions of $\chi^2/dof$ and best fit value of $\lambda$ for the N$^2$LO + N$^3$LL fits to ALEPH, DELPHI and SLD data.
The $\lambda$ and $\chi^2/dof$ obtained with the default theory setting is denoted by a black dot.
While in principle the best fit value of $\lambda$ and the goodness of the fit should not be impacted by (reasonable) choices of the theory parameters, we notice that
at N$^2$LO + N$^3$LL the cross section is still quite sensitive to the theory settings, in particular to the scale $\mu_M$, where we evaluate the QCD fragmentation function and start the DGLAP evolution. Changing
$\mu_M$ from $4.66$ GeV to $9.32$ GeV or $2.33$ GeV has the effect of both considerably shifting the best fit parameter, and changing the quality of the fit. The choice  $\mu_M = m_Q/2$
gives noticeably worse fits. In this case, the lowest value of $\chi^2/dof$ is 2.2, and the average $\chi^2/dof$ is much larger, $3.75$.
On the other hand, 86\% of the fits with $\mu_M = 4.66$ or $9.32$ GeV have $\chi^2/dof < 2$.

\begin{figure}
\includegraphics[width=15cm]{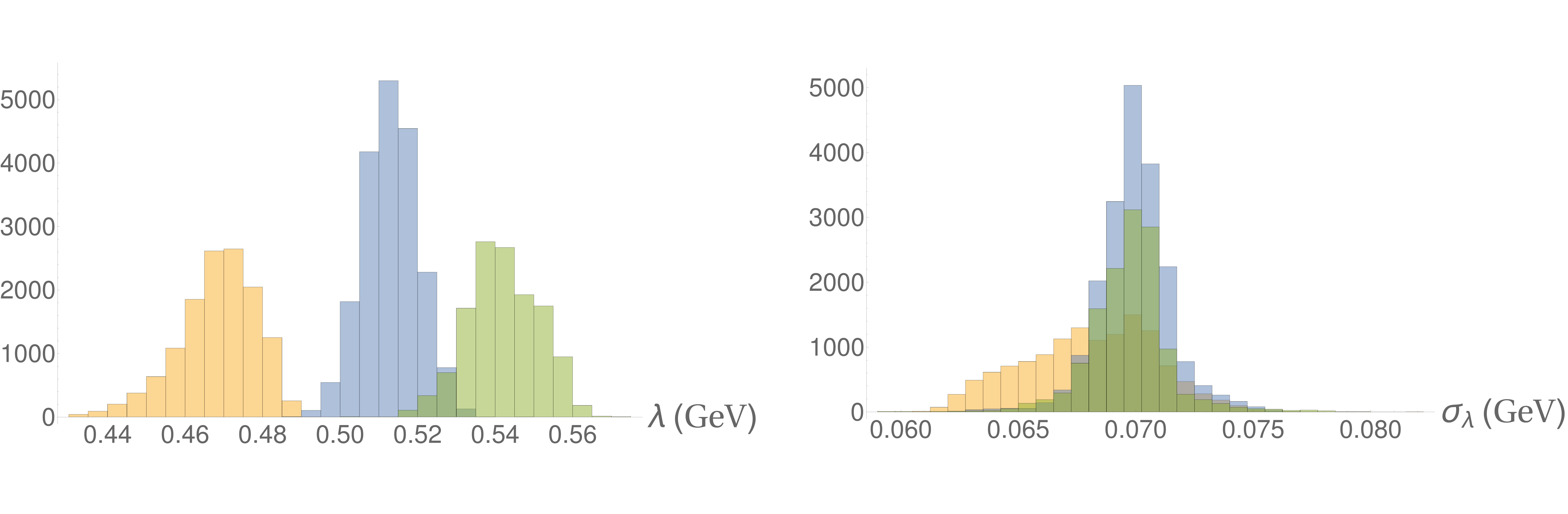}
\caption{Distributions of $\lambda$ and $\sigma_\lambda$ in the N$^2$LO + N$^3$LL fits. Color coding is the same as in Figure~\ref{chi2vslambda}.}
\label{Histo}
\end{figure}

In Fig. \ref{Histo} we show the distribution of $\lambda$ and of $\sigma_\lambda$, the statistical error on $\lambda$,  in the N$^2$LO + N$^3$LL fits.
We see that, for each value of $\mu_M$, varying the remaining theory parameters has little effect, 
resulting in a distribution with width of 10 MeV around the default values. 
However, the three distributions for different $\mu_M$ have little or no overlap, signaling a spurious dependence of the best fit value of $\lambda$ on the 
truncation of the perturbative expansion of the cross section.
The second histogram shows that the statistical error on $\lambda$ is roughly the same, at least for the values of $\mu_M$ which give good fits. 
Notice that the distance between the values of $\lambda$ obtained with different choices of $\mu_M$ is always within the statistical error on $\lambda$.

\begin{figure}[ht]
\centering
\subfigure[]{%
  \includegraphics[width=6.5cm]{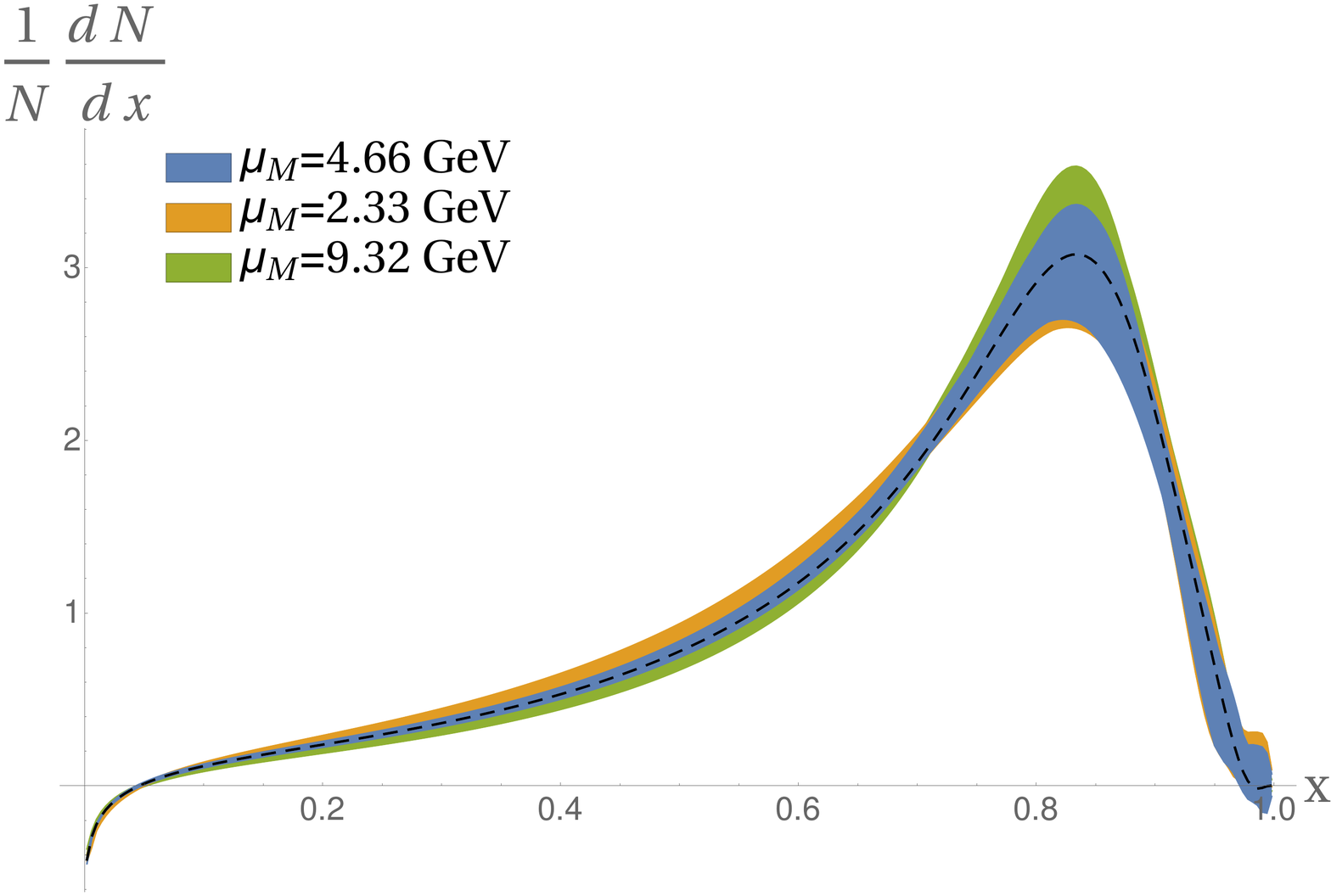}
  \label{envelopeMuM}}
\quad
\subfigure[]{%
  \includegraphics[width=6.5cm]{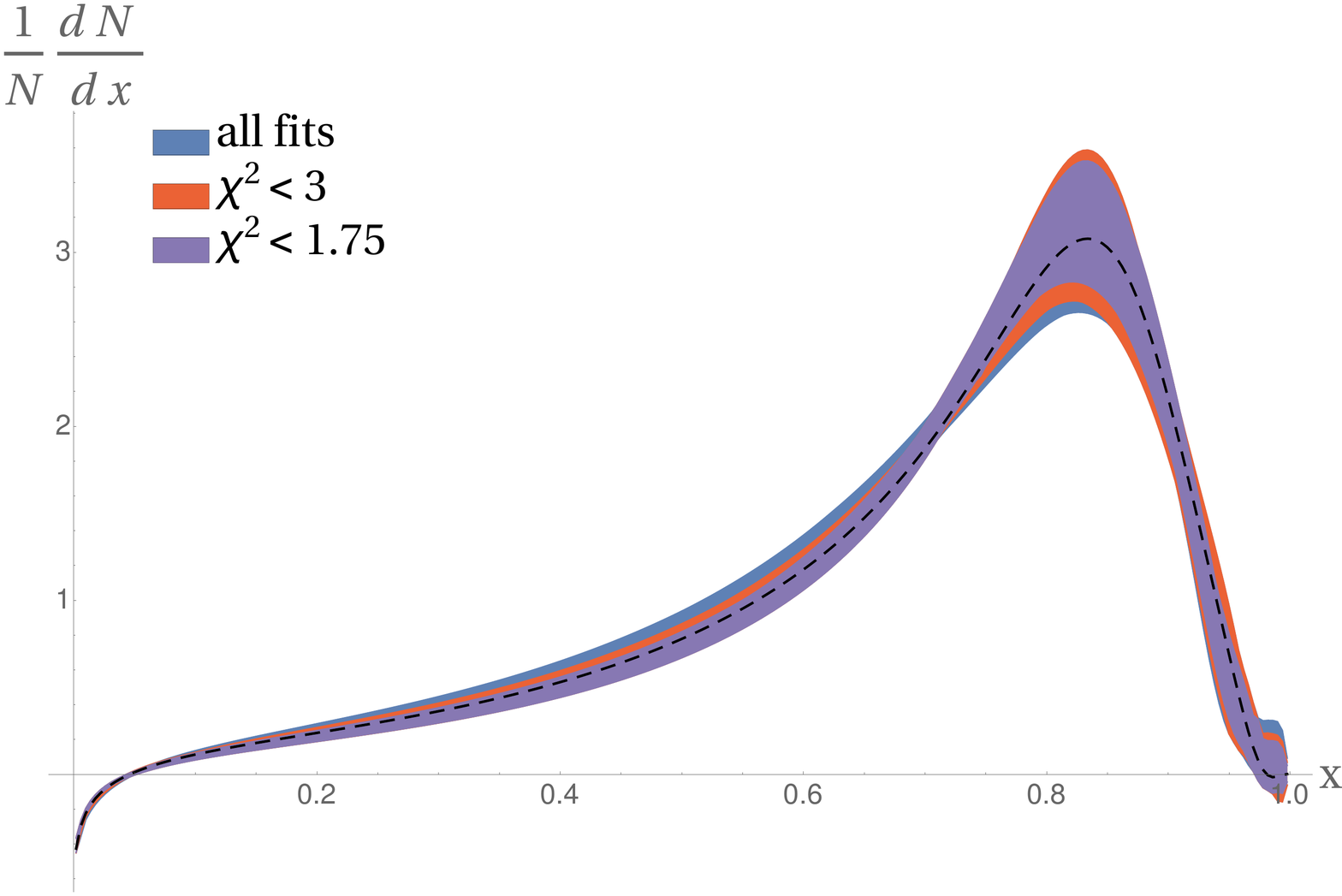}
  \label{envelopeChi2}}
\caption{ N${}^2$LO + N${}^3$LL differential cross section, including theory errors. The black dashed line denotes the curve obtained with default values of theory parameters. The left panel (a) displays the
effect of changing the values of $\mu_M$. The right panel (b) shows how excluding poorer fits reduces the width of the envelope of the curve.}
\label{envelope}
\end{figure}

In Fig. \ref{envelope} we show the differential cross section with theoretical error bands. 
The bands are obtained by considering, for each point in $x$,
the maximum and the minimum values of the differential cross section over all fit results.
The black dashed curve is obtained by setting the theory parameters to their default value.
In the left panel we highlight the dependence of the fit on the choice of $\mu_M $: the  blue band has $\mu_M = m_Q$, the yellow band has  $\mu_M = m_Q/2$, and the green band has $\mu_M = 2 m_Q$. 
It is clear that lowering the scale has the effect of raising the tail of the distribution, resulting in poorer fits. 
This feature is relatively independent of the model function.
In the right panel we highlight the effect of excluding fits with decreasing $\chi^2$: the (nearly hidden) blue band is the theoretical uncertainty obtained from all the fits, the red band is the theoretical uncertainty from fits with $\chi^2/dof < 3$ (which corresponds to including $80\%$ of all fits), and the purple band is the theoretical uncertainty from fits with $\chi^2/dof<1.75$ (which 
corresponds to including $50\%$ of all fits). The last choice excludes all fits with $\mu_M = 2.33$ GeV. 
Including only good fits reduces the width of the envelope, especially in the tail and intermediate $x$ region. The effect in the peak region is less important. 

\begin{figure}
\center
\includegraphics[width=12cm]{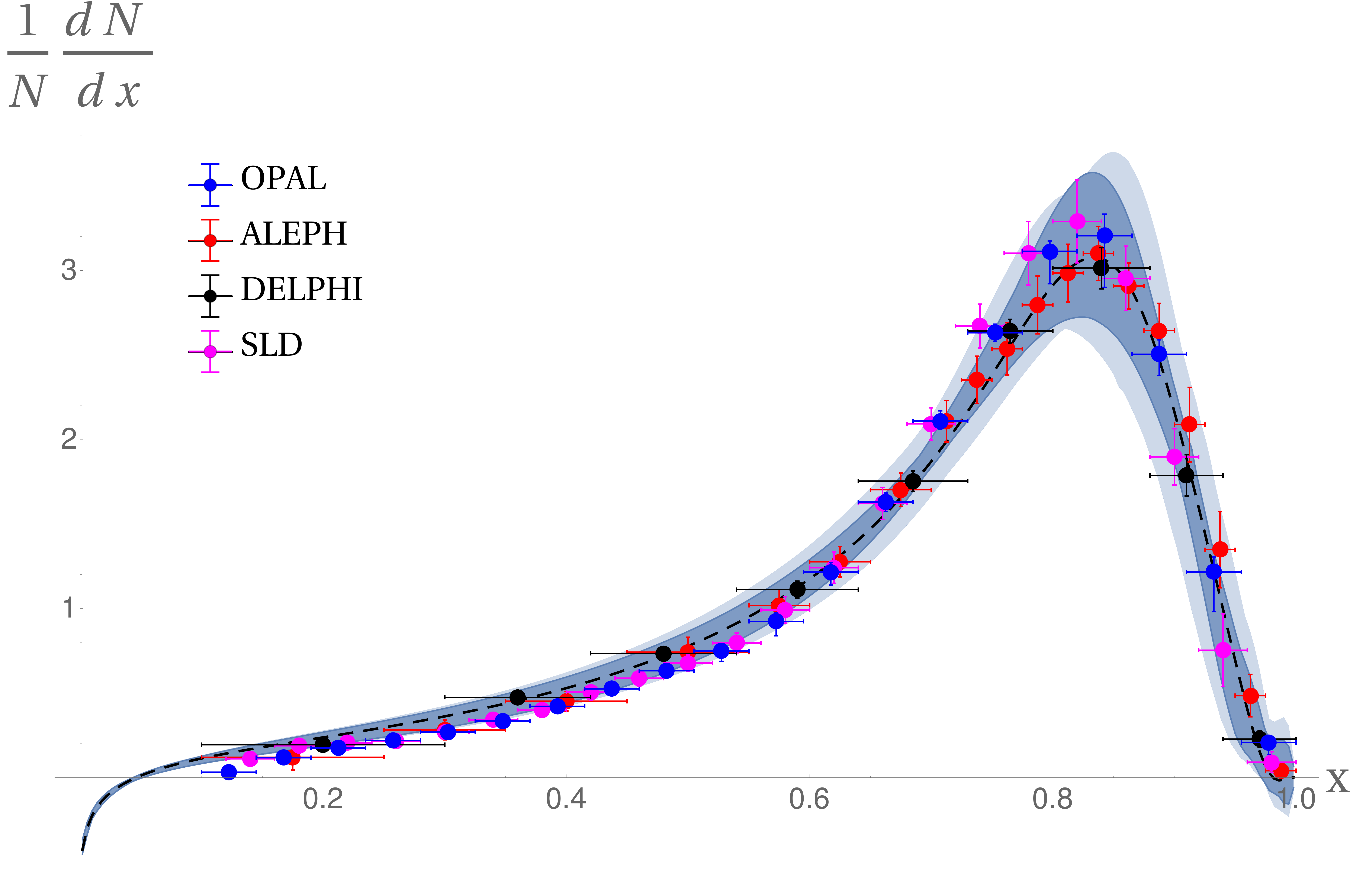}
\caption{Comparison of the N${}^2$LO + N${}^3$LL theoretical cross section to data. The black dashed line denotes the curve obtained with default values of theory parameters. 
The dark blue band denotes the theoretical uncertainty. The light blue band the combination of theoretical and statistical uncertainties.}
\label{envelopedata}
\end{figure}

\begin{figure}[ht]
\centering
\subfigure[]{%
  \includegraphics[width=6.5cm]{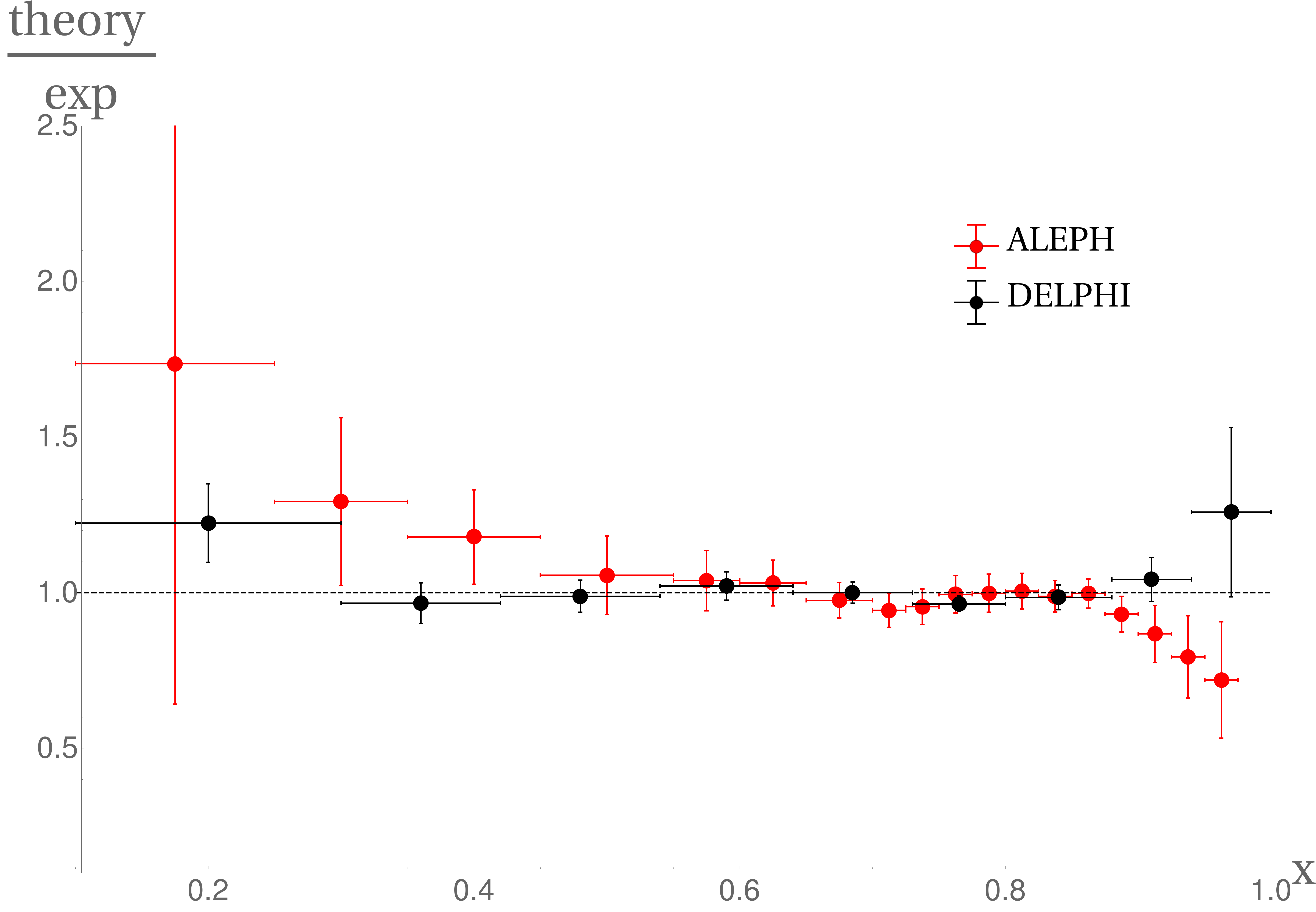}
  \label{ratioALEPH}}
\quad
\subfigure[]{%
  \includegraphics[width=6.5cm]{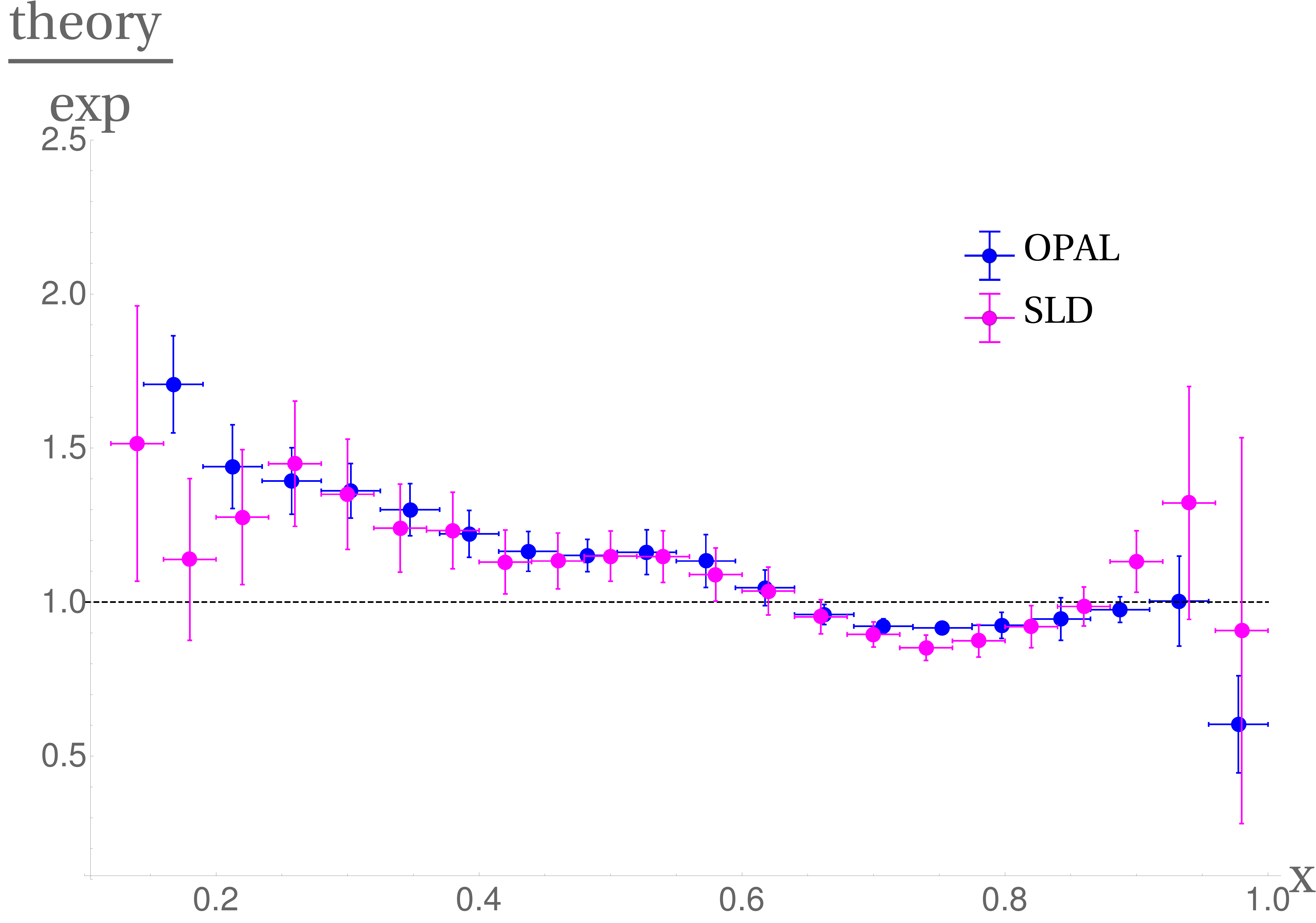}
  \label{ratioOPAL}}
\caption{
Ratio of the N${}^2$LO + N${}^3$LL theoretical cross section, obtained with default theory settings, to the data of the ALEPH and DELPHI experiments (left panel) 
and OPAL and SLD experiments (right panel). The error bars include only the experimental uncertainties of the data points.}
\label{ratiodata}
\end{figure}

Fig. \ref{envelopedata} compares the N$^2$LO + N$^3$LL theoretical cross section to data from the ALEPH, DELPHI, OPAL and SLD experiments.
The black dashed curve is obtained with the default theory setting.
Here the blue band includes all fits with $\chi^2 <3$. The light blue band shows the combination of the statistical and theoretical uncertainties. 
It is obtained by considering, for each point in $x$,
the maximum and the minimum values of the differential cross section over all fit results, with $\lambda$ set to the central value, or $\lambda \pm \sigma_\lambda$.
In Fig.~\ref{ratiodata} we show the ratio of the theoretical prediction, obtained with default theory settings, to the data of the ALEPH and DELPHI experiments (left panel) 
and OPAL and SLD experiments (right panel). The error bars in Fig.~\ref{ratiodata} include only the uncertainties of the data points.
The theoretical curves give a good description of the data. The broadness of the peak reflects differences between the experiments.
In the tail of the distribution, $x \lesssim 0.5$, the theory starts to overshoot some of the data points. While, with the exception of OPAL, the $\chi^2$ remains good, 
this effect might indicate the need to include power corrections in the matching of the QCD fragmentation function onto bHQET. 
In our theoretical framework, the tail of the distribution is not very sensitive to the nonperturbative model function $S_{H/Q}^{hadr}$. 
As discussed in Section \ref{sec:np}, the convolution with the model $S_{H/Q}^{hadr}$ corrects the partonic QCD fragmentation function $d_{\textrm{ns}}$
by including a series of  power corrections of order $\mathcal O(\LambdaQCD/m_Q)$. Therefore, the tail region of the differential cross section is a genuine QCD prediction, 
and cannot be easily adjusted by changing the parameters or the functional form of the hadronization model. 
Nonetheless, power corrections can be relevant, as they may be as large as $\lambda/m_Q \sim 10\%$. 
While the convolution with the hadronization model includes some of the power corrections, the small discrepancy we see in the tail suggests that it is important to systematically include all of them.
We will further investigate the issue in future work.

In the far tail, $x \lesssim 0.05$, the differential cross section becomes unphysical. Here the approximation of massless $b$ quark
breaks down, and power corrections of order $m_Q^2/Q^2$ need to be included. Since there are no experimental points in this region, we neglect this class of power corrections.

\begin{table}
\begin{tabular}{||c | c  | c | c | c | c | c||}
\hline
Data Sets & $\lambda$ (GeV)  & $\sigma_{\textrm{exp}}$ (GeV)  & $\sigma_{\textrm{th}}$ (GeV)  & $(\chi^2/dof)_\textrm{def}$ & $\langle \chi^2/dof \rangle$  & Order \\
\hline 
 &  0.545 & 0.055& $^{+ 0.030}_{-0.077}$ & 2.8 & 4.1   & N${}^2$LO + N${}^3$LL  \\
All &  0.547 & 0.055& $^{+ 0.027}_{-0.064}$ & 2.8 & 4.0   & N${}^2$LO + N${}^2$LL$^\prime$  \\
 &  0.592 & 0.053& $^{+ 0.045}_{-0.099}$ & 5.4 & 8.6   & NLO + N${}^2$LL  \\
\hline 
ALEPH,   & 0.512 & 0.070 & $^{+ 0.060}_{-0.089}$ & 1.3 & 2.2 & N${}^2$LO + N${}^3$LL \\
 DELPHI,  & 0.513 & 0.070 & $^{+ 0.050}_{-0.081}$ & 1.3 & 2.1 & N${}^2$LO + N${}^2$LL$^\prime$ \\
 SLD & 0.553 & 0.071 & $^{+ 0.097}_{-0.092}$ & 3.3 & 4.9 & NLO + N${}^2$LL\\
\hline
\end{tabular}
\caption{Best fit value of $\lambda$, with statistical and theoretical errors. The procedure used to assess the theoretical error is described in the text. $(\chi^2/dof)_\textrm{def}$
and $\langle \chi^2/dof \rangle$ denote the $\chi^2$ obtained with default theory settings, and the average of the $\chi^2$ over all fits. 
}\label{TabFits}
\end{table}

In Table \ref{TabFits} we summarize the best fit values of $\lambda$, the statistical and theoretical errors in the two cases described at the beginning of the Section. 
The best fit is obtained by setting the theory parameters to their default values. The theoretical error is given by taking the  difference between the best fit $\lambda$, and  the maximum (or minimum) value of $\lambda$
obtained by varying the theory settings. The fifth and sixth columns of Table \ref{TabFits} give the $\chi^2/dof$ in the case of default theory settings, and, as a measure of the quality of the fits when varying 
theory settings, the average $\chi^2/dof$ for the 45927 settings we considered.

\subsection{Convergence}\label{convergence}

\begin{figure}[t]
\centering
\subfigure[]{%
  \includegraphics[width=6.5cm]{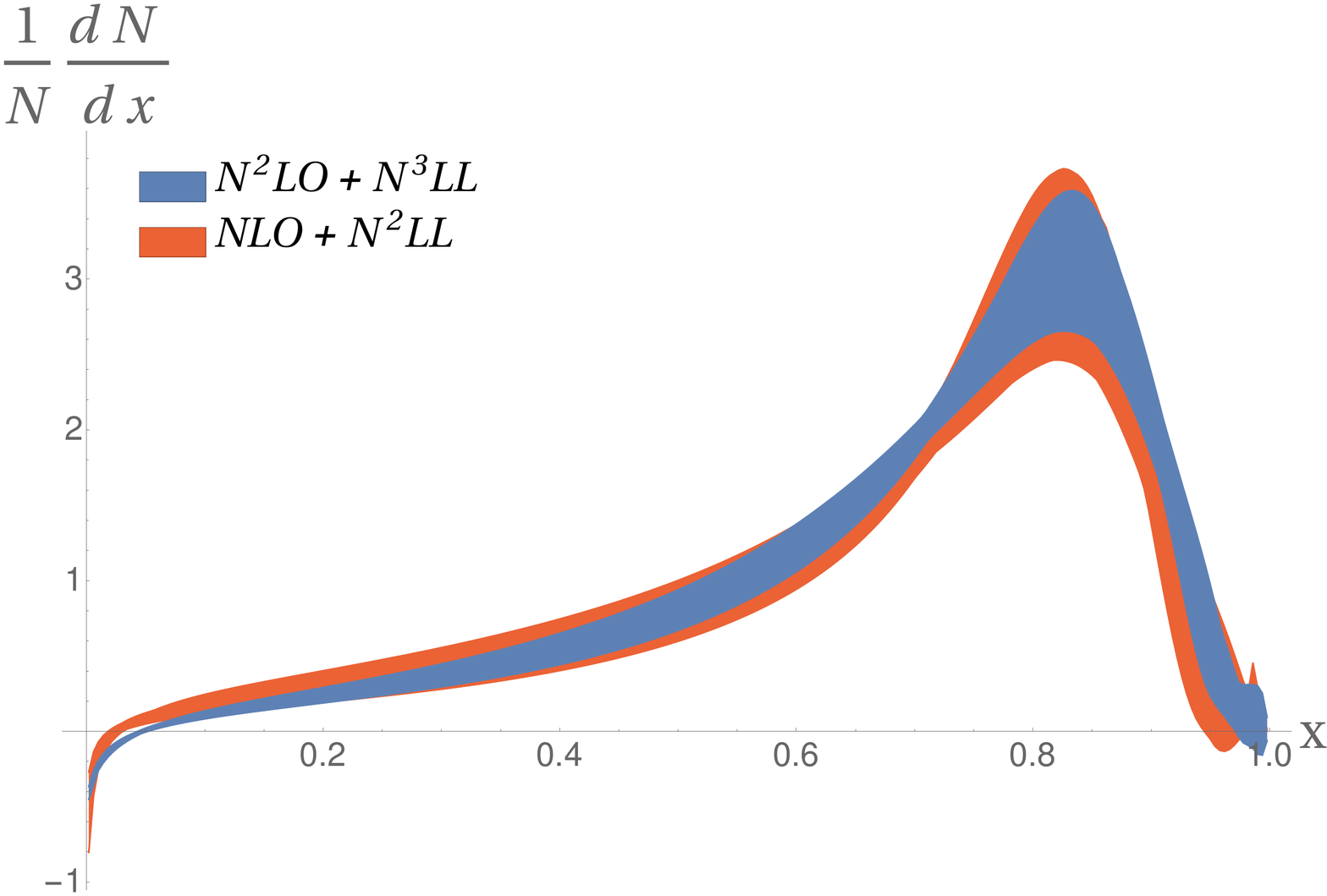}
  \label{NLOvsNNLOa}}
\quad
\subfigure[]{%
  \includegraphics[width=6.5cm]{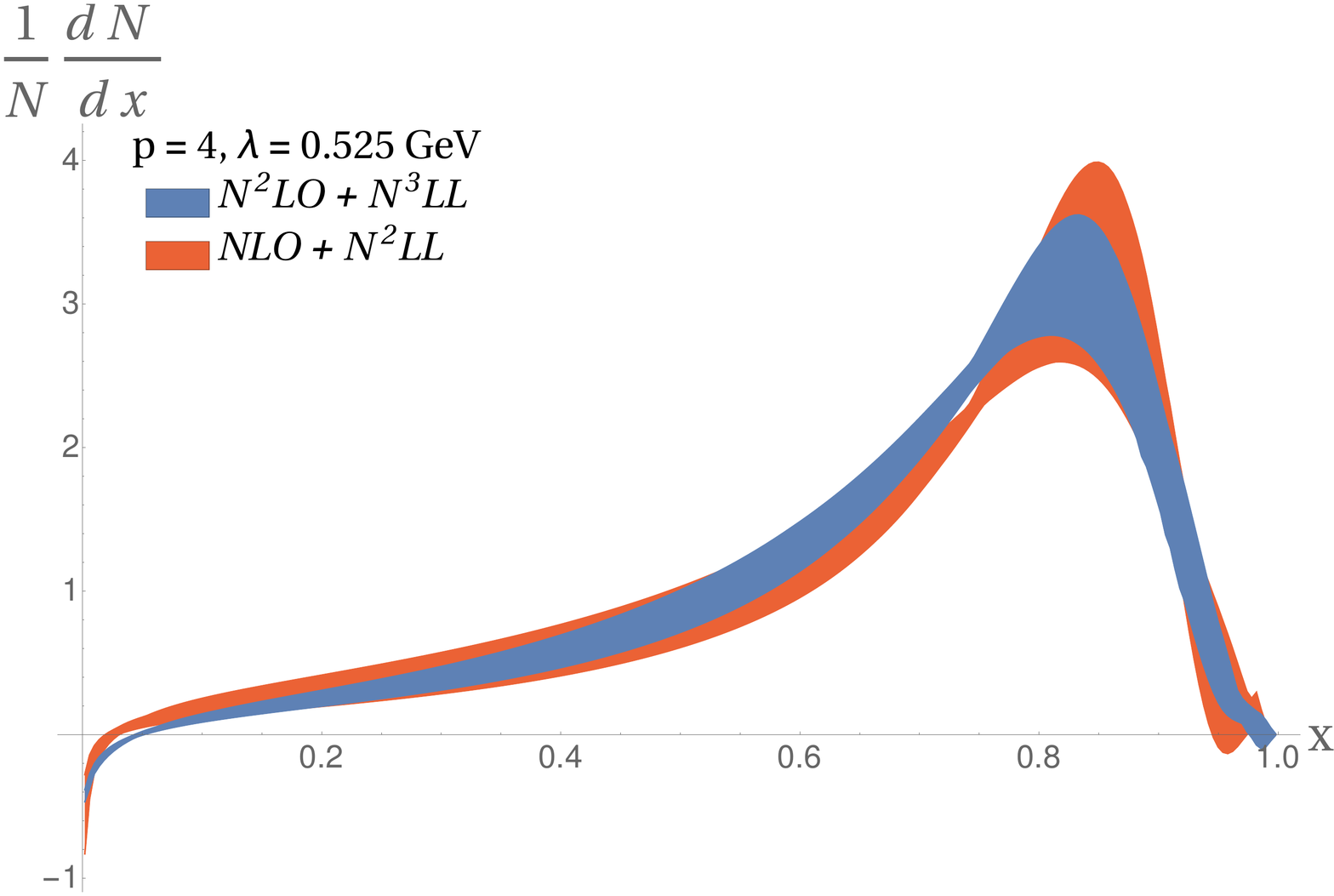}
  \label{NLOvsNNLOb}}
\caption{Comparison of the N${}^2$LO + N${}^3$LL and NLO + N${}^2$LL differential cross sections. In the left panel (a) we compare the results of the fitting procedure. 
In the right panel (b), we fix $p =4$ and $\lambda = 0.525$. The bands are obtained by varying the theory parameters in the ranges described in Table \ref{tab:theoryparams}. }
\label{NLOvsNNLO}
\end{figure}

To study the convergence of the perturbative series, we repeated the fits with lower order expressions,
namely N${}^2$LO + N$^2$LL$^\prime$ and NLO + N${}^2$LL. 
The ingredients included at each order are summarized in Section \ref{resum}.
As shown in Table \ref{TabFits}, the results at N${}^2$LO + N$^2$LL$^\prime$ are very close to the full analysis. This stresses the importance of including higher order 
corrections to the matching coefficients, in particular to the fragmentation and shape functions. Once the matching corrections are included, performing a complete N${}^3$LL resummation or limiting ourselves to N${}^2$LL makes little difference.

As Table \ref{TabFits} shows, the NLO + N${}^2$LL fits give larger values of $\lambda$, with  noticeably worse $\chi^2/dof$.
The NLO + N${}^2$LL cross section includes only terms that are strictly of $\mathcal O(\alpha_s)$, that is we discard  $\mathcal O(\alpha_s^2)$ terms from the product of the fragmentation function and short-distance cross section. 
We notice that including these terms, as was done in Ref. \cite{Cacciari:2005uk}, improves the agreement with the data. 
Since the full N${}^2$LO expressions are available \cite{Melnikov:2004bm,Mitov:2004du}, we decided not to include spurious 
$\mathcal O(\alpha_s^2)$ in the NLO cross section, and perform a complete N${}^2$LO analysis.

Notwithstanding the poor $\chi^2/dof$, we can use the NLO + N${}^2$LL fits to test the convergence 
of the perturbative expansion.  Fig. \ref{NLOvsNNLO} shows $\df\sigma/\df x$ at  NLO + N${}^2$LL (red band) and N${}^2$LO + N${}^3$LL (blue band). 
In the left panel, we show the results of the fits. The envelopes are obtained by considering, for each $x$,
the maximum and the minimum values of the differential cross section over all fit results.
We can see that the N${}^2$LO + N${}^3$LL band is narrower than and lies within the NLO + N${}^2$LL band. 
In the right panel, we show the comparison of the differential cross section at different orders, but for fixed value of $p = 4$ and $\lambda = 0.525$ GeV, so that the comparison in not contaminated by the effects of the experimental uncertainties and/or the poor agreement with data of the NLO cross section. Once again, it can be appreciated that the   
N${}^2$LO + N${}^3$LL band is significantly narrower than the NLO + N$^2$LL band, which indicates a reduction of the theory uncertainties. Furthermore,
the right panel of Fig. \ref{NLOvsNNLO} shows that, for fixed $\lambda$, the inclusion of N${}^2$LO corrections (in particular the corrections to the QCD fragmentation function)
causes an increase of the differential distribution in the region $x \in (0.5,0.7)$, compensated by a slightly lower peak. The N${}^2$LO shape follows the data more closely,  with a consequent improvement in $\chi^2$.

\subsection{Dependence on the value of $\alpha_s(m_Z)$}

\begin{figure}
\center
\includegraphics[width=12cm]{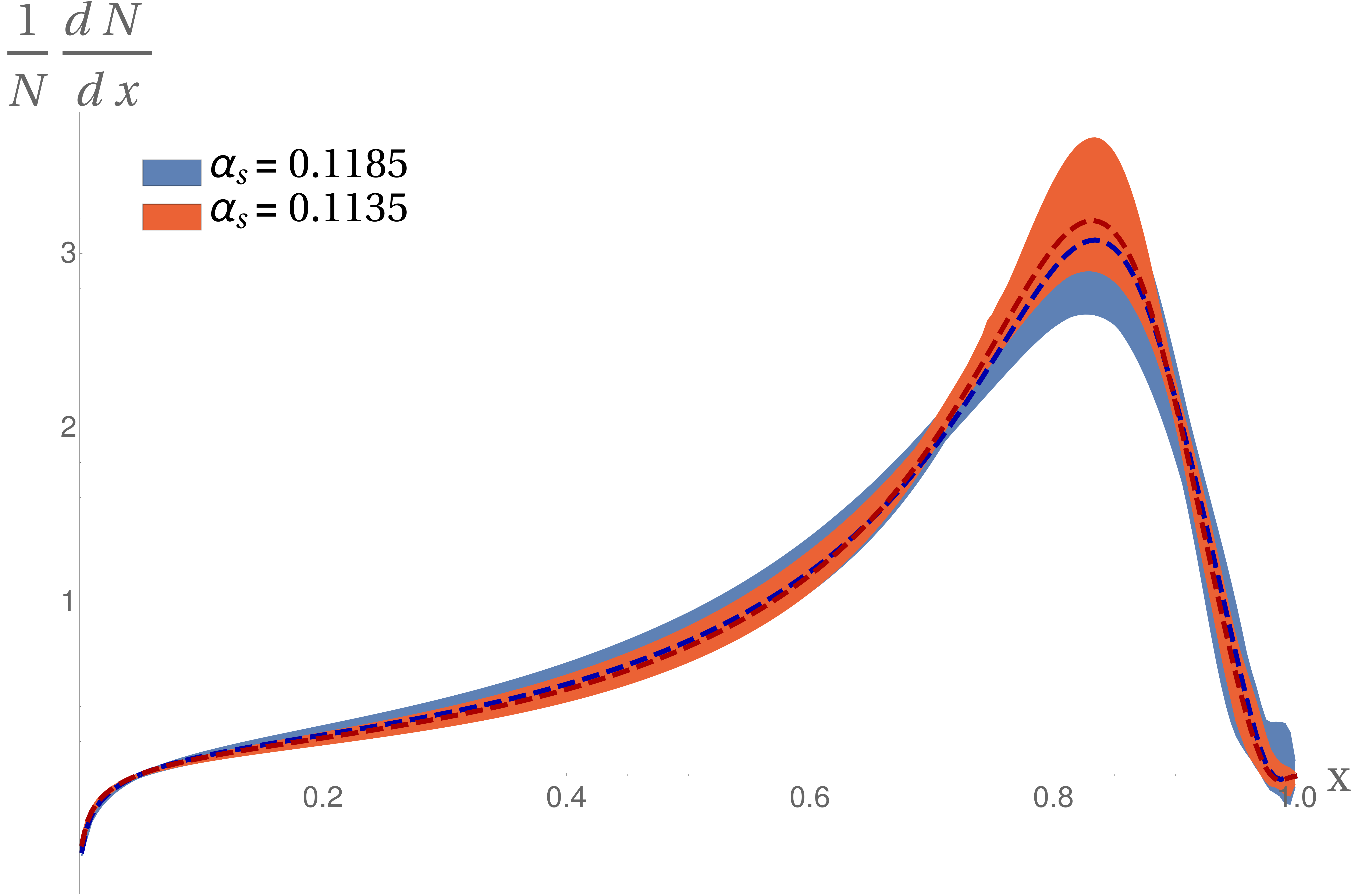}
\caption{Dependence of the N${}^2$LO + N${}^3$LL theoretical cross section on the value of $\alpha_s(m_Z)$.}\label{Envelopeas}
\end{figure}

\begin{table}
\begin{tabular}{||c |c| c | c | c | c ||}
\hline
data sets & $\lambda$ (GeV)  & $\sigma_{\textrm{exp}}$ (GeV)  & $\sigma_{\textrm{th}}$ (GeV)  & $(\chi^2/dof)_\textrm{def}$ & $\langle \chi^2/dof \rangle$   \\
\hline 
all  &  0.576 & 0.055 & $^{+ 0.018}_{-0.046}$ & 1.6 & 2.5    \\
\hline 
ALEPH, DELPHI \& SLD   & 0.552 & 0.070 & $^{+ 0.036}_{-0.051}$ & 1.0 & 1.6  \\
\hline
\end{tabular}
\caption{Best fit value of $\lambda$, with statistical and theoretical errors, and $\alpha_s(m_Z) = 0.1135$. 
The fits are performed with the N${}^2$LO + N${}^3$LL formulae. 
The procedure used to assess the theoretical error is described in the text. $(\chi^2/dof)_\textrm{def}$
and $\langle \chi^2/dof \rangle$ denote the $\chi^2$ obtained with default theory settings, and the average of the $\chi^2$ over all fits. }\label{TabFitsas}
\end{table}

For our best fit in Section~\ref{FitsA} we used the world average $\alpha_s(m_Z)=0.1185 \pm 0.0006$ of the Particle Data Group~\cite{Agashe:2014kda}. 
There are by now several extractions of $\alpha_s$ using event shapes in $e^+ e^-$ data that point to lower values of $\alpha_s(m_Z)$~\cite{Abbate:2010xh,Abbate:2012jh,Gehrmann:2012sc,Hoang:2014wka,Hoang:2015hka}.
To assess the impact of a lower value of $\alpha_s$, we repeated the analysis of Section~\ref{FitsA}, but with $\alpha_s(m_Z) = 0.1135 \pm 0.0011$ \cite{Abbate:2010xh}.

For simplicity, we did not vary $\Gamma_3$ and $\gamma_2^S$, which have little effects on the fits. We also did not include the error on $\alpha_s$ quoted in Ref. \cite{Abbate:2010xh}.
The variations of the other theory parameters produced a total of 1701 theory settings. In this case, we calculated the $\chi^2$ for $\lambda$ between $0.400$  and $0.700$ GeV, in steps of 0.025 GeV.
The results of the fits, with statistical and theoretical errors, the value of $\chi^2/dof$ for the default theory settings, and the average $\chi^2/dof$ for all the fits are shown in Table~\ref{TabFitsas}.
Comparing Tables  \ref{TabFitsas} and \ref{TabFits}, we see that the different value of $\alpha_s(m_Z)$ has three effects. First of all, a smaller $\alpha_s$ requires a larger value of $\lambda$.
Secondly, we find that the fits improve, even when including OPAL data. The effect is mainly due to a lower tail. Finally, the theory error is decreased as regards the discussion in Section \ref{FitsA}.

\begin{figure}
\center
\includegraphics[width=12cm]{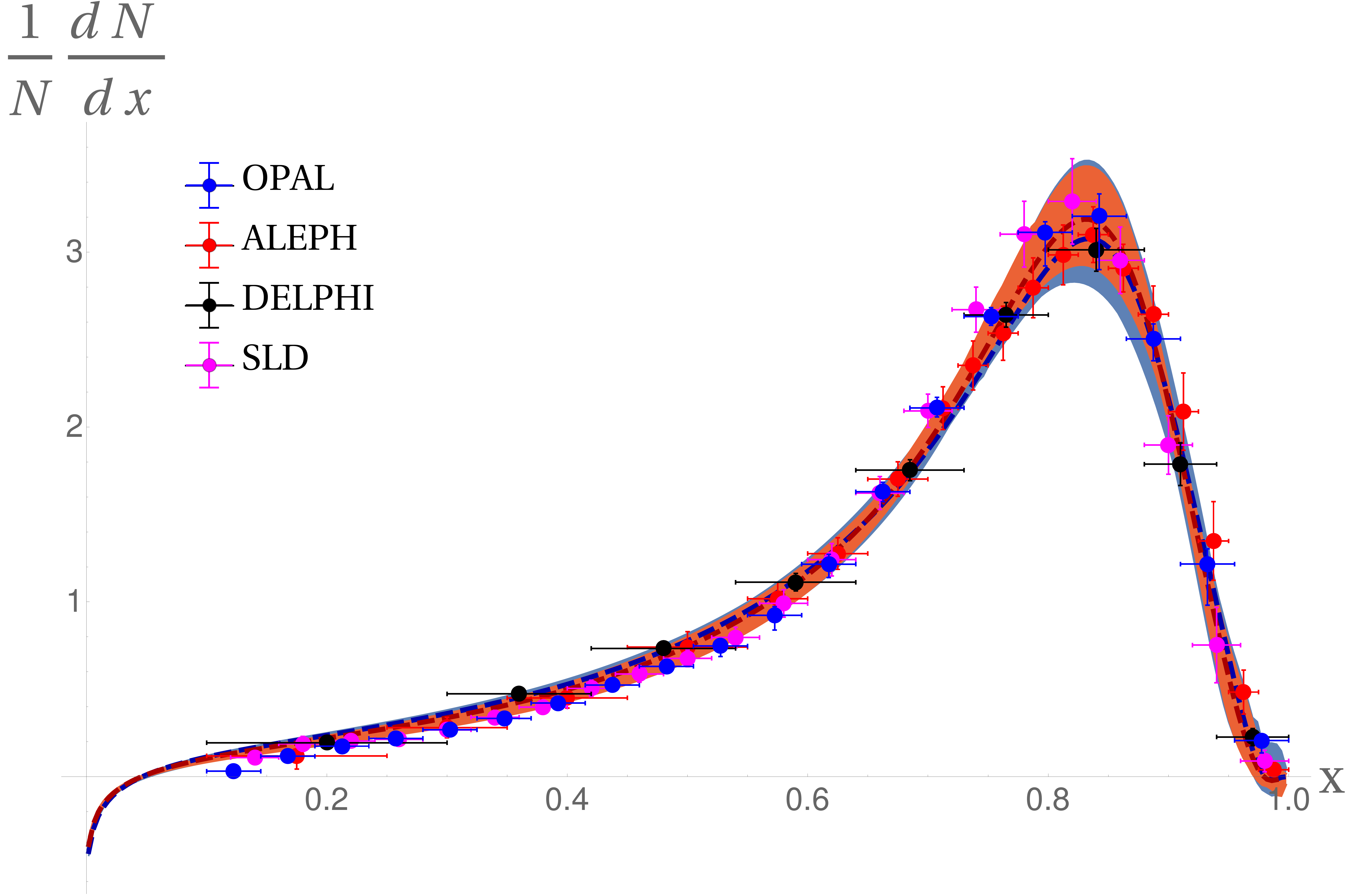}
\caption{Comparison of the N${}^2$LO + N${}^3$LL theoretical cross section  with $\alpha_s(m_Z) = 0.1135$ to data. The color code for the theoretical distributions is as in  Fig. \ref{Envelopeas}. The error bands include only theoretical uncertainties.}\label{Envelopeas_data}
\end{figure}

In Fig. \ref{Envelopeas} we show the theory bands obtained with $\alpha_s(m_Z) = 0.1185$ (blue band) and $\alpha_s(m_Z) = 0.1135$ (red band). The dashed blue and red lines are obtained with the default theory parameters.
The red band is narrower as a consequence of the better fits and smaller theory error, and it is higher in the peak and lower in the tail relative to the blue band.
Even if the effects are not dramatic, they  lead to better agreement with the data, as shown in Fig. \ref{Envelopeas_data}.

\subsection{Comparison to the literature}

\begin{figure}[t]
\centering
\subfigure[]{%
  \includegraphics[width=6.5cm]{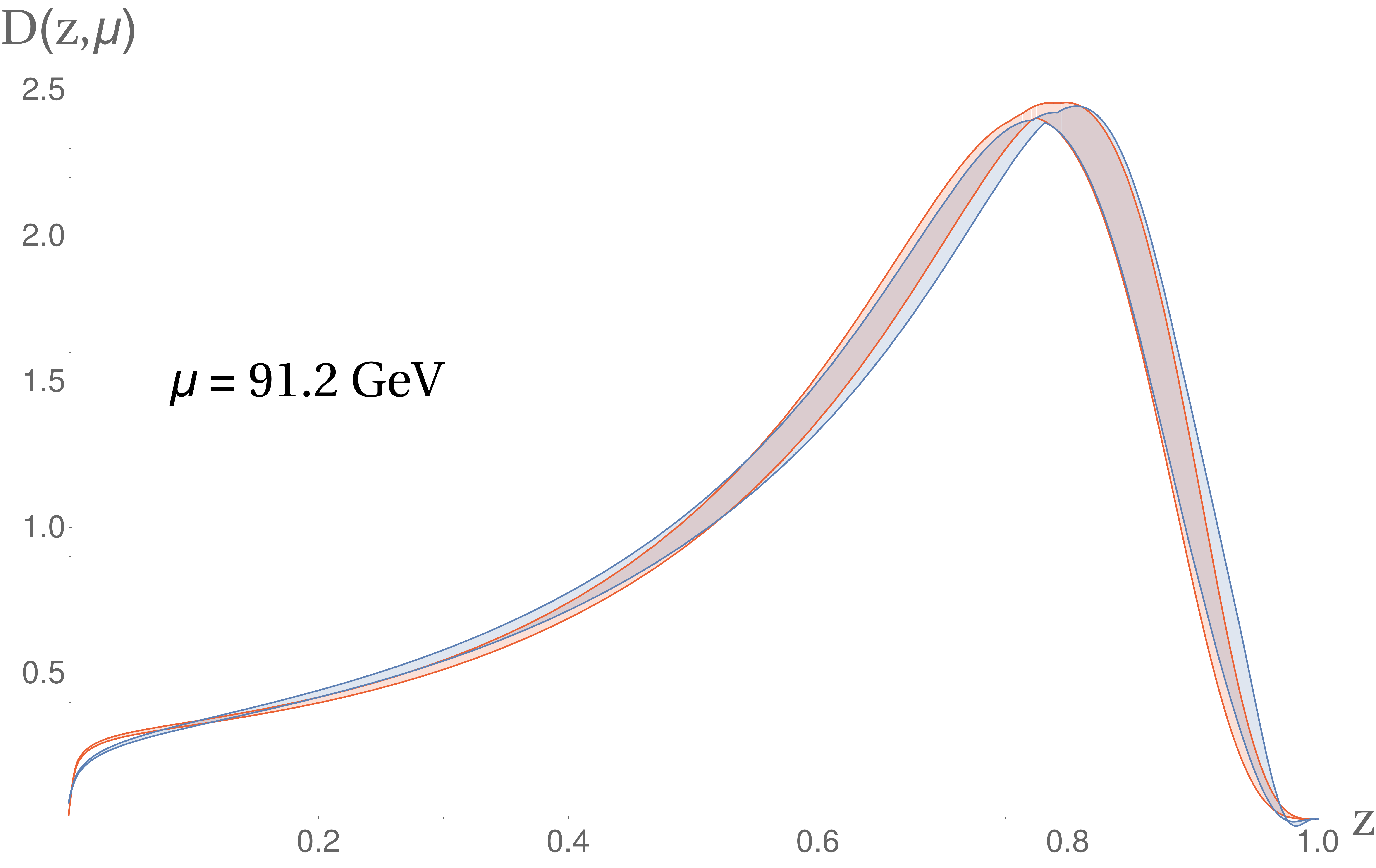}
  \label{Cacciaria}}
\quad
\subfigure[]{%
  \includegraphics[width=6.5cm]{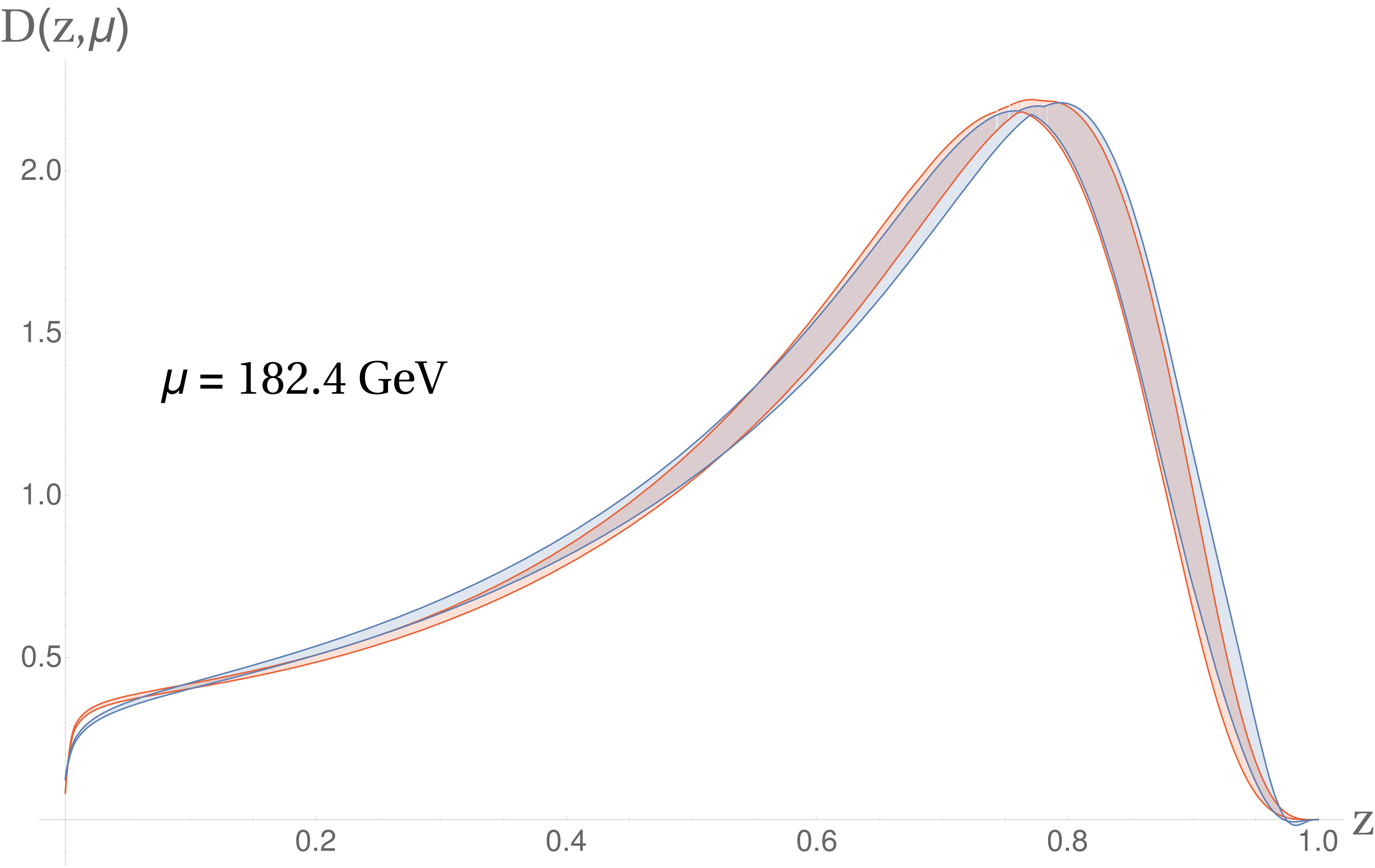}
  \label{Cacciarib}}
\caption{
Comparison of the heavy quark fragmentation function extracted in this paper (blue band) and in Ref. \cite{Cacciari:2005uk} (red band). 
In the left panel (a) the fragmentation function is evaluated at the factorization scale $\mu = 91.2$ GeV. In the right panel (b), at $\mu = 182.4$ GeV. 
The bands include statistical errors only.
}
\label{Cacciari}
\end{figure}

The most recent extraction of the $b$-quark fragmentation function from $e^+ e^-$ data has been performed by M.~Cacciari, P.~Nason and C.~Oleari in Ref. \cite{Cacciari:2005uk}. 
In this paper, the inclusive differential cross section for the production of a $b$-flavored hadron is considered at NLO. Using the time-like splitting functions at $\mathcal O(\alpha_s^2)$,
DGLAP logarithms are resummed at NLL. Soft logarithms of $1-x$ are resummed both in the hard coefficient and in the fragmentation function. The resummation is performed directly in Mellin space.
Non-perturbative effects are taken into account by performing a convolution of the partonic fragmentation function $d_{Q/Q}$ with a nonperturbative model, as in Eq. \eqref{FragFuncFull}.
The nonperturbative model chosen in Ref. \cite{Cacciari:2005uk} is
\begin{equation}
\tilde S_{H/Q}^{hadr}(z) = \frac{\Gamma(2+a+b)}{\Gamma(1+a) \Gamma(1+b)}z^{a} (1-z)^b.
\end{equation}
The parameters $a$ and $b$ were determined by fitting to data from the ALEPH~\cite{Heister:2001jg} and SLD~\cite{Abe:2002iq} experiments. The best fit parameters are $a = 24 \pm 2$ and $b = 1.5 \pm 0.2$,
with $\chi^2/dof = 2.3$. 

Away from the endpoint $x\sim 1$, the main novelty of our work is the inclusion of N${}^2$LO corrections to the hard coefficient \cite{Rijken:1996vr,Rijken:1996npa,Rijken:1996ns} and to the perturbative heavy quark fragmentation function \cite{Melnikov:2004bm,Mitov:2004du}, and the use of the three-loop time-like splitting functions \cite{Mitov:2006ic,Moch:2007tx,Almasy:2011eq} in the solution of the DGLAP equations. This allows us to reach N${}^2$LO + N${}^2$LL accuracy in the $x<1$ region. 

In the endpoint, using SCET and bHQET techniques, we are able to extend the resummation of soft logarithms of $1-x$ by an additional order, reaching (approximate) N${}^3$LL, in the counting delineated in Section \ref{resum}.  
Without going into a detailed comparison of the relation between resummation in perturbative QCD and SCET (which is carefully discussed elsewhere, for example in Ref. \cite{Bonvini:2012az}
in the context of threshold resummation, and in Ref. \cite{Almeida:2014uva} in the case of event shapes), here we simply observe the very good agreement between the extraction of the HQFF 
carried out in this paper and in Ref. \cite{Cacciari:2005uk}. This can be seen in Figure \ref{Cacciari}, where we compare the fragmentation function we obtained  by fitting the N${}^2$LO + N${}^3$LL 
cross section to data from the ALEPH, DELPHI and SLD experiments (blue band) to the fragmentation function extracted in Ref. \cite{Cacciari:2005uk} (red band). 
The bands only include the statistical errors of the fits, which is determined by the experimental errors. The fragmentation functions are evaluated at the scales $\mu = 91.2$ GeV (left panel)
and $\mu = 182.4$ GeV (right panel). We verified that the agreement is good in a wide range of factorization scales. 
As we have discussed in Sections \ref{FitsA} and \ref{convergence}, the inclusion of an additional perturbative order allows for a reduction of the theoretical errors due to residual dependences on the scales $\mu_H$, $\mu_J$, $\mu_M$ and $\mu_S$. 

Another advantage of the EFT framework discussed in Sections \ref{sec:bHQET} and \ref{sec:np} is a more immediate physical interpretation of the parameters of the nonperturbative model, due to the relation to local HQET matrix elements. Furthermore, the bHQET expansion can be systematically extended to include power corrections of $\mathcal O(\LambdaQCD/m_Q)$. 
bHQET power corrections constitute a large theoretical uncertainty in the determination of the HQFF, possibly as large as $5-10\%$. A separate investigation of these effects is needed.


\section{Conclusion}\label{conclude}

In this paper we have derived  a factorization theorem for the single inclusive production of heavy flavored hadrons in $e^+ e^-$ annihilation, 
using SCET and bHQET. 
We split the differential cross section into the tail region comprising moderate values of $x$ and the peak region where $x$ approaches 1. In each region we use a hierarchy of EFTs to systematically control theory errors, sum logarithms, and organize perturbative corrections. We then ``sew'' the two regions together using a prescription that smoothly goes from one region to the other. A crucial ingredient for this is the use of profile functions which allow us to change scales in various parts of the calculation, with the result that certain resummations are turned on and off depending on the value of $x$.
The EFT approach allows us to achieve a clean separation of the perturbative and non-perturbative aspects of the fragmentation of a heavy quark. The non-perturbative information is parametrized by 
the hadronic shape function $S_{H/Q}^{hadr}$, whose moments are related to  matrix elements of local bHQET operators. We eliminate renormalon ambiguities in the factorization of long- and short-distance physics contribution to the shape function by a suitable renormalon subtraction.

Using state-of-the art results for the fixed order expression in Eqs. \eqref{QCDfinal} and \eqref{eq:finalfact},
and for the anomalous dimensions in Eqs. \eqref{DGLAP}, \eqref{rgeH.0}, \eqref{rgeM.0}, \eqref{anomJ}, \eqref{anomS},
we evaluate the cross section at N${}^2$LO, with N${}^2$LL resummation of logarithms of the ratio of the heavy quark mass $m_Q$ and the center-of-mass energy $Q$, and N${}^3$LL resummation of logarithms of $1-x$ in the endpoint. 
By fitting the theoretical cross section to $e^+ e^-$ annihilation data at the $Z$ pole, we extract the $b$-quark fragmentation function at N${}^2$LO, one order higher than in the existing literature. 
We repeat the fits at NLO + N${}^2$LL and find that, by going to higher order, the size of theoretical errors is reduced.
As shown in Fig. \ref{Cacciari}, the fragmentation function we extract is in good agreement with previous extractions \cite{Cacciari:2005uk}.

One of the advantages of the EFT approach is a systematic control of the theoretical uncertainties, which stem from missing orders in the perturbative expansions, and from missing power corrections in the EFT.
We study the former by varying the scales $\mu_H$, $\mu_J$, $\mu_M$ and $\mu_S$. We find that at N${}^2$LO the cross section still has a noticeable dependence on $\mu_M$, the scale at which the heavy quark fragmentation functions is evaluated and the DGLAP evolution is started. This dependence leads to a 15\% theoretical uncertainty on the fit parameter $\lambda$, as big as the statistical uncertainty. The cross section is much less 
sensitive to the remaining scale variations, which induce an error on $\lambda$ of a few percents.
The most important power corrections originate from the bHQET expansion, and are of order $\LambdaQCD/m_Q \sim 10\%$. Though the fits to the $e^+ e^-$ data are in general very good, the inclusion of these power corrections might be important to achieve a better description of the tail of the distribution, where our prediction slightly overshoots the data.

The $b$-fragmentation function extracted in this work can be used in high precision calculations of $B$-meson production in other processes, 
such as hadronic collisions \cite{Cacciari:1998it}, and top quark decays  \cite{Campbell:2014kua}. Given the copious amounts of high quality data being produced by the experimental collaborations at the LHC, 
such a study is of the foremost interest. 

The N${}^2$LO $b$-fragmentation functions extracted in this work, tabled in the LHAPDF format \cite{Buckley:2014ana}, are available from the authors upon request.


\acknowledgments
The work of SF and MF was supported in part by the Director, Office of Science, Office of Nuclear Physics, of the U.S. Department of Energy under grant numbers DE-FG02-06ER41449 and DE-FG02-04ER41338. SF  and MF also acknowledges support from the DFG cluster of excellence ``Origin and structure of the universe''. MF was also supported in part by the US National Science Foundation, grant NSF-PHY-0969510 {\em the LHC Theory Initiative}, the Cluster of Excellence {\em Precision Physics, Fundamental Interactions and Structure of Matter\/} (PRISMA -- EXC 1098) and DFG grant NE~398/3-1. 
CK was supported by Basic Science Research Program through the National Research Foundation of Korea (NRF) funded by the Ministry of Science, ICT and Future Planning (Grants No. NRF-2014R1A2A1A11052687).
EM  acknowledge support by the US DOE Office of Nuclear Physics and by the LDRD program at Los Alamos National Laboratory.
EM thanks D.~Kang, Z.~Kang, A.~Hornig, Z.~Ligeti, and F.~Ringer for several interesting discussions. We thank in particular C.~Lee for detailed comments on the manuscript.
We thank P.~Pietrulewicz for pointing out a mistake in Eq. \ref{eq:c2thr} in the first version of the paper, and for illuminating discussions on the role of rapidity logarithms in the threshold coefficient 
$c_2^{\textrm{thr}}$.

\appendix

\section{Perturbative results}\label{Equations}

In this Appendix, we collect the fixed order expressions of the hard function, $H_Q$, mass coefficient, $C_m$, jet function, $J_{\bar n}$ and shape function, $S_{Q/Q}$, which enter the factorization
formula of the endpoint cross section in Eq. \eqref{eq:finalfact}. These functions are known to $\mathcal O(\alpha_s^2)$. 
In Section \ref{anomalous} we give the anomalous dimensions of these functions, and the solution of the RGEs.

\subsection{Fixed order results}

The hard function $H_Q$, which encodes dynamics at the hard scale and it is obtained by matching QCD onto \SCETa, is related to the quark time-like form factor,
which was computed up to three loops \cite{Baikov:2009bg}. 
For our analysis, it is enough to work  at $\mathcal O(\alpha_s^2)$.
At this order, $H_Q$ is given by \cite{Matsuura:1987wt,Matsuura:1988sm,Gehrmann:2005pd,Moch:2005id,Becher:2008cf}
\begin{eqnarray}\label{hard.1}
& & H_Q(Q,\mu)  = 1 + \frac{\alpha_s(\mu) C_F}{4\pi} \left( - 16 + \frac{7}{3} \pi^2 - 6 L_Q - 2 L_Q^2 \right)    \nonumber \\ 
& & + \left(\frac{\alpha_s(\mu) }{4\pi} \right)^2  C_F \left\{   
C_F \left(  \frac{511}{4} - \frac{83 }{3} \pi^2 + \frac{67}{30} \pi^4 - 60 \zeta(3)\right)
+  T_F n_f \left( \frac{4085}{81} - \frac{182}{27} \pi^2 + \frac{8}{9} \zeta(3)\right)
\right. \nonumber \\ & & \left.  + C_A  \left(  - \frac{51157}{324} + \frac{1061}{54} \pi^2 - \frac{8}{45} \pi^4 + \frac{626}{9}\zeta(3) \right)
\right. \nonumber \\ & & \left. + L_Q \left( C_F \left(93 - 10 \pi^2 - 48 \zeta(3) \right) +  T_F n_f \left(  \frac{836}{27} - \frac{16}{9} \pi^2\right)
+  C_A \left( - \frac{2545}{27} + \frac{44}{9} \pi^2 + 52 \zeta(3)\right)
\right) \right. \nonumber \\ & & \left.
+ L_Q^2 \left(C_F \left(50  - \frac{14}{3} \pi^2\right) + T_F n_f \frac{76}{9} + C_A \left(-\frac{233}{9} + \frac{2}{3} \pi^2\right) \right) \right. \nonumber \\ & & \left.
+ L_Q^3 \left(  12 C_F + \frac{8}{9} T_F n_f  - \frac{22}{9} C_A \right) + 2 C_F L_Q^4 \right\},   
\end{eqnarray}
with $L_Q = \ln \frac{\mu^2}{Q^2}$. The color factors in Eq. \eqref{hard.1}  are $C_F = 4/3$, $C_A = 3$, and $T_F = 1/2$. $n_f$ is the number of light flavors,
$n_f =5$ above the bottom threshold.

The derivation of the one-loop matching coefficient between SCET$_\textrm{M}$ and bHQET was discussed in Section \ref{sec:SCETM}.
The two-loop expression for $C_m$ can be obtained by comparing the singular terms of the perturbative fragmentation function in Ref. \cite{Melnikov:2004bm}
to the two loop shape function \cite{Neubert:2007je}, and it is given by
\begin{eqnarray}
& &C_m = 1 + \frac{\alpha_s}{4\pi} C_F \left( 4 + \frac{\pi^2}{6} + L_M + L_M^2\right) \nonumber \\ 
&& + \left(\frac{\alpha_s}{4\pi}\right)^2 \left\{
	C_F^2 \left(\frac{241}{8}+\pi ^2 \left(\frac{13}{3}-8 \log
	   (2)\right) -\frac{163 \pi ^4}{360} -6 \zeta (3) \right) 
 	 + C_F T_F n_f
	   \left(-\frac{1541}{162}-\frac{37 \pi ^2}{27} \right. \right. \nonumber \\ 
&& \left.  -\frac{52 \zeta (3)}{9}\right)+ 
	C_F C_A  \left(\frac{12877}{648}+\frac{755 \pi ^2}{108}+4 \pi ^2 \log (2)-\frac{47 \pi ^4}{180}+ \frac{89 \zeta (3)}{9}\right) \nonumber  \\ 
&& +
	L_M \left(C_F^2 \left(\frac{11}{2}-\frac{11 \pi^2}{6} + 24 \zeta (3) \right) -C_F T_F n_f \left(\frac{154}{27}+\frac{8 \pi ^2}{9}\right)  
	+ C_F C_A   \left(\frac{1165}{54}+\frac{28 \pi ^2}{9} \right.\right. \nonumber \\ 
 & & \left.\left. -30 \zeta (3)\right)  \right)
+ L_M^2 \left( C_F^2 \left(\frac{9}{2}+\frac{\pi ^2}{6}\right) -\frac{26 }{9} C_F T_F n_f  + \left(\frac{167}{18}-\frac{\pi ^2}{3}\right) C_F C_A  \right)  \nonumber \\ 
&& \left.
+ L_M^3 \left( C_F^2 -\frac{4 }{9} C_F T_F n_f  + \frac{11 }{9} C_F C_A \right)
+ \frac{1}{2}L_M^4 C_F^2
\right\}  \label{mass.1}
\end{eqnarray}
with $L_M = \log \frac{\mu^2}{m_Q^2}$, and $n_f = 4$, since we are below the bottom threshold.

The mismatch of $n_f$ is made up by the 2-loop matching coefficient at the flavor threshold, discussed in Section \ref{sec:flavorthr}
\begin{align}\label{eq:c2thr}
c_2^{thr}(z) &= L_M^2 \left(\frac{\delta (1-z)}{2}+\frac{2}{3} \left[\frac{1}{1-z}\right]_+\right)+ L_M \bigg(\left(-\frac{1}{6}-\frac{2 \pi ^2}{9}\right)
   \delta (1-z)-\frac{20}{9} \left[\frac{1}{1-z}\right]_+ \bigg)\nn\\
&+\left(\frac{2 \zeta
   (3)}{3}+\frac{3139}{648}-\frac{\pi ^2}{3}\right) \delta (1-z)+\frac{56}{27} \left[\frac{1}{1-z}\right]_+ .
\end{align}

To express the jet and shape function, we need to introduce the distributions
\begin{equation}
\int \df r \left[\frac{\theta(r)\log^n (r)}{r}\right]_+^{}  \varphi(r) =  \int_0^\infty \df r \frac{\log^n ( r)}{r}\Big( \varphi(r) - \theta( \kappa - r) \varphi (0)\Big) +   \frac{1}{n+1}\log^{n+1} ( \kappa) \varphi(0) \label{dist.2}, 
\end{equation}
where $r$ is a dimensionful variable, taking values in the $(0,+\infty)$ interval. 
In the case of the jet function, $r$ has dimension two, and represents the virtuality of the jet, while for the shape function, $r$ has dimension one.
$\kappa$ is an arbitrary cutoff, with the same dimensionality as $r$.  The expression in Eq. \eqref{dist.2} is independent of $\kappa$.
It is convenient to define
\begin{equation}
 \left[\frac{\theta( r)\log^n ( r/\mu^a)}{ r}\right]_+^{(\mu^a)}  \equiv   \left[\frac{\theta( r)\log^n ( r/\mu^a)}{r}\right]_+^{} + (-1)^{n+1} \frac{1}{n+1}\log^{n+1}(\mu^a) \delta( r) , \label{dist.3} 
\end{equation}
where $a$ is the dimension of the variable $r$.
The quark jet function has been computed to two loops in Ref. \cite{Becher:2006qw}. In terms of the distribution  \eqref{dist.3}, we can express the jet function as
\begin{eqnarray}\label{jet.1}
J(r) = J_{-1} \, \delta(r)  + \sum_{n = 0}^{\infty} J_n  \, \left[\frac{\theta(r)\log^n ( r/\mu^2)}{ r}\right]_+^{(\mu^2)},
\end{eqnarray}
with coefficients 
\begin{eqnarray}
J_{-1}  &=& 1 + \frac{\alpha_s C_F}{4\pi} \left( 7 - \pi^2 \right) + \left(\frac{\alpha_s}{4\pi}\right)^2  
\left[ C_F^2 \left( \frac{205}{8} - \frac{67 \pi^2}{6} + \frac{14 \pi^4}{15} - 18 \zeta(3) \right) \right. \nonumber \\
& &  \left. + C_F C_A  \left(  \frac{1417}{108} - \frac{7 \pi^2}{9} - \frac{17 \pi^4}{180} - 18 \zeta(3) \right) + C_F \beta_0 \left( \frac{4057}{216} - \frac{17 \pi^2}{9} - \frac{4 \zeta(3)}{3}  \right)
\right]  \nn \\
J_0   &=& - 3 \frac{\alpha_s C_F}{4\pi} - \left(\frac{\alpha_s}{4\pi}\right)^2  \left[ C_F^2 \left( \frac{45}{2} - 7\pi^2 + 8 \zeta(3) \right) + C_F C_A \left(\frac{73}{9} - 40 \zeta(3)\right) \right. \nonumber 
\\ &&  \left. + C_F \beta_0 \left( \frac{247}{18} - \frac{2\pi^2}{3}\right)  
\right]  \nn \\
J_1   &=& 4 \frac{\alpha_s C_F}{4\pi} + \left(\frac{\alpha_s}{4\pi}\right)^2  \left[ C_F^2 \left( 37 - \frac{20}{3}\pi^2  \right) + C_F C_A \left(\frac{16}{3} - \frac{4\pi^2}{3}\right)  + \frac{29}{3} C_F \beta_0   \right] \nn \\
J_2   &=& \left(\frac{\alpha_s}{4\pi}\right)^2  \left( 18 C_F^2 + 2 C_F \beta_0 \right) \nn \\ 
J_3   &=& \left(\frac{\alpha_s}{4\pi}\right)^2 8 C_F^2 \,
\end{eqnarray}
with $\beta_0$ the first coefficient of the QCD $\beta$ function.

The perturbative expression of the HQET shape function is also known to two loops \cite{Becher:2005pd}.
\begin{eqnarray}\label{shape.1}
S(\omega) = S_{-1} \, \delta\left(\frac{\omega}{\bar n \cdot v}\right)  + \sum_{n = 0}^{\infty} S_n  \, \left[\frac{\theta(\omega) \bar n \cdot v \, \log^n ( \omega/(\bar n \cdot v\mu))}{ \omega}\right]_+^{(\mu)}.
\end{eqnarray}
The coefficients of the expansion in Eq. \eqref{shape.1} are
\begin{eqnarray}
S_{-1}  &=& 1 - \frac{\alpha_s C_F}{4\pi} \frac{\pi^2}{6} - \left(\frac{\alpha_s}{4\pi}\right)^2 
\left[ C_F^2 \left( \frac{4 \pi^2}{3} + \frac{3 \pi^4}{40} - 32 \zeta(3) \right) \right.  \nn \\
& &  \left. + C_F C_A \left(  \frac{116}{27} + \frac{31 \pi^2}{9} - \frac{67 \pi^4}{180} + 18 \zeta(3) \right) + C_F \beta_0 \left( -\frac{2}{27} + \frac{5 \pi^2}{36} - \frac{5 \zeta(3)}{3}  \right)
\right]\nn \\
S_0   &=& - 4 \frac{\alpha_s C_F}{4\pi} + \left(\frac{\alpha_s}{4\pi}\right)^2  \left[ C_F^2 \left(  - \frac{14}{3}\pi^2 + 64 \zeta(3) \right) + C_F C_A \left(\frac{88}{9} + \frac{4 \pi^2}{3} - 36 \zeta(3)\right)   - \frac{4}{9}C_F \beta_0   \right]  \nonumber \\ 
S_1   &=&- 8 \frac{\alpha_s C_F}{4\pi} + \left(\frac{\alpha_s}{4\pi}\right)^2  \left[ C_F^2 \left( 16 - \frac{28}{3}\pi^2  \right) + C_F C_A \left(-\frac{32}{3} + \frac{8\pi^2}{3}\right)  - \frac{16}{3} C_F \beta_0   \right] \nn \\
S_2   &=& \left(\frac{\alpha_s}{4\pi}\right)^2  \left( 48 C_F^2 + 8 C_F \beta_0 \right) \nn \\ 
S_3   &=& \left(\frac{\alpha_s}{4\pi}\right)^2 32 C_F^2\, .
\end{eqnarray}

\subsection{Anomalous dimensions and solutions of the RGEs}\label{anomalous}

Here we give the anomalous dimension and the solution of the RGE for each ingredient of the endpoint factorization formula, Eq. \eqref{eq:finalfact}.
The hard coefficient $H_Q(Q,\mu)$ and the mass coefficient $C_m(m_Q,\mu)$ are renormalized multiplicatively, and their RGE is
\begin{eqnarray}
\frac{d}{d\ln \mu} H_Q(Q,\mu) &= &- 2 \left[ \Gamma_{\textrm{cusp}}(\alpha_s) \, \ln \frac{\mu^2}{Q^2} + 2 \gamma_H(\alpha_s) \right]  H_Q(Q,\mu),   \label{rgeH}\\
\frac{d}{d\ln \mu} C_m(m_Q,\mu) &=& \left[ \Gamma_{\textrm{cusp}}(\alpha_s) \ln \frac{\mu^2}{m_Q^2} + 2 \gamma_M(\alpha_s) \right] C_m(m_Q,\mu), \label{rgeM}
\end{eqnarray}
where  $\Gamma_{\textrm{cusp}}(\alpha_s)$ is the universal quark cusp anomalous dimension, while $\gamma_M(\alpha_s)$ and 
$\gamma_H(\alpha_s)$ are the nonuniversal non-cusp anomalous dimensions. 

The jet and shape function have convolution RGEs of the form
\begin{eqnarray}
\frac{d}{d\ln \mu} J_{\bar n} (Q r,\mu) &=& Q \int {\rm d} r^\prime \gamma_J(Q r - Q r^\prime, \mu)\, J_{\bar n}(Q r^\prime,\mu),  \label{rgeJ} \\
\frac{d}{d\ln \mu} S_{} (\omega,\mu) &=&  \int {\rm d} \omega^\prime \gamma_S(\omega - \omega^\prime, \mu)\, S(\omega^\prime,\mu), \label{rgeS}
\end{eqnarray}
with anomalous dimensions given by
\begin{eqnarray}
\gamma_J(r) &=& - 2 \Gamma_\textrm{cusp}(\alpha_s)\left[\frac{\theta(r)}{r}\right]_+^{(\mu^2)}  + \gamma_J(\alpha_s)  \delta(r)  \\
\gamma_S(\omega) &=&  2 \Gamma_\textrm{cusp}(\alpha_s)\left[\frac{\theta(\omega)}{\omega}\right]_+^{(\mu)}  + 2 \gamma_S(\alpha_s)  \delta(\omega). 
\end{eqnarray}
The solution of Eq. \eqref{rgeH} and \eqref{rgeM} can be expressed as 
\begin{eqnarray}
H_Q(Q,\mu)     = \left( \frac{\mu^2_0}{Q^2} \right)^{2 g(\mu,\mu_0)} \exp\left( 2 V(\mu,\mu_0)  + 4 K_H(\mu,\mu_0) \right) H_Q(Q,\mu_0)  \label{solH}\\
C_m(m_Q,\mu) = \left( \frac{\mu^2_0}{m_Q^2} \right)^{- g(\mu,\mu_0)} \exp\left( - V(\mu,\mu_0)  - 2 K_M(\mu,\mu_0) \right) C_m(m_Q,\mu_0)\label{solM}.
\end{eqnarray}
For the jet and shape function, the solution of the RGE involves a convolution with a renormalization group kernel,
\begin{eqnarray}
J_{\bar n}(Q r, \mu) &=& Q \int {\rm d} r^\prime J_{\bar n}(Q r^\prime, \mu_0) U_J(Q r - Q r^\prime, \mu,\mu_0) \label{solJ} \\
S_{Q/Q}(\omega, \mu) &=&  \int {\rm d} \omega^\prime S_{Q/Q}(\omega^\prime, \mu_0) U_S(\omega - \omega^\prime, \mu,\mu_0), \label{solS} 
\end{eqnarray}
and $U_J$ and $U_S$ are given by
\begin{eqnarray}
U_J( Q r ,\mu,\mu_0) &=&  \label{kernJ} 
\exp\left( - 2 V(\mu,\mu_0) - K_J(\mu,\mu_0) \right)
 \frac{ ( \mu_0^2 e^{   \gamma_E} )^{-2 g(\mu,\mu_0)} }
{\Gamma\left(  2 g (\mu,\mu_0)\right)} \,\left[\frac{\theta(Q r)}{Q r}\right]^{1-2 g(\mu,\mu_0)}  \\
U_S( \omega ,\mu,\mu_0) &=&  \label{kernS}
\exp\left(  V(\mu,\mu_0) - 2 K_S(\mu,\mu_0) \right)
 \frac{ ( \mu_0 \bar n \cdot v e^{  \gamma_E})^{2 g(\mu,\mu_0) }}
{\Gamma\left(  - 2 g (\mu,\mu_0)\right)} \,\left[\frac{\theta(\omega)}{\omega}\right]^{1 + 2 g(\mu,\mu_0)}\,.  
\end{eqnarray}
Expanding  the cusp and non-cusp anomalous dimension and the beta function as 
\begin{eqnarray}
& &\Gamma_{\textrm{cusp}}(\alpha_s)  = \sum_n  \Gamma_n  \left( \frac{\alpha_s}{4\pi} \right)^{n+1}, \qquad 
\gamma_{I}(\alpha_s) = \sum_n  \gamma_n^I  \left( \frac{\alpha_s}{4\pi} \right)^{n+1}, \nonumber \\ 
& & \beta(\alpha_s)  = -2 \alpha_s \, \sum_n  \beta_n  \left( \frac{\alpha_s}{4\pi} \right)^{n+1},
\end{eqnarray}
and introducing the variable $r = \alpha_s(\mu)/\alpha_s(\mu_0)$, we can express the universal functions $V(\mu,\mu_0)$ and $g(\mu,\mu_0)$, which depend on the cusp anomalous dimension, 
and the functions $K_I(\mu,\mu_0)$ as
\begin{eqnarray}
V(\mu_,\mu_0) &=&  \frac{\Gamma_0}{2 \beta_0^2} \frac{4 \pi }{\alpha_s(\mu_0)} \sum_{n=0}^{\infty}   \left(\frac{\alpha_s(\mu_0)}{4\pi}\right)^n  v^{(n)}(r),  \\
g(\mu,\mu_0) &=& \frac{\Gamma_0}{2\beta_0} \sum_{n = 0}^{\infty}   \left(\frac{\alpha_s(\mu_0)}{4\pi}\right)^n  g^{(n)}(r), \\
K_I(\mu,\mu_0) &=& \frac{\gamma_0^I}{2\beta_0} \sum_{n = 1}^{\infty} \left(\frac{\alpha_s(\mu_0) }{4\pi}\right)^{n-1} \kappa_I^{(n)}(r). 
\end{eqnarray}
To achieve  N${}^3$LL resummation, we need the expansion of $V$ and $g$ up to $n=3$, which are given by
\begin{eqnarray}
v^{(0)}(r) &=& 1 - \frac{1}{r} - \ln r \nonumber \\
v^{(1)}(r) &=& \left(\frac{\Gamma_1}{\Gamma_0} - \frac{\beta_1}{\beta_0}\right) ( 1 - r + \ln r)  +  \frac{\beta_1}{2\beta_0} \ln^2 r \nonumber \\
v^{(2)}(r) &=& \frac{1-r}{2} \left( \left( \frac{\beta_2}{\beta_0} - \frac{\beta_1^2}{\beta_0^2} - \frac{\Gamma_2}{\Gamma_0}  + \frac{\beta_1 \, \Gamma_1}{\beta_0 \, \Gamma_0}\right)(1-r)
- 2  \left(\frac{\beta_2}{\beta_0} - \frac{\beta_1\, \Gamma_1}{\beta_0\, \Gamma_0}\right)
\right) \nonumber \\ & &
+ \left(\frac{\beta_1^2}{\beta_0^2} - \frac{\beta_2}{\beta_0}\right) \ln r  + \left(\frac{\beta_1 \Gamma_1}{\beta_0 \Gamma_0} - \frac{\beta_1^2}{\beta_0^2} \right) r \ln r \nonumber \\
v^{(3)}(r) &=& \frac{1-r}{2} \left(  - \frac{\beta_3}{\beta_0} + \frac{\beta_1\beta_2}{\beta_0^2} + \frac{\beta_1}{\beta_0} \left(\frac{\Gamma_2}{\Gamma_0} - \frac{\beta_1\, \Gamma_1}{\beta_0\, \Gamma_0}\right) 
\right. \nonumber \\ & & \left. + \left(3 \frac{\beta_3}{\beta_0}   - 7 \frac{\beta_1\beta_2}{\beta_0^2}+ 4 \frac{\beta_1^3}{\beta_0^3} 
- 2 \frac{\Gamma_3}{\Gamma_0} + \frac{\beta_1\, \Gamma_2}{\beta_0\, \Gamma_0} + \frac{\Gamma_1}{\Gamma_0}\left(4 \frac{\beta_2}{\beta_0}- 3\frac{\beta_1^2}{\beta_0^2}\right)
 \right) \frac{1-r}{2}
\right.  \nonumber \\
& & \left.- \frac{2}{3} \left( \frac{\beta_3}{\beta_0} - 2 \frac{\beta_2\, \beta_1}{\beta_0^2} + \frac{\beta_1^3}{\beta_0^3}
- \frac{\Gamma_3}{\Gamma_0} + \frac{\beta_1\, \Gamma_2}{\beta_0\, \Gamma_0} + \frac{\Gamma_1}{\Gamma_0} \left(\frac{\beta_2}{\beta_0} - \frac{\beta_1^2}{\beta_0^2}\right)\right) (1-r)^2 
\right) 
\nonumber \\
& & + \frac{1}{2}\ln r \left( - \frac{\beta_3}{\beta_0} + 2 \frac{\beta_1 \beta_2}{\beta_0^2} - \frac{\beta_1^3}{\beta_0^3}
+ r^2 \frac{\beta_1}{\beta_0}\left( \frac{\beta_1^2}{\beta_0^2} - \frac{\beta_2}{\beta_0} + \frac{\Gamma_2}{\Gamma_0} - \frac{\beta_1\, \Gamma_1}{\beta_0\, \Gamma_0}\right)
\right), \label{Vexp}
\end{eqnarray}
and 
\begin{eqnarray}
g^{(0)}(r) &=& \ln r \nonumber \\
g^{(1)}(r) &=&  - (1-r) \left( \frac{\Gamma_1}{\Gamma_0}- \frac{\beta_1}{\beta_0} \right)  \nonumber \\
g^{(2)}(r) &=&  -\frac{1-r^2}{2} \left( \frac{\Gamma_2}{\Gamma_0} - \frac{\beta_1\, \Gamma_1}{\beta_0\, \Gamma_0} - \frac{\beta_2}{\beta_0} + \frac{\beta_1^2}{\beta_0^2}\right) \nonumber \\
g^{(3)}(r) &=&  -\frac{1-r^3}{3} \left( \frac{\Gamma_3}{\Gamma_0} - \frac{\beta_1\, \Gamma_2}{\beta_0\, \Gamma_0}
+ \frac{\Gamma_1}{\Gamma_0}\left(\frac{\beta_1^2}{\beta_0^2} - \frac{\beta_2}{\beta_0}\right)
- \frac{\beta_3}{\beta_0} + 2 \frac{\beta_2\beta_1}{\beta^2_0} - \frac{\beta_1^3}{\beta_0^3}\right). \label{gexp}
\end{eqnarray}
Eqs. \eqref{Vexp} and \eqref{gexp} involve the fourth coefficient of  the cusp anomalous dimension and of the QCD beta function, $\Gamma_3$ and $\beta_3$.
Similarly, we need the first three coefficients of the expansion of $K_I$, 
\begin{eqnarray}
\kappa_I^{(1)}(r) &=& \ln r \nonumber \\
\kappa_I^{(2)}(r) &=&  - (1-r) \left( \frac{\gamma_1^I}{\gamma_0^I}- \frac{\beta_1}{\beta_0} \right)  \nonumber \\
\kappa_I^{(3)}(r) &=&  -\frac{1-r^2}{2} \left( \frac{\gamma_2^I}{\gamma_0^I} - \frac{\beta_1\, \gamma_1^I}{\beta_0\, \gamma^I_0} - \frac{\beta_2}{\beta_0} + \frac{\beta_1^2}{\beta_0^2}\right),
\end{eqnarray}
which require the non-cusp anomalous dimension up to three loops, $\gamma_2^I$.

The four loop QCD beta function was given in \cite{vanRitbergen:1997va}
\begin{eqnarray}
\beta_0  &=& \frac{11}{3} C_A - \frac{4}{3} T_F  n_f  \nn \\
\beta_1  &=& \frac{34}{3} C_A^2 - 4 C_F T_F n_f  - \frac{20}{3} C_A T_F n_f \nn \\
\beta_2  &=& \frac{2857}{54} C_A^3 + 2 C_F^2 T_F n_f - \frac{205}{9} C_F C_A T_F n_f - \frac{1415}{27} C_A^2 T_F n_f \nonumber \\ & &+ \frac{44}{9} C_F T_F^2 n_f^2 + \frac{158}{27} C_A T_F^2 n_f^2 \nn \\
\beta_3  &=& \frac{149753}{6} +  3564 \zeta_3 - \left(\frac{1078361}{162} + \frac{6508}{27}\zeta_3\right) n_f \nonumber \\&&+ \left(\frac{50065}{162} + \frac{6472}{81}\zeta_3\right) n_f^2 + \frac{1093}{729} n_f^3.
\end{eqnarray}
We specialized the expression of $\beta_3$ to the QCD case, with $SU(3)$ color group. The general expression for  $SU(N_c)$ is given in Ref. \cite{vanRitbergen:1997va}.

\noindent
The first coefficients of the quark  cusp anomalous dimension are \cite{Korchemsky:1987wg,Korchemskaya:1992je,Moch:2004pa}
\begin{eqnarray}
\Gamma_0 &=& 4 C_F, \nonumber \\ \Gamma_1 &=& 4 C_F\, \left[ \left( \frac{67}{9} - \frac{\pi^2}{3} \right)\,C_A - \frac{10}{9}\, n_f \right]\,,  \nonumber \\
\Gamma_2 & = &  4 C_F  \left[ C_A^2 \left( \frac{245}{6} - \frac{134}{27} \pi^2 + \frac{11}{45} \pi^4 + 
     \frac{22}{3} \zeta(3) \right)  + 
   C_A T_F n_f \left(-\frac{418}{27} + \frac{40}{27} \pi^2 - \frac{56}{3} \zeta(3)\right)  \right. \nonumber \\ && \left. + 
   C_F T_F n_f \left(-\frac{55}{3} + 16 \zeta(3)  \right) - \frac{16}{27} T_F^2 n_f^2 \right].
\end{eqnarray}
We use the Pad\'{e} approximation  for the unknown coefficient  $\Gamma_3$,
\begin{align}
\Gamma_3(n_f) = (1+e_\Gamma)\frac{\Gamma_2^2}{\Gamma_1}\,,
\end{align}
where $e_\Gamma$ is one of the theory parameters we vary in our error analysis. We take $e_\Gamma$ from $-2$ to $2$. Note that $\Gamma_3$ depends on the number of flavors and hence is different below and above the flavor threshold. For the default $e_\Gamma=0$
\begin{align}
\Gamma_3(5) = 1553.06, \qquad
\Gamma_3(4) = 4313.26 \,.\nn
\end{align}
The non-cusp anomalous dimensions of the hard coefficient $H$ is
\begin{eqnarray}
\gamma_0^H &=& 3 C_F \nonumber \\
\gamma_1^H &=&  \frac{C_F}{2} \left[  \left(\frac{82}{9}  - 52 \zeta(3)\right) C_A  + \left(3 - 4 \pi^2 + 
      48 \zeta(3) \right) C_F + \left(\frac{65}{9} + \pi^2\right) \beta_0  \right]  \nonumber \\
\gamma_2^H &=&  
C_F \left[ C_A^2 \left(\frac{66167}{324}-\frac{686 \pi ^2}{81}-\frac{302 \pi^4}{135}  -\frac{782 \zeta (3)}{9}+\frac{44 \pi ^2 \zeta (3)}{9}+136 \zeta (5) \right)   \nonumber \right. \\ & & \left. 
          +C_A C_F \left(\frac{151}{4}-\frac{205 \pi ^2}{9}-\frac{247 \pi^4}{135} + \frac{844 \zeta (3)}{3}+\frac{8 \pi ^2 \zeta (3)}{3}+120 \zeta (5)\right)        \nonumber \right. \\ & & \left. 
          +C_F^2   \left(\frac{29}{2}+3 \pi ^2+\frac{8 \pi ^4}{5}+ 68 \zeta(3)-\frac{16 \pi ^2 \zeta (3)}{3}-240 \zeta (5)\right)                                    \nonumber \right. \\ & & \left.
          + \beta_0\, C_A \left(-\frac{10781}{108}+\frac{446 \pi ^2}{81}+\frac{449 \pi ^4}{270} -\frac{1166 \zeta (3)}{9}\right) 
          + \beta_0^2 \left(-\frac{2417}{324}+\frac{5 \pi^2}{6} + \frac{2 \zeta (3)}{3} \right) 								     \nonumber \right.  \\ && \left.  
          + \beta_1 \left(\frac{2953}{108}-\frac{13 \pi ^2}{18}-\frac{7 \pi ^4}{27} + \frac{128 \zeta (3)}{9} \right)\right],
\end{eqnarray}
while for the jet function \cite{Becher:2008cf,Becher:2006qw}
\begin{eqnarray}
\gamma_0^J &=& 6 C_F \nonumber \\
\gamma_1^J &=&  C_F \left[  \left(\frac{146}{9}  - 80 \zeta(3)\right) C_A  + \left(3 - 4 \pi^2 + 
      48 \zeta(3) \right) C_F + \left(\frac{121}{9} + \frac{2}{3}\pi^2\right) \beta_0  \right]  \nonumber \\
\gamma_2^J &=&  
2 C_F \left[ C_A^2 \left(\frac{52019}{162}-\frac{841 \pi ^2}{81}-\frac{82 \pi^4}{27} -\frac{2056 \zeta (3)}{9}+\frac{88 \pi ^2 \zeta (3)}{9}+232 \zeta (5) \right)   \nonumber \right.  \\ && \left.   
            +C_A C_F \left(\frac{151}{4}-\frac{205 \pi ^2}{9}-\frac{247 \pi^4}{135} +  \frac{844 \zeta (3)}{3}+\frac{8 \pi ^2 \zeta (3)}{3}+120 \zeta (5) \right)    \nonumber \right.  \\ && \left.
            +C_F^2 \left(\frac{29}{2}+3 \pi ^2+\frac{8 \pi^4}{5} + 68 \zeta(3)-\frac{16 \pi ^2 \zeta (3)}{3}-240 \zeta (5)  \right) 				     \nonumber \right.  \\ && \left.
            +\beta_0 C_A \left(-\frac{7739}{54}+\frac{325 \pi ^2}{81}+\frac{617 \pi ^4}{270}-\frac{1276 \zeta (3)}{9}\right)
            +\beta_0^2 \left(-\frac{3457}{324}+\frac{5 \pi^2}{9}+\frac{16 \zeta (3)}{3}\right) 									     \nonumber \right.  \\ && \left. 
            +\beta_1 \left(\frac{1166}{27}-\frac{8 \pi ^2}{9}-\frac{41 \pi ^4}{135}+ \frac{52 \zeta (3)}{9}\right)\right].
\end{eqnarray}
The non-cusp anomalous dimensions $\gamma_M$ and $\gamma_S$ are known at two loop \cite{Neubert:2007je,Becher:2005pd}  
\begin{eqnarray}
\gamma_0^S &=& 2 C_F \nonumber \\
\gamma_1^S &=&  -C_F \left[C_A \left(\frac{110}{27}+\frac{\pi ^2}{18}-18 \zeta (3)\right)+  T_F n_f \left(\frac{8}{27}+\frac{2 \pi ^2}{9}\right) \right], \nonumber
\\
\gamma_0^M &=&  C_F \nonumber \\
\gamma_1^M &=&  C_F  \left[C_F \left( \frac{3}{2} - 2 \pi^2 + 24 \zeta(3)\right) + C_A \left(\frac{373}{54}  + \frac{5}{2} \pi^2 - 30 \zeta(3)\right) 
- T_F n_f \left( \frac{10}{27} + \frac{2}{3} \pi^2 \right)
\right]\,. \nonumber \\
\end{eqnarray}
The three-loop non-cusp anomalous dimension $\gamma_2^S$ and $\gamma_2^M$ are not known.
Their sum is  constrained to be equal to the $x \rightarrow 1$ limit of the anomalous dimension of the fragmentation function. From Ref.  \cite{Moch:2004pa}, we get 
\begin{eqnarray}\label{gamma2MS}
\gamma_2^M + \gamma_2^S &=& 
  C_F^3 \left(\frac{29}{2}+3 \pi ^2+\frac{8 \pi ^4}{5} + 68 \zeta (3)-\frac{16 \pi ^2 \zeta (3)}{3}-240 \zeta (5) \right)  \nonumber \\ &&
+ C_A^2 C_F \left(-\frac{1657}{36}+\frac{2248 \pi ^2}{81}-\frac{\pi ^4}{18} -\frac{1552 \zeta (3)}{9}+40 \zeta (5) \right)  					\nonumber  \\ && 
+C_A C_F^2 \left(  \frac{151}{4}-\frac{205 \pi ^2}{9}-\frac{247 \pi ^4}{135} +  \frac{844 \zeta (3)}{3}+\frac{8 \pi ^2 \zeta (3)}{3}+120 \zeta (5) \right) \nonumber  \\ && 
+C_A C_F  T_F n_f \left(40-\frac{1336 \pi ^2}{81}+\frac{2\pi ^4}{45} +  \frac{400 \zeta (3)}{9}\right)  \nonumber \\ && 
+C_F^2  T_F n_f \left(-46+\frac{20 \pi ^2}{9}+\frac{116 \pi ^4}{135} -\frac{272 \zeta(3)}{3} \right) \nonumber \\
& & +C_F  T_F^2 n_f^2 \left(-\frac{68}{9}+\frac{160 \pi ^2}{81}-\frac{64 \zeta (3)}{9}\right).
\end{eqnarray}
In our error analysis, we use
\begin{align}
\gamma_2^S(n_f) = (1+e_\gamma)c_S\frac{(\gamma_1^S)^2}{\gamma_0^S}\,, \\
\gamma_2^M(n_f) = c_M\frac{(\gamma_1^M)^2}{\gamma_0^M} - e_\gamma c_S\frac{(\gamma_1^S)^2}{\gamma_0^S}\,,
\end{align}
with one common theory parameter $e_\gamma$ varied between $-2$ and $2$. The coefficients $c_S$ and $c_M$ are set by 
imposing Eq. \eqref{gamma2MS}  for both $n_f=4$ and $n_f=5$.
For the default $e_\gamma=0$ we get
\begin{align}
\gamma_2^s(5) = 1551.42 \,,\quad \gamma_2^s(4) = 1638.79 \,,\nn\\
\gamma_2^m(5) = -643.011 \,,\quad \gamma_2^m(4) = -402.858 \,.\nn
\end{align}

\subsection{Renormalon subtraction}\label{renormalon}

The renormalon subtracted perturbative shape function is given by \cite{Ligeti:2008ac}
\begin{equation}\label{ren}
\hat{S}_{Q/Q}(\hat\omega) = \left[ 1 + \delta m_Q \bar n \cdot v\, \frac{d}{d\hat\omega} + \left( \frac{(\delta m_Q)^2}{2} - \frac{\delta \lambda_1}{6}\right) \bar n \cdot v^2 \frac{d^2}{d\hat\omega^2} \right] S_{Q/Q}(\hat\omega).
\end{equation}
In the code, it is convenient to integrate by parts, and have the derivatives acting on the model $S_{H/Q}^{hadr}$.
As we remarked in Section \ref{sec:np}, Eq. \eqref{ren} was originally derived for the shape function in $B$ decays, but it can be applied to the fragmentation shape function. 
$\delta m_Q$ and $\delta \lambda_1$ are the shifts from infrared sensitive to infrared safe quantities. Here we used the 1S scheme for the heavy quark mass \cite{Hoang:1998ng,Hoang:1998hm}, and the  
``invisible scheme'', introduced in Ref.~\cite{Ligeti:2008ac}, for the $B$ meson kinetic energy.
At the order we are working
\begin{align}
\delta m_{Q} = R_{1S} \frac{C_F\alpha_s(\mu_S)}{8} \left( 1 
               + \frac{\alpha_s(\mu_S)}{\pi} \left( \left(\ln{\frac{\mu_S}{R_{1S}}}+\frac{11}{6}\right)\beta_0- \frac{4}{3}C_A  \right) \right)
\end{align}
 with $R_{1S}=m_Q^{1S}C_F\alpha_s(\mu_S)$ and  $m_Q^{1S} = 4.66$ GeV \cite{Agashe:2014kda}.
\begin{align}
\delta\lambda_1 = R^2\frac{C_F C_A\alpha_s^2(\mu_S)}{4\pi^2} \left( \frac{\pi^2}{3} - 1 \right),
\end{align}
where $R$ is a dimensionful quantity. We take $R=1$ GeV, and do not vary it in the analysis of theoretical errors. 
The number of flavors in $\beta_0$ is $n_f=4$. 

\section{Python code}\label{python}

Along with this publication, we release a python program for the DGLAP evolution of the $b$-quark fragmentation function \footnote{The program is available as ancillary file in the \texttt{arXiv} submission of this paper.}
The program was written for python $2.7$ and consists of five files.

The computation is executed by running the main file ``QCDcalc.py'' with the python interpreter. To use the parallel capabilities of the program the package SCOOP~\cite{SCOOP_XSEDE2014} has to be loaded together with python, \texttt{ python -m scoop QCDcalc.py }. This will automatically distribute the computation among all available local cpus. For more options and how to include remote machines we refer to the documentation of SCOOP~\cite{SCOOP_XSEDE2014}.
The file ``QCDcalc.py'' also contains a section ``OPTIONS'' where all theory parameters and some numerical switches are collected to configure the computation. In addition, ``QCDcalc.py'' contains routines to do the calculation described in Section~\ref{sec:fullx} and a routine for the solution of the DGLAP equation based on the brute force approach in Ref.~\cite{Miyama:1995bd}.

``from\_fortran'' is a python module created with f2py~\cite{f2py} from the Fortran code for the numerical evaluation of harmonic polylogarithms~\cite{Gehrmann:2001pz}, the Fortran code~\cite{Mitov:2006ic} of the exact 2-loop $\msbar$ non-singlet coefficient functions for the fragmentation function  $F_L$~\cite{Rijken:1996vr} and $F_T$~\cite{Rijken:1996ns}, the Fortran code for the exact 3-loop $\msbar$ non-singlet splitting functions $P_{NS}^{(2)}$~\cite{Moch:2002sn,Moch:2004pa} and the Fortran code for the differences between the time-like and space-like non-singlet splitting functions at second and third order in $\alpha_s$ from Ref.~\cite{Mitov:2006ic}.

``physics.py'' contains all physics expressions including the hard function, soft function and renormalization group evolution kernel. Hence, all equations used for the calculations in Section~\ref{sec:fullx} can be found in this file.

``convolution.py'' contains integration routines and a function to numerically calculate convolutions in momentum fraction space. These routines can be used for convolutions of expressions other than the ones in ``physics.py''.

``mytools.py'' contains some useful tools like a routine for parallelization using the package SCOOP~\cite{SCOOP_XSEDE2014} and a simple progress counter.


\bibliographystyle{JHEP} 
\bibliography{HQFFnew}

\end{document}